\documentclass{aa}

\pdfpageattr{/Group <</S /Transparency /I true /CS /DeviceRGB>>}

\def\GA{\mathrel{\raisebox{0.13\baselineskip}{\hbox{\rlap{\hbox{\lower0.41\baselineskip\hbox{$\sim$}}}\hbox{$>$}}}}}
\def\LA{\mathrel{\raisebox{0.13\baselineskip}{\hbox{\rlap{\hbox{\lower0.41\baselineskip\hbox{$\sim$}}}\hbox{$<$}}}}}

\def\apjl{ApJ}% Astrophysical Journal, Letters
\def\apjs{ApJS}% Astrophysical Journal, Supplement
\def\ao{Appl.~Opt.}% Applied Optics
\def\aap{A\&A}% Astronomy and Astrophysics
\usepackage{graphicx}
\usepackage{txfonts}
\usepackage{amsmath}	% Advanced maths commands
\usepackage{amssymb}	% Extra maths symbols
\usepackage{pdflscape}

\usepackage[dvipsnames]{xcolor}
\usepackage{xfrac}
\usepackage{algorithm2e}
\usepackage{multicol}
\usepackage{multirow}
\usepackage{bigstrut}
\usepackage[normalem]{ulem}

\usepackage[english]{babel}
\usepackage{microtype}

\usepackage{lipsum}

\newcommand{\norm}[1]{\left\lVert#1\right\rVert}
\newcommand{\oii}{\mbox{O\,{\scshape ii}}}
\newcommand{\oiii}{\mbox{O\,{\scshape iii}}}

\newcommand{\ciii}{\mbox{C\,{\scshape iii}}}
\newcommand{\civ}{\mbox{C\,{\scshape iv}}}

\newcommand{\nv}{\mbox{N\,{\scshape v}}}
\newcommand{\hii}{\mbox{H\,{\scshape ii}}}

\usepackage{supertabular}

\usepackage[caption=false]{subfig}
\usepackage{float}

\begin{document} 
\label{firstpage}
\selectlanguage{english}

\title{An atlas of MUSE observations towards twelve massive lensing clusters\footnote{Tables A.1 and A.2 are available in electronic form
at the CDS via anonymous ftp to cdsarc.u-strasbg.fr (130.79.128.5)
or via \url{http://cdsweb.u-strasbg.fr/cgi-bin/qcat?J/A+A/}}}

\titlerunning{MUSE Lensing Clusters survey}
\authorrunning{Richard et al.}

%   \author{G. Wuchterl
%           \inst{1}
%           C. Ptolemy\inst{2}\fnmsep\thanks{Just to show the usage
%           of the elements in the author field}
%           }
\author{
Johan Richard\inst{1},
Ad\'ela\"\i de Claeyssens\inst{1},
David Lagattuta\inst{1,2,3},
Lucia Guaita\inst{4,5},
Franz Erik Bauer\inst{4,6,7}, % https://orcid.org/0000-0002-8686-8737
Roser Pello\inst{8,9},
David Carton\inst{1},
Roland Bacon\inst{1},
Genevi\`eve Soucail\inst{8},
Gonzalo Prieto Lyon\inst{4,6},
Jean-Paul Kneib\inst{10,9},
Guillaume Mahler\inst{1,11}, %0000-0003-3266-2001
Benjamin Cl\'ement\inst{1,10},
Wilfried Mercier\inst{8},
Andrei Variu\inst{10},
Am\'elie Tamone\inst{10},
Harald Ebeling\inst{12}, Kasper B. Schmidt\inst{13}, 
Themiya Nanayakkara\inst{14,15},
Michael Maseda\inst{14},
Peter M. Weilbacher\inst{13},
Nicolas Bouch\'e\inst{1},
Rychard J. Bouwens\inst{14},
Lutz Wisotzki\inst{13},
Geoffroy de la Vieuville\inst{8},
Johany Martinez\inst{1},
Vera Patr\'\i cio\inst{1}
}

\institute{Univ Lyon, Univ Lyon1, Ens de Lyon, CNRS, Centre de Recherche Astrophysique de Lyon UMR5574, 69230, Saint-Genis-Laval, France\\
              \email{johan.richard@univ-lyon1.fr}
\and
% %DURHAM
Centre for Extragalactic Astronomy, Durham University, South Road, Durham DH1 3LE, UK
\and
Institute for Computational Cosmology, Durham University, South Road, Durham DH1 3LE, UK
\and
% % PUC
Instituto de Astrof{\'{\i}}sica and Centro de Astroingenier{\'{\i}}a, Facultad de F{\'{i}}sica, Pontificia Universidad Cat{\'{o}}lica de Chile, Casilla 306, Santiago 22, Chile
\and
N\'ucleo de Astronom\'ia, Facultad de Ingenier\'ia, Universidad Diego Portales, Av. Ej\'ercito 441, Santiago, Chile 
\and
% % MAS
Millennium Institute of Astrophysics (MAS), Nuncio Monse{\~{n}}or S{\'{o}}tero Sanz 100, Providencia, Santiago, Chile
\and
% % SSI
Space Science Institute, 4750 Walnut Street, Suite 205, Boulder, Colorado 80301
\and
% % IRAP
Institut de Recherche en Astrophysique et Plan\'etologie (IRAP), Universit\'e de Toulouse, CNRS, UPS, CNES, 14 Av. Edouard Belin, F-31400 Toulouse, France
\and
% % LAM
Aix Marseille Universit\'e, CNRS, CNES, LAM (Laboratoire d’Astrophysique de Marseille), UMR 7326, 13388, Marseille, France
\and
% EPFL
Institute of Physics, Laboratory of Astrophysics, Ecole Polytechnique F\'ed\'erale de Lausanne (EPFL), Observatoire de Sauverny, 1290 Versoix, Switzerland.
\and
Department of Astronomy, University of Michigan, 1085 South University Ave, Ann Arbor, MI 48109, USA
\and
Institute for Astronomy, University of Hawaii, 2680 Woodlawn Dr, Honolulu, HI 96822, USA
\and
 Leibniz-Institut für Astrophysik Potsdam (AIP), An der Sternwarte 16, 14482, Potsdam, Germany
 \and
 Leiden Observatory, Leiden University, P.O. Box 9513, 2300 RA Leiden, The Netherlands
 \and
 Centre for Astrophysics and Supercomputing, Swinburne University of Technology, Hawthorn, VIC 3122, Australia
 }

   \date{Received YYYY MMMMM DD accepted YYYY MMMMM DD}

% \abstract{}{}{}{}{} 
% 5 {} token are mandatory
 
  \abstract
{Spectroscopic surveys of massive galaxy clusters reveal the properties of faint background galaxies thanks to the  magnification provided by strong gravitational lensing.}
{We present a systematic analysis of integral-field-spectroscopy observations of 12 massive clusters, conducted with the Multi Unit Spectroscopic Explorer
 (MUSE). All data were taken under very good seeing conditions ($\sim$0\farcs6) in effective exposure times between two and 15 hrs per pointing, for a total of 125 hrs. Our observations cover a total solid angle  of $\sim$23 arcmin$^2$ in the direction of clusters, many of which were previously studied by the MAssive Clusters Survey (MACS), Frontier Fields (FFs), Grism Lens-Amplified Survey from Space (GLASS) and Cluster Lensing And Supernova survey with Hubble (CLASH) programmes. The achieved emission line detection limit at 5$\sigma$ for a point source varies between (0.77--1.5)$\times$10$^{-18}$ erg\,s$^{-1}$\,cm$^{-2}$ at 7000\AA.}
{We present our developed strategy to reduce these observational data, detect continuum sources and line emitters in the datacubes, and determine their redshifts. We constructed robust mass models for each cluster to further confirm our redshift measurements using strong-lensing constraints, and identified a total of 312 strongly lensed sources producing 939 multiple images.}
{The final redshift catalogues contain more than 3300 robust redshifts, of which 40\% are for cluster members and $\sim$30\% are for lensed Lyman-$\alpha$ emitters. Fourteen percent of all sources are line emitters that are not seen in the available {\it HST} images, even at the depth of the FFs ($\sim29$ AB). We find that the magnification distribution of the lensed sources in the high-magnification regime ($\mu{=}$ 2--25) follows the theoretical expectation of $N(z)\propto\mu^{-2}$. The quality of this dataset, number of lensed sources, and number of strong-lensing constraints enables detailed studies of the physical properties of both the lensing cluster and the background galaxies. The full data products from this work, including the datacubes, catalogues, extracted spectra, ancillary images, and mass models, are made available to the community.}{}


   \maketitle

%%%%%%%%%%%%%%%%%%%%%%%%%%%%%%%%%%%%%%%%%%%%%%%%%%

%%%%%%%%%%%%%%%%% BODY OF PAPER %%%%%%%%%%%%%%%%%%

\section{Introduction}

% Cluster lensing
% Advantages of magnification 
% MUSE+lensing
% This paper describes....

Strong gravitational lensing by massive galaxy clusters leads to the magnification of sources lying behind them, and this amplification can reach very large factors in the cluster cores ($\mu{\sim}$5--10, \citealt{2011A&ARv..19...47K}), and even higher factors for images in the vicinity of the so-called critical lines \citep{1998MNRAS.298..945S}. For this reason, massive clusters are sometimes referred to as nature's telescopes since the combined power of large diameter telescopes and gravitational magnifications provide us with the best views of background galaxies in the distant Universe. Since the first spectroscopic confirmation of a giant gravitational arc was reported \citep{1988A&A...191L..19S}, high resolution images from the \textit{Hubble Space Telescope} (HST) have significantly contributed to the success of lensing clusters, with the discovery of a high density of multiple images in deep observations (e.g.  \citealt{1996ApJ...471..643K,2005ApJ...621...53B,2015MNRAS.452.1437J}). Indeed, multiply-imaged systems give us the most precise constraints on the mass distribution in the cluster cores, and consequently the magnification factors. 

Multi-object spectrographs on 8-10m class telescopes have helped start large spectroscopic campaigns to confirm the lensed nature of very distant galaxies and the identification of multiple images (e.g. \citealt{2001A&A...378..394C,2011ApJS..193....8B}). 
However these large spectroscopic campaigns were largely limited by the crowding of galaxy clusters in their central regions, which leads to strong contamination between cluster galaxies and background sources as well as an inefficient use of multi-object spectrographs. In addition, the redshift distribution of lensed sources peaks at $z>1.5$ in the redshift desert, where only the brightest UV-selected galaxies can be confirmed in the optical \citep{2007ApJ...668..643L}. Because of these limitations, typically only a small number of multiply-imaged systems (typically $<10$) have been spectroscopically confirmed in a given cluster with such instruments (e.g. \citealt{2010MNRAS.404..325R}). Several observing campaigns have focused on near-infrared spectroscopy to avoid the redshift desert and complement optical observations (e.g. HST grism \citealt{2015ApJ...812..114T}, or Keck/MOSFIRE \citealt{2015ApJ...813...37H}).

The advent of the Multi Unit Spectroscopic Explorer (MUSE, \citealt{2010SPIE.7735E..08B}) on the Very Large Telescope (VLT) has revolutionised the study of strong lensing galaxy clusters. MUSE is a panoramic integral field spectrograph, fully covering a 1 arcmin$^2$ field of view with spectroscopy in the optical range (475-930nm). Together with its very high throughput (40\% end-to-end including thetelescope and atmosphere), medium resolution (R$\sim$3000), and fine spatial sampling (0\farcs2), MUSE is very well-suited for crowded field spectroscopy \citep{2018A&A...618A...3R}, and more specifically in galaxy cluster cores. Its pairing with the ground-layer Adaptive Optics (AO) system in 2017 \citep{2017Msngr.170...20L} has improved its observing efficiency even further.  

The versatile capabilities of MUSE on galaxy cluster fields were demonstrated almost immediately: in commissionning \citep{2015MNRAS.446L..16R}, science verification \citep{2015A&A...574A..11K} and regular observations (e.g. \citealt{2016MNRAS.457.2029J,2016ApJ...822...78G,2017A&A...600A..90C,2017MNRAS.469.3946L,2018MNRAS.473..663M,2019MNRAS.485.3738L}). Most notably, it has been very successful to follow up on the Frontier Field clusters (hereafter FFs; \citealt{2017ApJ...837...97L}), a programme initiated by STScI to get deep observations of six massive clusters with the \textit{Hubble} ({\it HST}) and \textit{Spitzer} space telescopes. But MUSE was also very successful to follow up on clusters with shallower {\it HST} images (e.g. \citealt{2019MNRAS.483.3082J,2019ApJ...873...96M,2020A&A...635A..98R}).

This success has pushed several teams to analyse MUSE observations of known massive clusters such as the Cluster Lensing and Supernova Survey with Hubble (CLASH, \citealt{2012ApJS..199...25P}) programme \citep{ 2017arXiv170309239R,2019AandA...632A..36C,2020arXiv200610700J}. Indeed, the richness of the MUSE spectroscopic datasets have a strong legacy aspect, for example to identify small-scale gravitational lenses in the clusters \citep{2020Sci...369.1347M}, or  cross-match with multiwavelength observations of the same fields (e.g. with ALMA; \citealt{2017A&A...604A.132L}; \citealt{Fujimotosubmitted} submitted).%Fujimoto et al. 2020 submitted).

In this paper, we present a full analysis of 12 lensing clusters, totalling more than 125 hours of exposure time with the MUSE instrument. These observations are in majority taken as part of the MUSE Guaranteed Time Observations (GTO) programme, but are complemented by additional MUSE datasets publicly available on the same clusters or following a similar target selection. We have benefited from many years of developments in MUSE analysis tools as part of the MUSE GTO programmes \citep{2017A&A...608A...1B,2017A&A...608A...2I,2019ASPC..521..545P} to improve the data reduction, source detection and analysis. The results of this analysis are made available in the form of a public data release.

This paper is organised as follows. Section \ref{sec:obs} presents the target selection, observations and data reduction for the MUSE and ancillary datasets used in our analysis. Section \ref{sec:building} describes the construction of the spectroscopic catalogues contained in the data release and Sect.~\ref{sec:massmodels} describes the mass models we use to estimate the magnification and source properties. We provide an overview of the full spectroscopic catalogue and a few science highlights in Sect.~\ref{sec:sources}, and give our conclusions in Sect.~\ref{sec:conclusion}.

Throughout the paper, we assume a standard $\Lambda$-Cold Dark Matter (CDM) cosmological model with $\Omega_{M}=0.3$, $\Omega_{\Lambda}=0.7$, and $H_{0}=70$\,km\,s$^{-1}$\,Mpc$^{-1}$ whenever necessary. 
At the typical redshift of the lensing clusters  ($z=0.4$), 1\arcsec\ covers a physical distance of 5.373\,kpc. All magnitudes are given in the  AB system.

\section{Observations and data reduction}
\label{sec:obs}
\subsection{Cluster sample}
%The observations we use in our analysis are targetting the cores of massive galaxy clusters, and are taken both under the MUSE Lensing Cluster GTO programme or from existing archival data on a similar cluster selection.
%The observations used in our analysis target the cores of massive galaxy clusters and are gathered from both the MUSE Lensing Cluster GTO programme and from existing archival data that utilize a similar cluster selection.
%Cluster fields for the MUSE GTO programme were selected based on their strong-lensing magnification of background sources, in particular LAEs at $z>3$ which were expected to be detected within the MUSE spectral range.

%We started from a master list of massive X-ray luminous clusters mainly from the ROSAT Brightest Cluster Sample \citep[BCS,][]{1998MNRAS.301..881E,2000MNRAS.318..333E} and the MAssive Cluster Survey (MACS, \citealt{2001ApJ...553..668E}), as well as its southern counterpart (SMACS). Valuable follow-up imaging with the \textit{Hubble Space Telescope} (HST) was performed in particular for MACS and SMACS clusters for {\it HST} SNAPshot programmes GO-10491, -10875, 11103, -12166, and -12884 (PI Ebeling), allowing the identification of strong-lensing features in the form of arcs and arclets in almost every single target (e.g. \citealt{2007ApJ...661L..33E,2016MNRAS.457.1399R}).

The observations used in our analysis focus on the cores of massive galaxy clusters from both the MUSE Lensing Cluster GTO programme and available archival programmes that target similar clusters. Cluster fields in the MUSE GTO programme were chosen based on their strong-lensing efficiency at magnifying background sources, in particular Lyman-$\alpha$ Emitters (LAEs) at $z>3$ expected to be detected within the MUSE spectral range. 

We compile our sample from a master list of massive X-ray luminous clusters, mainly from the {\it ROSAT} Brightest Cluster Sample \citep[BCS,][]{1998MNRAS.301..881E,2000MNRAS.318..333E} and the MAssive Cluster Survey (MACS, \citealt{2001ApJ...553..668E}), along with its southern counterpart, SMACS. Valuable follow-up imaging with the {\it HST} was performed in particular for MACS and SMACS clusters for {\it HST} SNAPshot programmes GO-10491, -10875, -11103, -12166, and -12884 (PI Ebeling), allowing the identification of strong-lensing features in the form of arcs and arclets in almost every single target (e.g. \citealt{2007ApJ...661L..33E,2016MNRAS.457.1399R}).

To build the sample, we chose targets based on the following selection criteria: firstly a cluster redshift $0.2<z_{\rm cl}<0.6$, to ensure that the main spectral signatures (K,H absorption lines and 4000 break) of cluster members are located in a low-background region of the MUSE spectrum; secondly, a wide range in Right Ascension (R.A.), to allow easier scheduling with respect to the rest of MUSE GTO observations, and a transit at low airmass ($<1.25$) as seen from Cerro Paranal Observatory, corresponding to declinations $-60<$ Dec $<+10$ degrees. Thirdly, we require at least one existing  high-resolution broad-band {\it HST} image (in either the F606W or F814W filter to ensure overlap with the MUSE spectral range), that shows bright arcs and multiple images; these images are critical to pinpoint the location of sources sufficiently bright in the continuum. Finally, we require a preliminary mass model based on {\it HST} images and spectroscopic confirmation of at least one multiply imaged system to roughly estimate the total mass in the cluster core. From this crude map the MUSE observations could be designed to efficiently cover the critical lines and multiple-image region.

For clusters observed in the framework of the MUSE GTO, the choice of the precise centre for the pointing, or the mosaic configuration, were determined in such a way that the observed area included as far as possible the tangential critical lines, in order to maximise the number of strongly magnified and multiple images (see also Sect.\,\ref{sec:mulimages}).

We present here the results for a set of 12 clusters selected through this process, all of which were analysed in a uniform manner. The combination of the aforementioned criteria makes our selection similar to the one used in other cluster surveys such as CLASH and the FFs. It is therefore not surprising that half of the clusters we selected %were studied previously by 
overlap with these two programmes.

The main properties of the 12 selected clusters are summarised in Table \ref{tab:clusters_table}. Additional clusters observed as part of the GTO programme will be analysed following the same procedure as described in this paper, and included in a future data release. 

\begin{table*}
    \centering
    \begin{tabular}{l|rrcll}
Cluster & R.A. & Dec. & $z_{\rm cl}$ & Notes & Model Reference\\
 & (J2000) & (J2000) & & & \\
\hline
Abell 2744  &  00:14:20.702  & $-$30:24:00.63  &  0.308 & MACS, FF & \citet{2015MNRAS.452.1437J}\\
Abell 370  &  02:39:53.122 & $-$01:34:56.14  &  0.375 & FF & \citet{2010MNRAS.404..325R}\\
MACS\,J0257.6$-$2209  &  02:57:41.070 & $-$22:09:17.70  &  0.322 & MACS & \citet{2018MNRAS.479..844R}\\
MACS\,J0329.6$-$0211  &  03:29:41.568 & $-$02:11:46.41 &  0.450 & MACS, CLASH & \citet{2012ApJ...747L...9Z} \\
MACS\,J0416.1$-$2403  &  04:16:09.144 & $-$24:04:02.95  &  0.397 & MACS, CLASH, FF & \citet{2014MNRAS.443.1549J} \\
1E 0657$-$56 (Bullet)  &  06:58:38.126 & $-$55:57:25.87  &  0.296 & & \citet{2016AandA...594A.121P}\\
MACS\,J0940.9+0744  &  09:40:53.698 & $+$07:44:25.31  &  0.335 & MACS & \citet{2016ApJ...831..152L}\\
MACS\,J1206.2$-$0847  &  12:06:12.149 & $-$08:48:03.37  &  0.438 & MACS, CLASH & \citet{2009MNRAS.395.1213E}\\
RX\,J1347.5$-$1145  &  13:47:30.617 & $-$11:45:09.51 &  0.451 & MACS, CLASH & \citet{2008AandA...481...65H}\\
SMACS\,J2031.8$-$4036 &  20:31:53.256 & $-$40:37:30.79  &  0.331 & MACS & \citet{2012MNRAS.427.1953C}\\
SMACS\,J2131.1$-$4019  &  21:31:04.831 & $-$40:19:20.92  &  0.442 & MACS & \citet{2018MNRAS.479..844R}\\
MACS\,J2214.9$-$1359  &  22:14:57.292 & $-$14:00:12.91  &  0.502 & MACS & \citet{2007ApJ...661L..33E}\\
    \end{tabular}
    \medskip\par
    \caption{Summary of selected lensing clusters. The reference WCS location corresponds to the brightest cluster member of the (sub-)cluster targetted by the MUSE observations. We mention in the notes whether the target belongs to the MACS, Frontier Fields \citep{2017ApJ...837...97L} or CLASH \citep{2012ApJS..199...25P} surveys, as well as other relevant publications about preliminary mass models or known lensed arcs in these clusters prior to the MUSE observations. The rightmost columns give the reference to the lens model used to design the MUSE observations.}
    \label{tab:clusters_table}
\end{table*}

\subsection{MUSE GTO survey}
The MUSE Lensing Cluster project (PI: Richard) is a multi-semester programme, which has been running since ESO semester P94 starting in Fall 2014. It targets the central regions of the massive lensing clusters introduced in the previous section, with an effective exposure time of $2-10$ hrs per pointing.  Data are acquired in Wide Field Mode (WFM), using both standard (WFM-NOAO-N) and adaptive-optics (WFM-AO-N) observations -- the latter configuration having been available since Fall 2017, following the commissioning of the ground layer Adaptive Optics (AO) correction. Although the AO mode improves the seeing, we note that WFM-AO-N datasets have a gap in the 5800-5980 \AA\ wavelength range due to the AO notch filter.  Additionally, while nearly all of the data are acquired in Nominal mode, covering the 4750 - 9350 \AA\ wavelength range, one exception is the Bullet cluster, for which observations were taken in Extended mode (WFM-NOAO-E) covering bluer wavelengths (See Sect.~\ref{sec:datared}).

\begin{figure*}
    \centering
    \includegraphics[width=19cm]{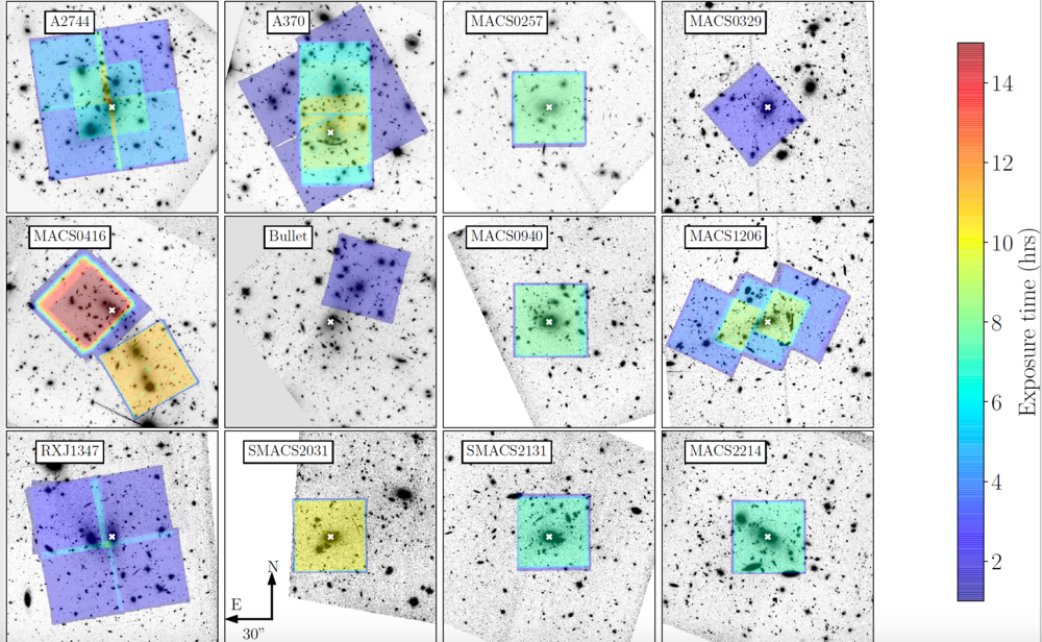}
    \caption{MUSE exposure maps overlayed on top of the {\it HST} / F814W images. The 12 clusters from Table~\ref{tab:clusters_table} are displayed in reading order, with the reference position from the same table marked as a white cross. }
    \label{fig:overview}
\end{figure*}

The location and orientation of each MUSE pointing has been chosen to maximise the coverage of the multiply-imaged regions (Fig.~\ref{fig:overview}), and also to guarantee the availability of suitable Tip-Tilt stars for observations taken in WFM-AO-N mode. In the case of FF and CLASH clusters, we adopted a coverage of larger mosaics of multiple contiguous MUSE pointings to increase the legacy value of these datasets, in coordination with non-GTO programmes (see below). 

The observations were split into blocks of $\sim$1-1.15 hr execution time. Each observing block consists in either 2$\times$1800 sec., 3$\times$900 sec or 3$\times$1000 sec. exposures. We included a small spatial dithering box ($<$0\farcs3) as well as 90 degrees rotations of the instrument between each exposure. Indeed this strategy has been shown to help reduce the systematics due to the IFU image slicer \citep{2015A&A...575A..75B}. A summary of all exposures taken is provided in Table \ref{tab:museobs}.

Standard calibrations have been used for this programme, including day-time instrument calibrations as well as standard star observations.
All exposures taken after the 2nd MUSE commissioning in June 2014 (i.e. all cluster fields presented here except for SMACS2031) include single internal flat-field exposures taken with the instrument as night calibrations. These short (0.35 sec.) exposures are used for an illumination correction and taken every hour, or whenever there is a sudden temperature change in the instrument. These calibrations are important to correct for time and temperature dependence on the flat-field calibration between each slitlet throughout the night. In addition, twilight exposures are taken every few days and are used to produce an on-sky  illumination correction between the 24 channels.

\begin{table*}
    \centering
    \begin{tabular}{l|l|l|l|l|l|l}
Cluster & Prog. ID & Obs. Date & Exp. Time (Mode) & PSF & T$_{\rm eff}$ & Notes \\
 & & [UT] & [s] & \arcsec & [hr] & \\
\hline
A2744  &  094.A-0115, 095.A-0181,  & 2014-09-21 -- 2015-11-09 & 40$\times$1800 (NOAO) & 0.61 & 3.5-7 & 2$\times$2 mosaic\\
& 096.A-0496 & & & & & \\
A370  & 094.A-0115, 096.A-0710  &  2014-11-20 -- 2016-09-28 & 4$\times$1800 (NOAO) &0.66 & 1.5-8.5 & 2$\times$2 mosaic\\
      &                         &                           & 37$\times$962 (NOAO) & & & \\
      &                         &                           & 24$\times$930 (NOAO) & & & \\
      &                         &                           & 3$\times$953 (NOAO) & & & \\
MACS0257  & 099.A-0292, 0100.A-0249, & 2017-09-20 -- 2019-09-28 & 6$\times$1000 (NOAO) &0.52 & 8 &  \\
  & 0103.A-0157 & & 24$\times$1000 (AO) & &   \\
MACS0329  &  096.A-0105 &  2016-01-09 -- 2016-01-29 & 6$\times$1447 (NOAO) &0.69 & 2.5 & \\
MACS0416 (N) &  094.A-0115, 0100.A-0763 & 2014-12-17 -- 2019-03-04 & 4$\times$1800 (NOAO) &0.53 & 17 \\
              &                          &                          & 6$\times$1670 (NOAO) & & \\
              &                          &                          & 27$\times$1670 (AO) & & \\
MACS0416 (S) &  094.A-0525  &  2014-10-02 -- 2015-02-24 & 50$\times$700 (NOAO) & 0.65& 11-15 &\\
              &                          &                          & 8$\times$667 (NOAO) & & \\
Bullet  & 094.A-0115 & 2014-12-18 & 4$\times$1800 (NOAO)$^{(1)}$ & 0.56& 2 &  \\
MACS0940  &  098.A-0502, 098.A-0502, & 2017-01-30 -- 2018-05-13 & 3$\times$1000 (NOAO) & 0.57& 8 & \\
          & 0101.A-0506 & & 30$\times$900 (AO)$^{(2)}$ & & \\
MACS1206  &  095.A-0181, 097.A-0269 & 2015-04-15 -- 2016-04-09 & 26$\times$1800 (NOAO)$^{(3)}$ &0.52 & 4-9 & 3$\times$1 mosaic\\
RXJ1347  &  095.A-0525, 097.A-0909  & 2015-07-16 -- 2018-03-21 & 8$\times$1475 (NOAO) &0.55 & 2-3 &2$\times$2 mosaic\\
              &                          &                          & 18$\times$1345 (AO) & & \\
SMACS2031 & 60.A-9100$^{(4)}$ & 2014-04-30 -- 2014-05-07 & 33$\times$1200 (NOAO)  & 0.79& 10 & \\
SMACS2131  &  0101.A-0506, 0102.A-0135, & 2018-08-13 -- 2019-09-30 & 30$\times$900 (AO) & 0.59 & 7 & \\
& 0103.A-0157 &  & & & & \\
MACS2214  & 099.A-0292,0101.A-0506,  & 2017-09-21 -- 2019-10-22 & 3$\times$1000 (NOAO) & 0.55& 7 & \\
  & 0103.A-0157, 0104.A-0489 & & 27$\times$900 (AO) & & & \\
\hline
    \end{tabular}
$^{(1)}$Taken in extended mode (WFM-NOAO-E).
$^{(2)}$1 exposure stopped after 740 sec.

$^{(3)}$1 exposure stopped after 1405 sec.
$^{(4)}$MUSE commissioning run .
\medskip\par
    \caption{Summary of MUSE observations analysed on the 12 clusters. Note that the two datasets on MACS0416 cluster (N)orth and (S)outh have been treated separately. From left to right: MUSE dataset, ESO Programs, range of observing dates, list of exposure times and instrument modes, final image quality of the combined datacube (FWHM at 700 nm), range of effective exposure time per spaxel, notes on multiple pointings (see also Fig.~\ref{fig:overview}). }
    \label{tab:museobs}
\end{table*}

\subsection{MUSE archival data}

The importance of getting deep MUSE exposures on the FFs has led us to coordinate a joint effort between the GTO programme and additional programmes towards MACS0416 and Abell 370 (ESO programmes 094.A-0525 and 096.A-0710 respectively, PI: Bauer). Overall, the same  strategy as for the GTO campaign has been followed for these observations, and were combined with GTO data when overlapping. Similarly, additional exposure time in the northern part of the FF cluster MACS0416 has been obtained as part of ESO programme 0100.A-0764 (PI: Vanzella, \citealt{2020MNRAS.494L..81V}), and combined with the GTO observations.

Finally, we include in our analysis two CLASH clusters observed as part of ESO programmes 095.A-0525, 096.A-0105, 097.A-0909 (PI: Kneib) on MACS0329 and RXJ1347, again with very similar science goals and observing strategy to the GTO programme. A first analysis on these datasets with only partial exposure time was presented in \citet{2019AandA...632A..36C} and we present here a more detailed analysis of the full datasets after homogeneous reduction as for the other fields.

\subsection{Ancillary {\it HST} data}
\begin{table*}
    \centering
    \begin{tabular}{l|l | l | l}
Cluster & HST Program(s) & Filters  & Reference \\
\hline
MACS\,J0257.6$-$2209  & 12166, 14148 & F606W, F814W, F110W, F160W & \citet{2018MNRAS.479..844R}\\
1E 0657$-$56 (Bullet)  & 10863, 11099, 11591 & F435W, F606W, F775W, F814W, F850LP, F110W, F160W & \citet{2016AandA...594A.121P}\\
MACS\,J0940.9+0744  & 15696 & F606W, F814W, F125W, F160W & This Work \\
SMACS\,J2031.8$-$4036 & 12166, 12884 & F606W, F814W & \citet{2018MNRAS.479..844R}\\
SMACS\,J2131.1$-$4019  & 12166 & F814W, F110W, F140W & \citet{2018MNRAS.479..844R}\\
MACS\,J2214.9$-$1359  & 9722, 13666 & F555W, F814W, F105W, F125W, F160W & \citet{2011MNRAS.410.1939Z}\\
    \end{tabular}
    \medskip\par
    \caption{Details of the HST programs and filters used for the 6 clusters which are not part of the FF or CLASH samples.}
    \label{tab:hstobs}
\end{table*}
\label{sec:hst}

We make use of the available high-resolution WFPC2, ACS/WFC, and WFC3-IR images in the optical / near-infrared covering the MUSE observations. Six clusters in our sample are included in the CLASH \citep{2012ApJS..199...25P} and FF \citep{2017ApJ...837...97L} surveys, for which High Level Science Products (HLSP) incorporating all observations taken in 12 and 6 filters respectively have been aligned and combined. We use the HLSP images provided by the Space Telescope at the respective repositories for CLASH\footnote{\url{https://archive.stsci.edu/missions/hlsp/clash/}} and FFs\footnote{\url{https://archive.stsci.edu/missions/hlsp/frontier/}}.

For the remaining six clusters, Snapshot and GO programmes on {\it HST} were obtained as part of the MACS survey (PI: Ebeling) as well as follow-up {\it HST} programmes (PIs: Bradac, Egami). The details of the {\it HST} programmes and available bands are given in Table \ref{tab:hstobs}, and were used in previous strong-lensing work. We make use of the reduced {\it HST} images available in the Hubble Legacy Archive (HLA).

Following the MUSE observations on MACS0940, we have obtained as part of {\it HST} programme 15696 (PI: Carton) new images with ACS and WFC3-IR at a higher resolution and going much deeper than the previous WFPC2 snapshot in F606W. Three orbits were obtained in F814W (totalling 7526 sec), one orbit in F125W (totalling 2605 sec) and 1.5 orbits in F160W (totalling 3900 sec). We have aligned and drizzled the individual calibrated exposures for each band using the {\tt astroDrizzle} and {\tt TweakReg} utilities \citep{2011ApJS..197...36K}.

All HLSP datasets were already calibrated for absolute astrometry. The {\it HST} images for the remaining 6 clusters were aligned against one another and with respect to star positions in J2000 selected from the Gaia Data Release 2 
\citep{2018A&A...616A...1G} catalogue. The accuracy of the absolute astrometry is $<$0\farcs06.

\subsection{MUSE Data reduction}
\label{sec:datared}

For consistency, data from each cluster are reduced using a common pipeline, which processes the raw exposures retrieved from the ESO archive into a fully-calibrated combined datacube. This sequence largely follows the main standard steps described in \citet{2020A&A...641A..28W} as well as the MUSE Data Reduction Pipeline User Manual\footnote{\url{https://www.eso.org/sci/software/pipelines/muse/}}.  However, we make some modifications due to the crowded nature of lensing cluster fields, which contain extended bright objects. We summarise each step below. While the specific version of the Data Reduction Pipeline used on each cluster ranged between v2.4 to v2.7, there are only minor changes between these versions, and these do not affect the quality of the resulting datacubes.

The first step in the pipeline provides basic calibrations. Raw calibration exposures are combined and analysed to produce a master bias, master flat and trace table (which locates the edges of the slitlets on the detectors), as well as the wavelength solution and Line Spread Function (LSF) estimate for each observing night. These calibrations are then applied on all the raw science exposures to produce a \textit{pixel table} propagating the information on each detector pixel without any interpolation. A bad pixel map is used to reject known detector defects, and we make use of the geometry table created once for each observing run to precisely locate the slitlets from the 24 detectors over the MUSE FoV. %The
Twilight exposures and night-time internal flat calibrations (when available) are used for additional illumination correction.

The second step of the data reduction makes use of the {\tt muse\_scipost} module of the data reduction pipeline to process science pixel tables. The calibrations performed in this step include flux calibration using standard star exposures taken at the beginning of the night, telluric correction, auto-calibration (detailed below), sky subtraction, differential atmospheric refraction correction, relative astrometry and radial velocity correction to barycentric velocity. In the case of WFM-AO-N observations with adaptive optics, laser-induced Raman lines are also fit and subtracted. 

Each calibrated pixel table is then resampled onto a datacube with associated variance, regularly sampled at a spatial pixel (hereafter spaxel) scale of 0\farcs2 and a wavelength step of 1.25 \AA , between 4750 (4600 for WFM-NOAO-E) and 9350 \AA. The drizzling method from {\tt muse\_scipost} is used in the resampling process, which also performs a rejection of cosmic-rays. A white light image is constructed for each datacube by inverse-variance weighted averaging all pixel values along the spectral axis. Individual white light images are used to measure spatial offsets between individual exposures. This is done either through the use of the {\tt muse\_exp
\_align} task of the data reduction pipeline, or by locating point sources against an {\it HST} image in the F606W or F814W filter, in the case of large MUSE mosaics. The measured offsets are then used to produce fully aligned (in World Coordinate System, or WCS) datacubes for each exposure.

Once resampled to the same WCS pixel grid, all datacubes are finally combined together outside of the pipeline using the MPDAF\footnote{\url{https://mpdaf.readthedocs.io/en/latest/}} \citep{2019ASPC..521..545P} software task {\tt CubeList.combine}, or its equivalent {\tt CubeMosaic.pycombine} for large mosaics. This allows one to perform an inverse-variance weighted average over a large number of datacubes. A 3-5\,$\sigma$ rejection (depending on the number of exposures) of the input pixels was applied in the average, to remove remaining defects and cosmic rays, totalling $\sim$5-10\% of all pixels in each exposure.  We visually inspected each individual white light image and manually rejected very few obvious cases of light contamination, issues in telescope tracking, etc. from the combination. We also masked 4 cases of satellite and asteroid trails by selecting and masking the relevant spaxels across the cubes. We estimated variations in atmospheric transmission by comparing the total flux of bright isolated sources between individual exposures and provided the correction factors as input to {\tt CubeList.combine} or {\tt CubeMosaic.pycombine}. 

\begin{figure}
    \centering
    \includegraphics[width=9cm]{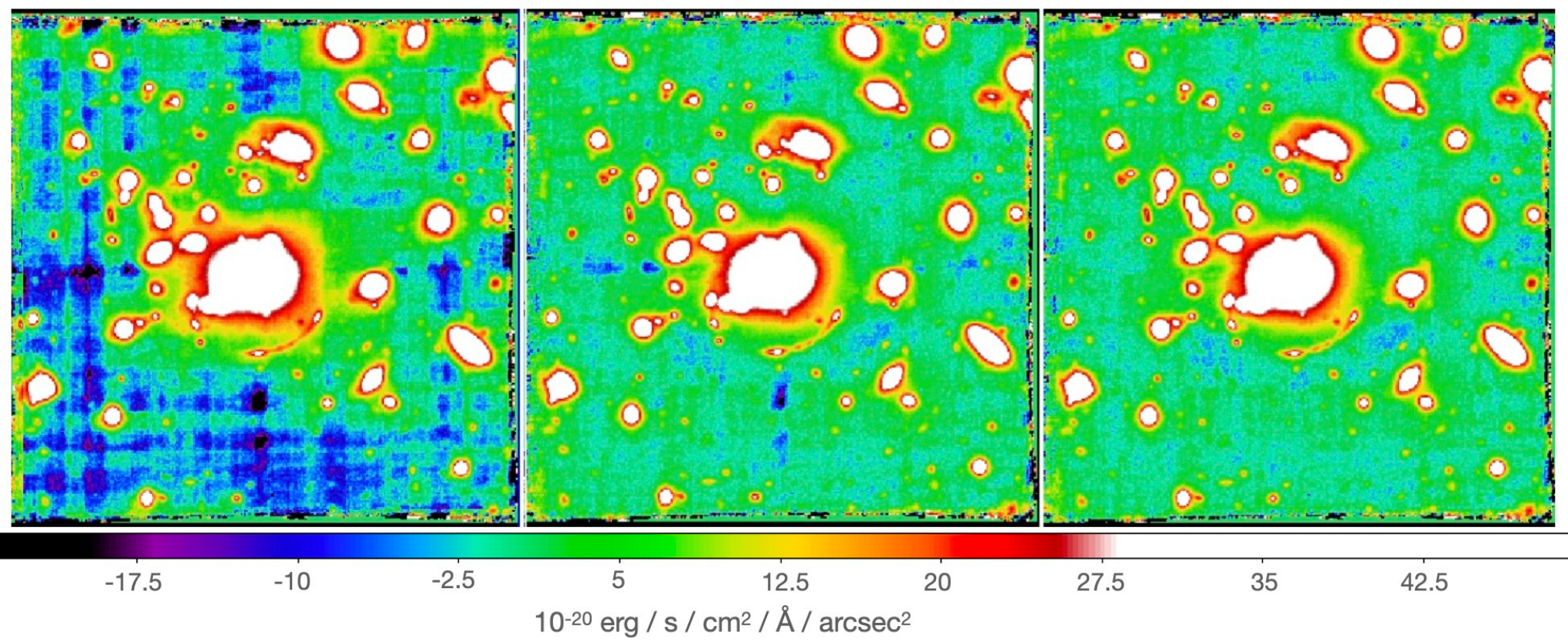}
    \caption{Effect of auto-calibration on the white light image (example shown for MACS0940). From left to right: without auto-calibration, using auto-calibration with tuned object masks, with additional interstack masks. See Sect.~\ref{sec:datared} for more details.}
    \label{fig:datared}
\end{figure}

Low level systematics due to flat-field residuals (at the $\sim$1\% level) remain between each slitlet after applying the internal flat-field calibrations and nightime illuminations. They appear prominently as a weaving pattern over the FoV in the white light image (Fig.~\ref{fig:datared}, left panel). In v2.4 (and subsequent versions) of the data reduction pipeline, an optional self-calibration can be performed as part of {\tt muse\_scipost}, which makes use of the overall sky signal within each slitlet to correct for these flat-field systematics. This procedure has been shown to work accurately in MUSE deep fields where very bright continuum sources are scarce \citep{2017A&A...608A...1B}. However the very bright extended cluster light haloes in the MUSE lensing fields, in particular in the vicinity of the Brightest Cluster Galaxies (BCGs) strongly bias this measurement when the procedure is applied automatically (Fig.~\ref{fig:datared}), due to the extent of the Intra-Cluster light (ICL). We have the possibility to give as input to the pipeline a sky mask for the self-calibration procedure, to help the algorithm identify spaxels with a clean background. We constructed this  mask in two steps, by thresholding a deep {\it HST} continuum image (smoothed by a $\sim$0\farcs6 FWHM gaussian filter) or a MUSE white-light image  created from a first combined datacube, and then applying it to produce the self-calibrated exposures. The threshold was determined by inspecting the mask to avoid using a too small fraction of the spaxels in a given slitlet for the background level estimation. In a few very crowded regions we make use of the multiple rotations in the observations to compute the self-calibration corrections only in the most favourable cases, and provide them as a user table (see more details in \citealt{2020A&A...641A..28W}). When deemed necessary, another iteration on the sky mask was performed on the new combined cube to improve the self-calibration further. An illustration of the resulting combined white light image and sky mask is presented in the middle panel of Fig.~\ref{fig:datared}. In the case of the SMACS\,J2031 cluster field, which is the only cluster lacking for night-time illumination corrections during the commissioning run, the improvements due to self-calibration are even more significant (see \citealt{2019MNRAS.489.5022C}).

The white-light image of the self-calibrated combined datacube shows some imperfections in the sky background, in the form of negative 'holes' in the interstacks between the MUSE slices, in particular within the top channels of the instrument. These can typically be corrected in empty fields by producing a \textit{super-flat} calibration (Bacon et al. in prep.) combining all exposures in the instrument frame of reference while masking objects. However this method is not suited for crowded cluster fields. Instead, we aligned all the individual white light exposures in the instrument referential frame by inverting all WCS offsets and rotations between exposures, then detected deviant spaxels with strongly negative flux over the average combined image. This produces a small mask containing $\sim0.6\%$ of the FoV,  which we use to mask pixels in the datacubes at all wavelengths prior to combination. This results in the removal of the majority of the imperfections as seen in the white-light image (Fig.~\ref{fig:datared}, right panel).

The combined datacube is finally processed using the Zurich Atmospheric Purge (ZAP, \citealt{2016MNRAS.458.3210S}) v2, which performs a subtraction of remaining sky residuals based on a Principal Component Analysis (PCA) of the spectra in the background regions of the datacube. This has been shown to remove most of the systematics in particular towards the longer wavelengths (see e.g. \citealt{2017A&A...608A...1B} in the UDF). We provide as input to ZAP an object mask by inverting the sky mask described above. Although the number of eigenvectors removed is automatically selected by ZAP, we have performed multiple tests adjusting this number to check that the chosen value did not affect the extracted spectra of faint continuum and line emitting sources.

\subsection{MUSE Data Quality}
\label{sec:quality}

The reduced datacube is given as a FITS file with two extensions, containing the $f_\lambda$ and associated variance over a regular 3D grid. The spectral resolution of the observations
varies from R=2000 to R=4000, with a spectral range between 4750 and 9350 \AA. To ensure that cubes are properly Nyquist sampled, we set the wavelength grid to 1.25 \AA\ pixel$^{-1}$. The final
spatial sampling is 0.2 arcsec pixel$^{-1}$ in order to properly sample
the Point Spread Function (PSF).

We have cross-checked the relative astrometry between the MUSE cubes and the {\it HST} images presented in Sect.~\ref{sec:hst}, by cross-matching bright sources in the two datasets. We measure a global rms of $\sim$0\farcs12, that is to say about half a MUSE spaxel.

In order to estimate the spatial Point Spread Function, which varies as a function of wavelength, we have followed the MUSE Hubble Ultra Deep Field approach, described by \citet{2017A&A...608A...1B}. From the MUSE datacube we produce pseudo-HST images in the bandpasses matching the {\it HST} filters (Table \ref{tab:hstobs}) and overlapping with the MUSE spectral range. These images are compared with pseudo-MUSE images created from {\it HST} and convolved with a \citet{1969A&A.....3..455M} PSF at fixed $\beta=2.5$. The FWHM of the Moffat function is optimised by minimising the residuals in Fourier space. This method allows to use all objects in the field to measure the PSF.

The wavelength dependence on the PSF is then adjusted by assuming a linear relation:

\begin{equation}
FWHM(\lambda)= C\,(\lambda-\lambda_0)+FWHM(\lambda_0)\ .
\label{eq:psf}
\end{equation}

The slope $C$ of this relation depends on the use of Adaptive Optics, we find values ranging between $-1.9\times10^{-5}$  and $-3.6\times10^{-5}$ arcsec\,\AA$^{-1}$, which are similar to the ones from the MUSE Ultra Deep Fields \citep{2017A&A...608A...1B} and MUSE-Wide \citep{2019A&A...624A.141U}. We report for each dataset the FWHM of the PSF at 7000 \AA\ in Table \ref{tab:museobs}. In general the datacubes have been taken in very good seeing conditions, translating into a typical FWHM$_{7000}$ value between 0\farcs55 and 0\farcs65. The only exception is SMACS2031 (FWHM$_{7000}$=0\farcs79) which was observed during MUSE commissioning. The same MUSE vs {\it HST} comparison allows us to cross-check the photometric accuracy of the MUSE cubes, and we find an agreement within 5-10\% between the two datasets.

\begin{figure}
    \centering
    \includegraphics[width=7.5cm]{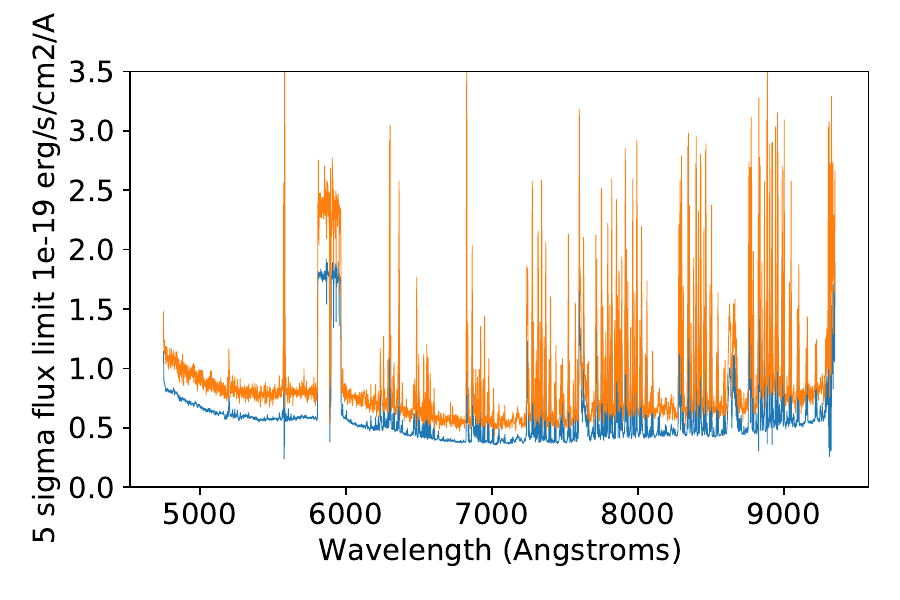}
    \caption{5$\sigma$ continuum flux limit estimated in empty sky region of the MACS0940 datacube (8 hr exposure time), including (orange curve) or not (blue curve) the correction for correlated noise. The increased noise level within the AO filter (around 5890 Angstroms) is due to the shorter effective exposure time in WFM-NOAO-N instrument mode (50 min).}
    \label{fig:noise}
\end{figure}

The variance extension of the cube is estimated during the data reduction and gives an estimate of the noise level which is generally underestimated due to correlation in the noise introduced during the final 3D resampling process (see e.g. \citealt{2017A&A...606A..12H,2019A&A...624A.141U,2020A&A...641A..28W}). To properly assess the noise properties of each cube as a function of wavelength, we make use of objects-free regions (as defined with the sky masks mentioned in Sect.\ref{sec:datared}) and measure the variance of the flux level integrated over 1\arcsec\ square apertures. When compared with the direct estimate from the variance extension, the two measurements show a simple constant correction factor between 1.3 and 1.5 on the standard deviation as a function of wavelength (e.g. Fig.~\ref{fig:noise}). We apply this constant correction factor to the noise estimation of the datacube.

The relative noise levels presented in Fig.~\ref{fig:noise} are very similar in all datacubes (with the exception of the AO filter region which depends on the relative fraction of AO vs NOAO exposure time), and roughly scale with the square root of the effective exposure time. The 5$\sigma$ surface brightness limits (averaged in a 1 arcsec$^2$ region) between skylines at 7000 \AA\ is 4.8$\times$10$^{-19}$ and 2.5$\times$10$^{-19}$ erg\,s$^{-1}$\,cm$^{-2}$\,\AA$^{-1}$\ arcsec$^{-2}$ for cubes with 2 hrs and 8 hrs exposure time respectively. This translates into a 5$\sigma$ line flux limit of 1.5$\times$10$^{-18}$ and 7.7$\times$10$^{-19}$ erg\,s$^{-1}$\,cm$^{-2}$ at 7000 \AA\ when averaged over a 5$\times$5 spaxels aperture and 3 wavelength planes (6.25 \AA).

\section{Construction of MUSE spectroscopic catalogues}
\label{sec:building}
\subsection{Overview}

The sheer number of volume pixels (voxels) in a 1 arcmin$^2$ MUSE datacube ($\sim 4 \times 10^8$) makes it challenging to locate and identify all the extragalactic sources; this is especially true when observing crowded fields like galaxy clusters. Prior knowledge of source locations from {\it HST} images provides crucial external information to help distinguish and deblend overlapping sources. Our goal is to extract a spectrum and estimate the redshift for all sources down to a limiting signal-to-noise ratio (described in subsequent subsections), either in the stellar continuum or via emission lines. Additional prior information on the redshift of background sources can be obtained for strongly-lensed multiple images, where well-constrained models allow. We have therefore developed a common process to analyse the MUSE datacubes incorporating all these elements.

\begin{figure}
    \centering
    \includegraphics[angle=0,width=10cm]{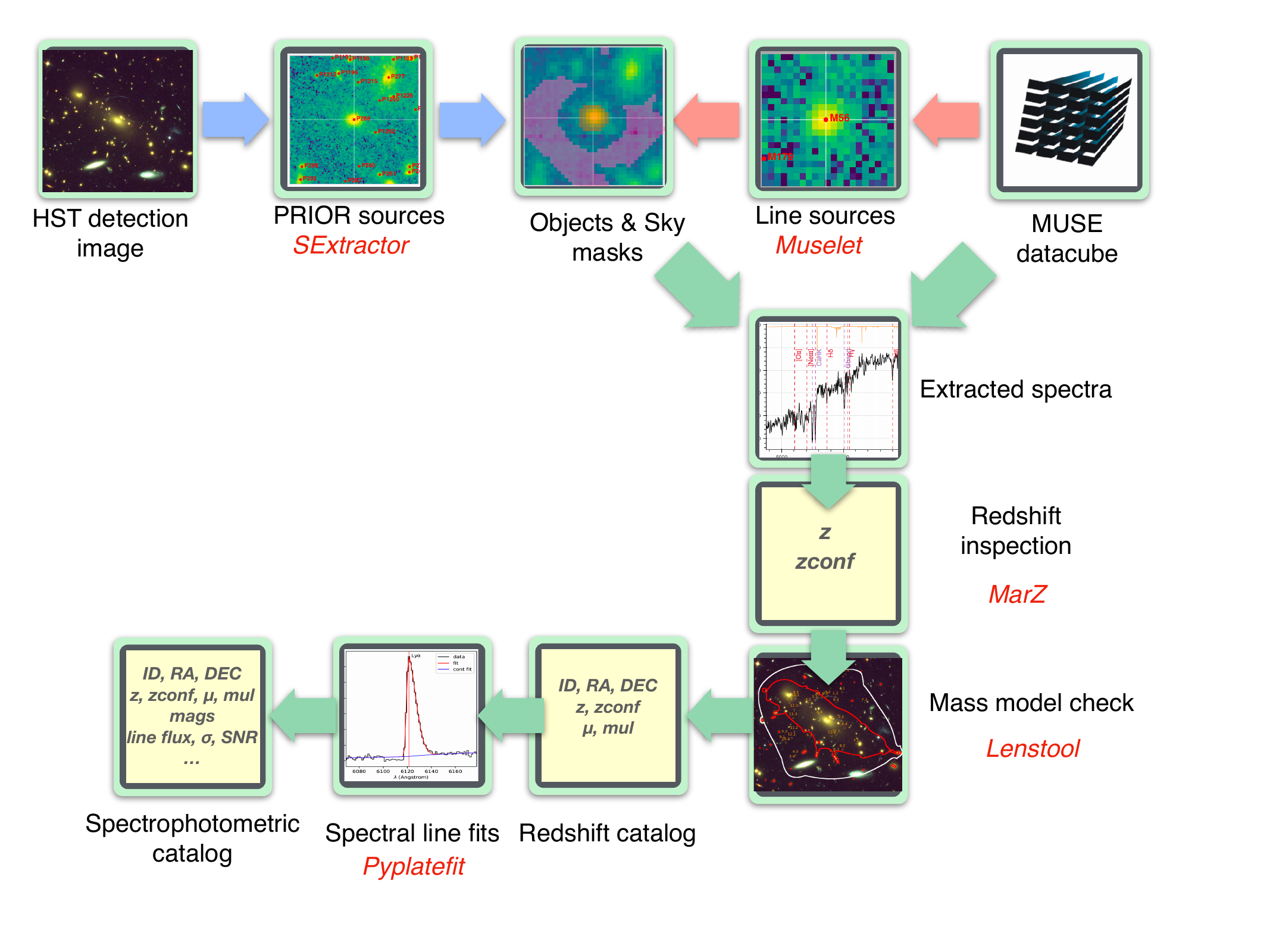}
    \caption{Flowchart presenting the process of building the MUSE spectrophotometric catalogues, based on the input {\it HST} images and MUSE datacube. Sect.~\ref{sec:building} presents each step of the procedure in detail.}
    \label{fig:flowchart}
\end{figure}

The general concept of this procedure has been described previously in \citet{2018MNRAS.473..663M} and \citet{2019MNRAS.485.3738L}, but we have further refined and partially automated the methods for the subsequent cluster observations presented here, in order to create more uniform and robust catalogues. We summarise each step of the process in the flowchart shown in Fig.~\ref{fig:flowchart}, and provide details on each step in the following subsections.

\subsection{HST prior sources}

We make use of all the available high-resolution {\it HST} imaging in each field (see Sect.~\ref{sec:hst}) to produce input photometric catalogues of continuum sources that overlap with the MUSE FoV to identify regions for subsequent spectral extraction. Since our goal is to measure the spectra of typically faint compact background sources, usually embedded within the bright extended haloes of cluster members, we first produce a \textit{median-subtracted} version of each image by subtracting a 1\farcs5$\times$1\farcs5 running median filter box. The WCS-aligned, median-subtracted {\it HST} images are cropped to the common MUSE area and further combined into an inverse-variance-weighted \textit{detection image}, using {\sc SWarp} \citep{2002ASPC..281..228B}.

This procedure was used in previous works searching for faint high-redshift dropouts \citep{2015ApJ...800...18A,2018MNRAS.479.5184A} in deep \textit{Hubble} FF images, and significantly improves the detection of background sources while still keeping the central cores of all bright cluster members. The negative impact of the median subtraction is a reduction of the isophotal area for bright extended sources in the segmentation map, which reduces the total flux in their extracted MUSE spectra. However, as these sources all have high signal-to-noise ratio in the MUSE datacube, this has no impact on the spectroscopic redshift measurements, and we adjusted the size of the median-box filter in order to mitigate this effect.

The detection image is given as input to the {\sc SExtractor} software \citep{1996A&AS..117..393B} used in \textit{dual} mode to measure the photometry over each {\it HST} band (aligning each band image with {\sc Montage})\footnote{\url{http://montage.ipac.caltech.edu/}}. We make use of the {\it HST} weight maps to account for variation of depth over the FoV, and selected typical {\sc SExtractor} parameter values {\tt detect\_minarea=10-15} pixels and {\tt detect\_thresh=1.2-1.5}. These values were optimised for each cluster to account for variations in depth and pixel scale of the {\it HST} images.

The photometry performed by {\sc SExtractor} in each band is merged together into a photometric catalogue of {\it HST} sources, dubbed \textit{prior sources} through the rest of the paper. {\sc SExtractor} also produces a segmentation map where each spaxel in the MUSE FoV is flagged with a source ID number.

During the detection process, {\sc SExtractor} tends to split large clumpy sources such as foreground disk galaxies or very elongated arcs into multiple entries in the catalogue. Such cases are visually identified as part of the source inspection and redshift estimation process (see Sect.~\ref{sec:inspection}), and the corresponding photometry and segmentation map are merged together to produce a single \textit{prior source} located at the {\it HST} light barycenter.

\subsection{Line-emitting sources}

Independent from the {\it HST} continuum sources, we produced a catalogue of line-emitter candidates identified directly from the MUSE datacubes. This was performed by running the {\tt muselet} software, which is part of the MPDAF python package \citep{2019ASPC..521..545P}. {\tt muselet} first creates a `pseudo-narrow-band' MUSE datacube by replacing each wavelength plane of the cube with a continuum-subtracted narrow-band image. %The narrow-band \textcolor{cyan}{emission?} is created through a weighted average of the 5 neighbouring wavelength planes of the cube centered at this wavelength, 
%The narrow-band \textcolor{cyan}{emission?} is created through a weighted average of 5 neighbouring wavelength planes of the cube (the original wavelength plane and its four nearest neighbours, 
To generate the narrow-band emission for a given wavelength, we take a weighted average of five planes of the cube (the original wavelength plane and its four nearest neighbours), using a fixed FWHM Gaussian weight function corresponding to 150 km\,s$^{-1}$ for a line centred at 7000\AA. This weighting scheme improves the signal-to-noise ratio of the narrow-band at the centre of a typical emission line of similar FWHM. The subtracted continuum is estimated by an inverse-variance weighted average of two 25 \AA\ wide wavelength windows located directly bluewards and redwards of the narrow-band. We have found that these parameters are generally appropriate for an automatic subtraction of the local continuum.

The full set of narrow-band images is run through {\sc SExtractor} to detect candidate spectral lines at each wavelength. %Because we expect diffuse extended emission, in particular for Lyman-$\alpha$ emitters, we have selected typical {\sc SExtractor} detection parameters {\tt detect\_minarea=12} pixels and {\tt detect\_thresh=1.2}, as well as a 5$\times$5 pixels tophat spatial filter of FWHM=0.8\arcsec.
Because we expect some galaxies to show diffuse extended emission, especially Lyman-$\alpha$ emitters, we tune our {\sc SExtractor} detection parameters ({\tt detect\_minarea=12} pixels; {\tt detect\_thresh=1.2}), and spatial smoothing filter (a 5$\times$5 pixel tophat with FWHM$=$0\farcs8) to better identify these objects.
By construction, all bright lines will be detected in adjacent wavelength planes, and therefore they are automatically merged into a master list of individual line candidates. We specifically masked three 4\AA-wide wavelength regions in the cube corresponding to bright sky lines at 5577 and 6300 \AA\ as well as the strongest Raman line at 6825 \AA, to prevent artefacts in the detections. Sets of lines spatially coincident (within 1 seeing disk) are then grouped together into {\tt muselet} \textit{sources} and a segmentation map is produced. A first redshift estimate is also provided by {\tt muselet} based on an automatic match with a list of bright emission lines. 

The MPDAF v2.0 of the {\tt muselet} software was used for a first set of clusters (Abell 2744, Abell 370, MACS1206, MACS0416S). MPDAF v3.1 introduces improvements in the merging and weighting schemes of the {\tt muselet} sources, and was used for the rest of the clusters. 

\subsection{Spectral extraction}
\label{sec:extract}
In order to characterise the detected \textit{prior} and {\tt muselet} sources, we adopt the following procedure to extract 1D spectra, which optimises the signal-to-noise of the detections while being adapted to crowded fields. A weight image is created for each source by taking the flux distribution over the segmentation maps produced as part of the detection process. In the case of \textit{prior} sources this flux distribution is taken from the combination of all {\it HST} filters, and the 2D weight map is convolved by the MUSE PSF model described in Sect.~\ref{sec:quality}. To account for the wavelength variation of the PSF (Eq.~\ref{eq:psf}), weight images are created for each source at 10 wavelength planes (i.e. at $\sim500$\,\AA\ intervals) over the MUSE spectral range and interpolated into a 3D weight cube. In the case of {\tt muselet} sources, the segmentation map used is the one of the brightest emission line of the object. Since we do not account for wavelength variations in the intrinsic flux profiles of galaxies (e.g. colour gradients), the weighted extractions, while optimal for the distant compact sources, are not spectrophotometrically accurate for the extended cluster members. For this reason we also extract unweighted spectra.

The weighted spectrum is computed following the optimal extraction algorithm from \citet{1986PASP...98..609H}:

\begin{equation}
    %F(\lambda)=\frac{\sum\limits_{x,y} C(x,y,\lambda)W(x,y,\lambda)/V(x,y,\lambda)}
    %{\sum\limits_{x,y} W^2(x,y,\lambda)/V(x,y,\lambda)}
    F(\lambda)=\frac{\sum\limits_{x,y} C_{x,y,\lambda}\,W_{x,y,\lambda}\,/V_{x,y,\lambda}}
    {\sum\limits_{x,y} W^2_{x,y,\lambda}/V_{x,y,\lambda}}\ ,
    \label{eq:extraction}
\end{equation}

where $C_{x,y,\lambda}$ and $V_{x,y,\lambda}$ respectively correspond to the pixel flux and associated variance at a specific location in the cube, and $W_{x,y,\lambda}$ is the optional weight cube.

In addition, we estimate a local background spectrum around each source, produced by averaging the spectra from MUSE spaxels outside of all detected sources in the field. This local background estimate removes large-scale contamination from bright sources, such as stars and cluster members, as well as potential systematics in the background level remaining from the data reduction. We define spaxels containing sky and intra-cluster light by convolving the combined {\it HST} 2D weight map with the central wavelength MUSE PSF. From this, the darkest 50\% of spaxels are considered as background-spaxels. However, for each object extracted we compute the local background, by aggregating the nearest $\sim$500 background-spaxels surrounding it (with a minimum Manhattan distance of 0\farcs4 from the object). To retain locality, we do not aggregate beyond 4\farcs0, even if we have not reached the desired $\sim$500 spaxel target. This ensures a good trade-off between aggregating sufficient sky spaxels to achieve a high signal-to-noise estimate of the background, without aggregating so many spaxels that they are no longer locally relevant.

Therefore, for every object we compute four sets of spectra: weighted (with and without) local background subtraction and unweighted (with and without) local background subtraction. In general weighted spectra with local background subtraction are those most optimal for source identification and redshift measurement. But for the bright very extended cluster members, the unweighted spectra without local background subtraction would provide superior spectrophotometric accuracy.

\subsection{Source inspection}
\label{sec:inspection}

To determine the redshift of extracted sources, we adopt a semi-automatic approach similar to the one used to produce the redshift catalogue in the MUSE observations over the \textit{Hubble} Ultra Deep Field \citep{2017A&A...608A...2I}. We make use of the {\tt Marz} software \citep{2016ascl.soft05001H} which looks for the best redshift solutions by performing a cross-correlation with a set of pre-defined spectral templates. The templates used in our analysis are described in detail in Appendix B of \citet{2017A&A...608A...2I}. They include a variety of stellar spectra from \citet{2014MNRAS.441.2440B} and deep MUSE spectra from blank fields \citep{2015A&A...575A..75B,2017A&A...608A...1B}.

%All {\tt muselet} sources and every \textit{prior} source down to a limiting S/N in the continuum were inspected individually by at least 3 coauthors. The limiting magnitude was adjusted based on our first assessments to prevent reviewing too low S/N sources, and selected to be F814W$_{AB}$=25+1.25\ log(T/2.0) with $T$ the maximum exposure time over the field in hours. This corresponds to a typical median S/N of $\sim$1.5 per pixel over the MUSE spectral range. This means that redshifts for fainter continuum sources were measured through {\tt muselet} detections.
All {\tt muselet} sources and every \textit{prior} source down to a limiting (continuum-level) signal-to-noise ratio (S/N) have been inspected individually by at least 3 coauthors. Based on our first assessments, we adjust the \emph{prior} magnitude limit for each cluster as a function of MUSE exposure time, given by F814W$_{AB}$=25+1.25\ log(T/2.0), with $T$ the maximum exposure time over the field in hours. This allows us to avoid reviewing sources with too low signal, and corresponds to a typical median S/N of $\sim$1.5 per pixel over the MUSE spectral range. Any \emph{prior} source in our final catalogue below this limit is therefore measured from a corresponding {\tt muselet} detection.

%LIST OF INSPECTORS
%JRI,ACL,DLA,DCA,RPE,GSO,WME,GDV,FBA,LGU,BCL,GPR,JMA,GMA,VPA,JPK,ATA,AVA.

%A specific tool was designed within the GTO consortium (Bacon et al.\, in prep.) to give the user a large overview of all {\it HST} and MUSE white light images, extracted spectra (and their variations) shown against matching line from the five best {\tt Marz} redshift solutions, zoom-ins on specific spectral lines and narrow-band images, as well as navigate quickly through all neighbouring sources. This tool was adapted to the needs of the lensing cluster fields.
To inspect each source in detail, we use a graphical interface tool designed within the MUSE collaboration (Bacon et al.\, in prep.) that lets a user simultaneously view all {\it HST} and MUSE white light images of the object, the extracted spectrum (and its variations) compared to features of the five best {\tt Marz} redshift solutions, and zoom-ins on specific spectral lines and narrow-band images. The tool allows us to navigate quickly between all neighbouring sources, to check for contaminations or object blending. This tool has been adapted to the needs of the lensing cluster fields.

More specifically, the following information is inspected for each source:
\begin{itemize}
    \item[$\bullet$] Spectroscopic redshift and confidence. All {\tt Marz} solutions are assessed, and the user can also choose a different value manually. When a redshift solution is found, a confidence level ($zconf$) is selected according to its level of reliability:
    \begin{itemize}
        \item Confidence 1: redshift based on a single low-S/N or ambiguous emission line, or several very low S/N absorption features. These sources are given in the public catalogue for comparison with other measurements, but are not used in the subsequent analysis. 
        \item Confidence 2: redshift based on a single emission line without additional information, several low S/N absorption features, or whose redshift confidence is increased by the identification of multiply-imaged systems (see Sect.~\ref{sec:mulimages}).
        \item Confidence 3: redshift based on multiple spectral features, or with additional information on a high S/N emission line (e.g. very clear asymmetry of the line, {\it HST} non-detection bluewards of the line).
    \end{itemize}
      \item[$\bullet$] Association between \textit{prior} and {\tt muselet} sources. The two detection methods  
      will most of the time identify the same sources, and we keep the \textit{prior} sources by default when a clear match is found with {\tt muselet} emission lines. It is important to note that significant spatial offsets can be found between the two sources, for example in the case of star-forming/photoionised gas trailing behind infalling galaxies in the cluster, or Lyman-$\alpha$ emission which is physically offset from the underlying UV continuum emission.
      \item[$\bullet$] Merging of sources. Clumpy star-forming galaxies or very elongated lensed arcs tend to be separated into multiple \textit{prior} sources. We identify them and perform a new extraction of the merged source as in Sect.~\ref{sec:extract} by aggregating their segmentation maps and combining their photometry.
      \item[$\bullet$] Defects and artefacts. We identify stellar spikes or cosmic ray residuals in the \textit{prior} catalogue, as well as artefacts found by {\tt muselet} in the cube due to strongly varying spectral continuum (typically low-mass stars or cluster members).
\end{itemize}  

The resulting inspection information is stored in a database, and since each spectrum was reviewed independently by at least three users, the results are cross-matched during a consolidation phase, where a consensus is reached in case of disagreement between the users, and the entry is written in the catalogue.

\subsection{Assessment of multiple images}
\label{sec:mulimages}
All sources with a reliable redshift in the catalogue are tested with the corresponding {\sc Lenstool} mass model of the cluster (see Sect.~\ref{sec:massmodels}) to predict multiple images. As the precision of the lens model is typically 0\farcs$5--0$\farcs9 over the area of multiple images, and the density of objects in a given redshift plane is low, we easily find %good matching systems
good matches for systems of images at the same redshift which are predicted to arise from the same source. We manually inspect the corresponding spectra of each matching candidate, as we expect precisely the same redshift and spectral features from multiple images. This assessment serves two purposes: (a) identifying multiple images in order to pinpoint truly independent sources, and (b) cross-checking the redshift measurements for background lensed sources. 

Indeed, the main spectral features identified for background galaxies through the redshift inspection are typically bright emission lines ([\oii], \ciii], \civ, Lyman-$\alpha$) or absorption features (rest-frame UV ISM absorptions or Lyman-$\alpha$ break). Ambiguous redshifts are typically limited to an uncertainty between [\oii] and Lyman-$\alpha$ as a single emission line, which produce very different lensing configurations between $z\sim0.5$--1.5 and $z\sim3$--7. The multiple-image assessment allows us to correct a very small number of erroneous redshifts (typically less than 3 images per cluster), and more importantly to boost the redshift confidence level to $zconf=2$ for faint line emitters with $zconf=1$ showing a clear matching counterimage predicted by the model.

\begin{figure}
    \centering
    \includegraphics[width=9cm]{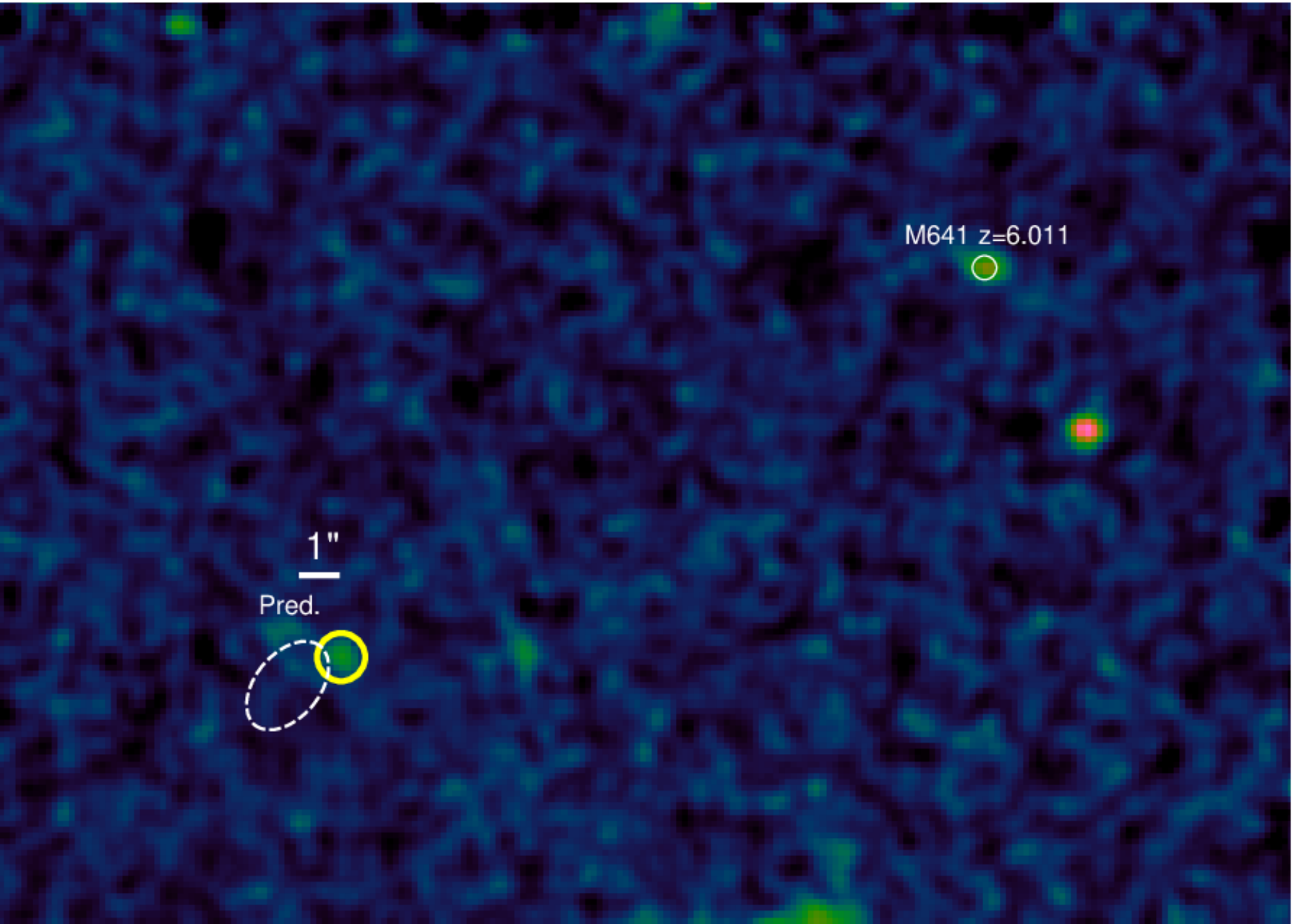}
    \caption{Example of faint Lyman-$\alpha$ emission (large yellow circle) detected near the position of the predicted counterimage from {\tt muselet} source 641 in cluster MACS0329, at $z=6.011$ (small white circle). The image shows a smoothed narrow-band image at the peak wavelength of the emission. The dotted ellipse denotes the estimated 1-$\sigma$ error region of the predicted location.}
    \label{fig:counterimage}
\end{figure}

As new multiple images help us refine and improve the lensing model, this is an iterative process: we start by using only spectroscopically confirmed images from the most secure systems (having confidence levels 2 or 3) as constraints in the model, and make further predictions from the refined model. We systematically search for spectral features in all predicted counter-images, even when they are not included in the original catalogue, solving all possible inconsistencies. This process allows us to identify a small number (up to 5-10 per cluster) of images as faint emission lines in the narrow-band datacube, which were not identified originally by {\tt muselet}. We include these images as additional entries in the catalogue by manually forcing their extraction, adopting a point source extraction aperture in the procedure described in Sect.~\ref{sec:extract}. A typical example of faint counter image discovered during the assessment is presented in Fig.~\ref{fig:counterimage}.

Remaining strongly-lensed sources not matched to a given system are $zconf=1$ objects for which we are able to explain the lack of counterimage by either (a) its much lower magnification, (b) its strong contamination by a cluster member or foreground source, or (c) it being outside the MUSE FoV. Conversely, when the counterimage is predicted to be bright and isolated, but is not seen in the observations, we choose to entirely remove the $zconf=1$ source from the redshift catalogue due to the large uncertainty in the measured redshift.

Multiple images are identified in a dedicated column ({\tt MUL}) of the final spectroscopic catalogue: we adopt the usual notation X.Y where X identifies a system of multiple images and Y is a running number identifying all images for a given system.

\subsection{Cross-checks with public spectroscopic catalogues}

We performed a comparison of our spectroscopic catalogues with available spectroscopy from VLT and HST in the literature, including some of the same MUSE observations included in our sample. 

\subsubsection{Abell 2744 and Abell 370}\label{subsec:A2744_A370}

First versions of the MUSE spectroscopic catalogues, using the same datasets but based on an earlier version of our data reduction and analysis process, were presented in \citet{2018MNRAS.473..663M} for Abell 2744 and in \citet{2019MNRAS.485.3738L} for Abell 370. Both catalogues were cross-checked with available observations from the GLASS (\citealt{2014ApJ...782L..36S,2015ApJ...812..114T}). The datacubes included in these publications were reduced and analysed using an earlier version of the %pipeline processing 
reduction pipeline and MPDAF, %without 
which did not include autocalibration or the interstack masks, and which used older versions of {\tt ZAP} and {\tt MUSELET}. In light of the newly processed datacube and source inspection catalogues, we reviewed all of the low-confidence sources ($zconf=1$ or 2) from the public catalogues and rejected 18 sources. We corrected the redshifts for 9 sources, typically faint [\oii] and Lyman-$\alpha$ emitters or sources in the redshift desert.

\subsubsection{SMACS2031}

An early catalogue from the MUSE commissioning data taken on this cluster was presented in \citet{2015MNRAS.446L..16R} together with a cluster mass model. In view of the significant improvements in the data reduction since these observations were taken (see also \citealt{2019MNRAS.489.5022C}) we fully re-reduced and analysed the MUSE observations following our common process. We identify three additional multiply-imaged systems as faint Lyman-$\alpha$ emitters (forming a total of nine images), and also find two counterimages for existing systems. One published redshift has been corrected.

\subsubsection{MACS0416}

A first analysis of the MUSE datasets in MACS0416 was presented by \citet{2017A&A...600A..90C} together with the CLASH/VLT redshifts from \citet{2016ApJS..224...33B}. The data they analysed includes the full MUSE coverage in the southern part (MACS0416S, programme ID 094.A-0525) but only the GTO coverage in the northern part (MACS0416N, programme ID 094.A-0115). In the southern part, we find small redshift differences for five sources in common and at low confidence ($zconf=1$). One confirmed Lyman-$\alpha$ emitter (CLASHVLTJ041608.03-240528.1 at $z=4.848$) was not detected with {\tt muselet} in our catalogue but we confirmed it from the narrow-band cube at low confidence. In the same region, we identify nine new multiply-imaged systems producing 22 images in the MUSE data, which were not used in previous published models. 

In the northern part, the MUSE data used in our analysis is much deeper and we find no discrepancy for the sources in common with \citep{2017A&A...600A..90C}.  \citet{2020MNRAS.494L..81V} present one source at $z=6.63$ from the deep observations, which was also identified in our {\tt muselet} catalogue. The deeper data provide us with additional redshifts (not used in previous published models) for 15 systems and 33 multiple images in that region, including 11 systems newly discovered with MUSE as faint Lyman-$\alpha$ emitters.

Following the submission of this manuscript, \citet{2020arXiv200908458V} presented an analysis of the same MACS0416 MUSE fields. They identify a counter-image for 6 multiply imaged systems in the north, which were not originally included in our analysis, and we confirm them in our extracted spectra at low to medium confidence level ($zconf=1-2$). In addition, they present one additional triply imaged system in the overlap region between the two MUSE fields, which we add in our catalogues for completeness as system 100. In comparison with their catalogue of multiple images, we confirm 16 additional multiply imaged systems with MUSE redshifts in our analysis, including 7 in the deep northern part.

Finally, as in Sect.~\ref{subsec:A2744_A370}, we cross-check the agreement between our redshift measurements and the GLASS catalogue from \citep{2016ApJ...831..182H} for the sources in common, and find no redshift discrepancies.

\subsubsection{MACS1206}

\citet{2017A&A...607A..93C} presented an analysis of the same MUSE data taken in the field of the cluster MACS1206 as in our sample. Their full spectroscopic catalogue is not publicly available but we have cross-checked the redshifts and multiply-imaged systems with our own analysis. We recover all of the systems at the same spectroscopic redshifts as used in their strong-lensing model. In addition, we identify eight new systems (named 29 to 36) of two to four images each (total 21 images) in our analysis of the datacubes. These 21 images are in their large majority Lyman-$\alpha$ emitters, generally diffuse and extended, that were found by {\tt muselet}. The redshifts were further confirmed by predictions from the well-constrained lens model (rms$_{\rm model}=0$\farcs52).

\subsubsection{RXJ1347}

\citet{2019AandA...632A..36C} presented an analysis of early observations taken in the RXJ1347, restricted to the upper-left quadrant of the mosaic taken in WFM-NOAO-N. We compare our spectroscopic redshifts with the public catalogue available on VizieR, as well as the constraints presented in their lens model, based on 8 multiply-imaged systems including 4 with spectroscopic confirmation (totalling 9 images). All the published sources are in common with our spectroscopic catalogue and there are no redshift discrepancies. The full dataset on this cluster gives us a significant increase in the number of spectroscopic redshifts and multiple systems: we identify 25 additional systems with a spectropic redshift, producing a total of 119 images.
We compared our sample with the public redshift catalogue from GLASS on this cluster and did not identify any discrepancies between common redshifts. We boosted the confidence level for 2 MUSE sources with matching lines with GLASS observations. 

\subsubsection{MACS0329}
\citet{2019AandA...632A..36C} presented an analysis of the same MUSE data taken in the field of the cluster MACS0329. We compare our spectroscopic redshifts with the public catalogue available on VizieR, and identify one source at $z=2.919$ with diffuse Lyman-$\alpha$ emission which we did not detect in our {\tt muselet} catalogue. We confirm from our lensing predictions that it is indeed a multiply imaged candidate associated with the halo of our {\tt muselet} source 129 showing the same spectral line. We treat these images as a multiply-imaged candidate, but their precise location is too uncertain to be included as multiple images in our lensing model.

In addition, we identify two clear pairs of multiply imaged Lyman-$\alpha$ emitters which we include as strong lensing constraints (our systems 4 and 7) but which were not used in the strong lensing model from \citet{2019AandA...632A..36C}. In both cases, one image from the system was in common between the two spectroscopic catalogues. We also identify 11 low confidence ($zconf=1$) Lyman-$\alpha$ emitters not present in their catalogue, which either do not predict or for which we cannot rule out the presence of a counterimages, as well as a high confidence ($zconf=3$) [\oii] emitter at $z=0.963$.

\subsection{Spectral analysis}

Due to their different physical origins, the spectral features identified in each source as part of the cross-correlation (e.g. K,H Ca lines in cluster members, nebular emission lines, ISM absorption lines, Lyman-$\alpha$ emission) %give a non-uniformity in the accuracy and precision 
or through manual redshift measurement 
do not provide uniform constraints on the accuracy and precision of the redshift estimates. In addition, the cross-correlation carried out by {\tt Marz} can be systematically off, %and does not 
and the results do not provide us with a reliable error on the redshift measurement. 

Therefore, to properly distinguish between the various redshift estimates and get a better estimate on the redshift error, we make use of the {\sc pyplatefit} tool developed for the MUSE deep fields (Bacon et al. in prep.) on the weighted, sky-subtracted 1D spectra. {\sc pyplatefit} is a simplified python version of the original {\sc Platefit} IDL routines developed by \citet{2004ApJ...613..898T} and \citet{2004MNRAS.351.1151B} for the SDSS project. It performs a global continuum fitting of the spectrum, and then fits individual emission and absorption lines using a Gaussian profile (or asymmetric Gaussian for the Lyman-$\alpha$ line). Multiple redshift estimates are measured for each family of spectral features (nebular emission lines, Balmer absorption lines, Lyman-$\alpha$, ..), allowing for small velocity offsets (within 150 km/s, but up to 500 km/s for Lyman-$\alpha$ emission). 

\begin{figure}
    \centering
    \includegraphics[width=9cm]{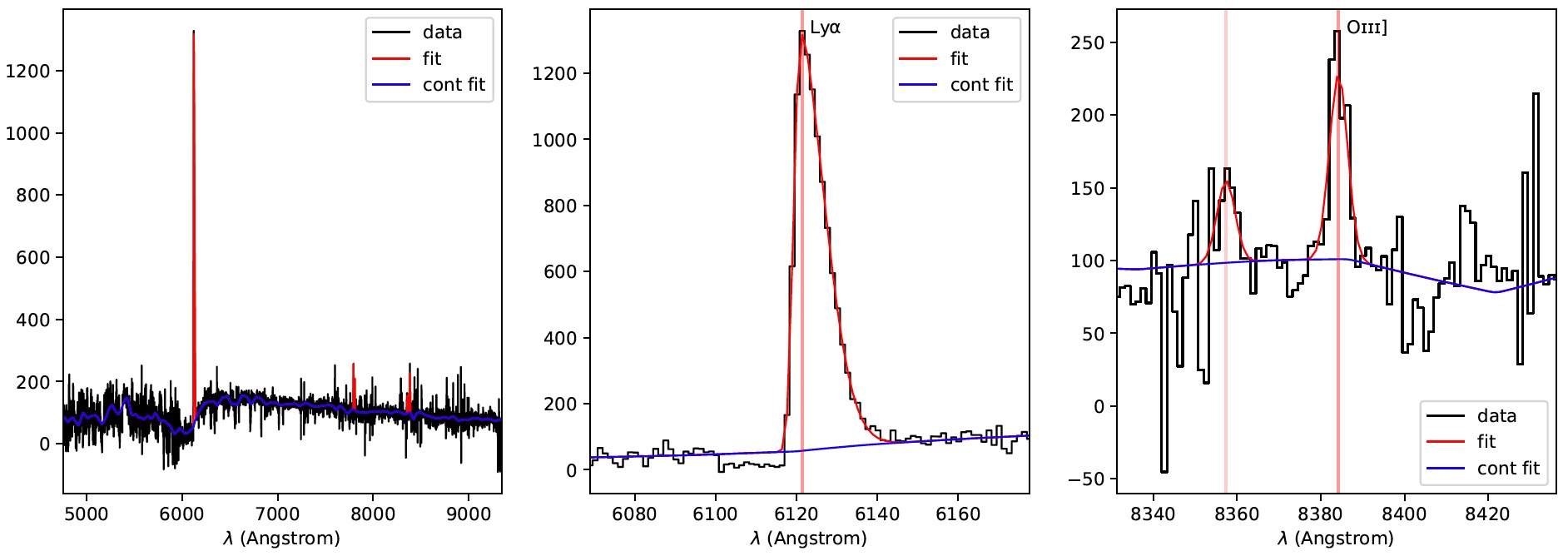}
    \includegraphics[width=9cm]{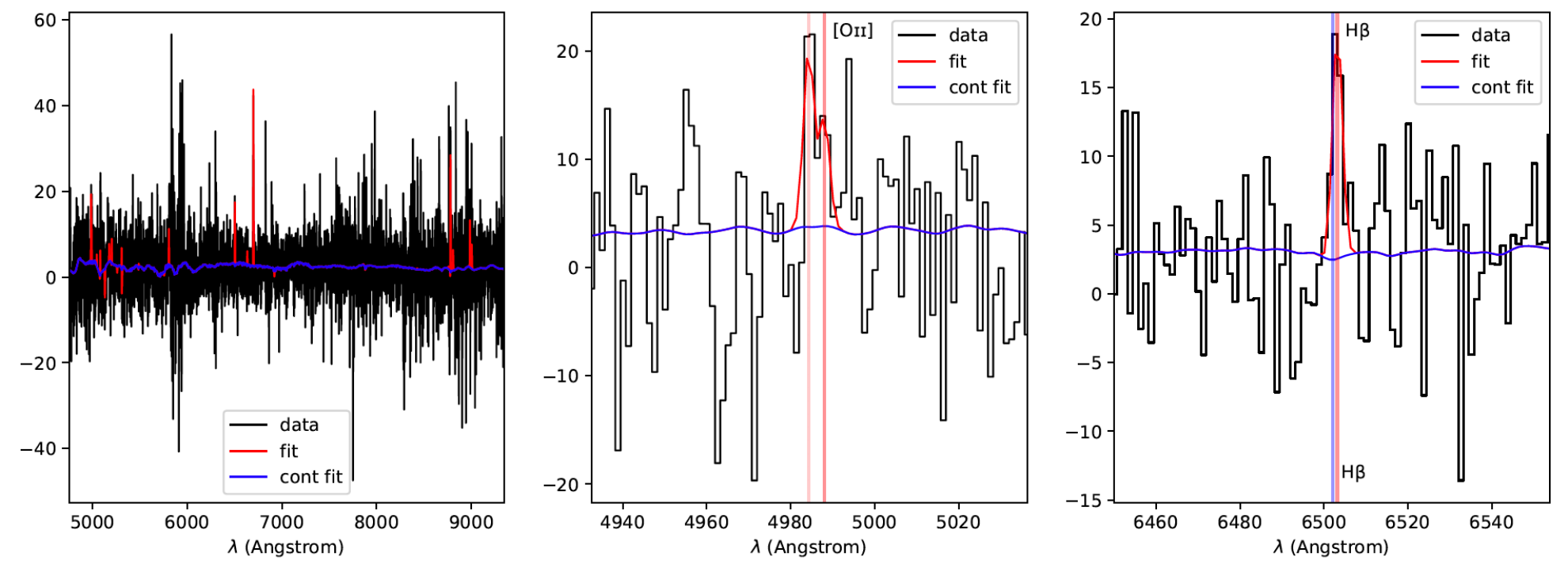}
    \caption{Examples of {\sc pyplatefit} spectral fitting, for two  sources in our catalogues. Top row shows the case of the bright $z=4.03$ galaxy in MACS0940 \citep{2019MNRAS.489.5022C} with $zconf=3$ and the bottom row a {\tt MUSELET}-selected galaxy at $z=0.337$ and $zconf=2$ in the same cluster field (M364). The left panel gives an overview of the spectrum, while the middle and right panels zoom in on specific spectral line fits. The best-fit spectrum and global continuum fit are plotted in red and blue, respectively.}
    \label{fig:pyplatefitl}
\end{figure}

The line profile fitting is performed using the {\sc lmfit} python module. The best values and errors on the redshift and spectral line parameters (flux, signal-to-noise ratio, equivalent width) are computed using a bootstrap technique. Two examples of this spectral line fitting are presented in Fig.~\ref{fig:pyplatefitl}.

\subsection{Full redshift catalogue}

We provide a summary of all spectroscopic measurements for each cluster, in the form of two tables available online (examples for the first entries are presented in Tables~\ref{tab:spectable} and \ref{tab:speclinestable}).
A main table  gives the relevant information for each source: location, redshift, and confidence, as well as photometric measurements, magnification, and multiple image identification. A companion table lists the measurements obtained by the {\sc pyplatefit} spectral fits for each source.

% Master catalogue, show an example with a few entries 

% Magnification, multiple image

% Data Release

Figure~\ref{fig:fullspec1} presents the spectroscopic catalogues overlaid on the colour {\it HST} images of each cluster, showing the spatial distribution of each redshift category with respect to the MUSE Fields as well as the region where we expect the multiple images.

Table \ref{tab:specsummary} summarises the redshift measurements both for all sources and for the high-confidence ($zconf>1$) sources. There is a clear trend for clusters having shallower MUSE data (e.g. MACS0329, Bullet) to have a lower fraction of secure redshifts ($zconf>1$). We present a redshift histogram for all sources in Fig.~\ref{fig:histz}, which is discussed further in Sect.~\ref{sec:sources}. Fourteen percent of all redshift measurements from the MUSE datacubes (and 12\% of high confidence detections, with $zconf>1$) are sources purely detected from their line emission with {\tt muselet}, that is to say they cannot be securely associated with an {\it HST} source from the photometric catalogues. While the number of such \textit{line-only} sources in the final catalogue strongly depends on the depth of the {\it HST} images (which varies from 26-29 AB for a point source depending on the field / filter), we have identified a few such sources even at the depth of the {\it HST} FF images ($\sim29$ AB). They correspond to high equivalent width Lyman-$\alpha$ emitters, comparable to sources discovered in the MUSE deep fields \citep{2017A&A...608A...1B,2017A&A...608A..10H,2018ApJ...865L...1M}.
\label{sec:lineonly} Low-redshift line-only sources are typically associated stripped gas and jellyfish galaxies in the clusters, or high equivalent width emission lines from compact galaxies.

\begin{table*}
\begin{tabular}{|c|c|c|c|c|c|c|c|c}
\hline\hline
Cluster & Nz & $z_\mathrm{min}$--$z_\mathrm{max}$ & $z<z_\mathrm{min}$& $z_\mathrm{min}<z<z_\mathrm{max}$ &
$z_\mathrm{max}<z<1.5$ & $1.5<z<2.9$ & $z>2.9$ \\
\hline
A2744 & 506 (\textbf{471})  & 0.280--0.340 & 28 (\textbf{28})  & 158 (\textbf{153})  & 115 (\textbf{113})  & 29 (\textbf{23})  & 176 (\textbf{154}) \\
A370 & 572 (\textbf{546})  & 0.340--0.410 & 62 (\textbf{60})  & 244 (\textbf{227})  & 148 (\textbf{144})  & 20 (\textbf{17})  & 98 (\textbf{98}) \\
MACS0257 & 215 (\textbf{183})  & 0.300--0.345 & 8 (\textbf{8})  & 94 (\textbf{85})  & 28 (\textbf{25})  & 4 (\textbf{3})  & 81 (\textbf{62}) \\
MACS0329 & 147 (\textbf{107})  & 0.430--0.470 & 18 (\textbf{14})  & 68 (\textbf{55})  & 20 (\textbf{16})  & 4 (\textbf{4})  & 37 (\textbf{18}) \\
%MACS0416 & 530 (\textbf{429})  & 0.380--0.420 & 26 (\textbf{20})  & 177 (\textbf{148})  & 79 (\textbf{66})  & 68 (\textbf{48})  & 180 (\textbf{147}) \\
MACS0416 & 523 (\textbf{421})  & 0.380--0.420 & 26 (\textbf{20})  & 176 (\textbf{147})  & 78 (\textbf{65})  & 71 (\textbf{50})  & 172 (\textbf{139}) \\
%BULLET & 130 (\textbf{105})  & 0.250--0.330 & 19 (\textbf{18})  & 63 (\textbf{55})  & 26 (\textbf{15})  & 4 (\textbf{2})  & 18 (\textbf{15}) \\
BULLET & 130 (\textbf{105})  & 0.250--0.330 & 20 (\textbf{19})  & 63 (\textbf{55})  & 26 (\textbf{15})  & 4 (\textbf{2})  & 17 (\textbf{14}) \\
MACS0940 & 216 (\textbf{175})  & 0.320--0.355 & 8 (\textbf{8})  & 67 (\textbf{61})  & 53 (\textbf{44})  & 3 (\textbf{2})  & 85 (\textbf{60}) \\
MACS1206 & 442 (\textbf{415})  & 0.405--0.460 & 22 (\textbf{20})  & 186 (\textbf{171})  & 119 (\textbf{116})  & 24 (\textbf{21})  & 91 (\textbf{87}) \\
RXJ1347 & 542 (\textbf{450})  & 0.420--0.485 & 77 (\textbf{69})  & 152 (\textbf{138})  & 107 (\textbf{97})  & 24 (\textbf{17})  & 182 (\textbf{129}) \\
SMACS2031 & 158 (\textbf{138})  & 0.325--0.360 & 13 (\textbf{11})  & 60 (\textbf{57})  & 19 (\textbf{17})  & 4 (\textbf{4})  & 62 (\textbf{49}) \\
SMACS2131 & 187 (\textbf{157})  & 0.410--0.480 & 18 (\textbf{16})  & 92 (\textbf{76})  & 33 (\textbf{30})  & 6 (\textbf{3})  & 38 (\textbf{32}) \\
MACS2214 & 189 (\textbf{159})  & 0.480--0.520 & 21 (\textbf{17})  & 81 (\textbf{71})  & 34 (\textbf{32})  & 8 (\textbf{2})  & 45 (\textbf{37}) \\
\hline
%Total & 3834 (\textbf{3335})  & -- & 320 (\textbf{289})  & 1442 (\textbf{1297})  & 781 (\textbf{715})  & 198 (\textbf{146})  & 1093 (\textbf{888}) \\
Total & 3827 (\textbf{3327})  & -- & 321 (\textbf{290})  & 1441 (\textbf{1296})  & 780 (\textbf{714})  & 201 (\textbf{148})  & 1084 (\textbf{879}) \\

\hline
\end{tabular}
\caption{\label{tab:specsummary}Summary of all spectroscopic measurements. For each cluster we provide the total number of redshifts measured as well as in separate redshift bins: in the foreground, in the cluster, and in the background. The cluster redshift limits are defined from the velocity distributions and given as $z_\mathrm{min}-z_\mathrm{max}$. Numbers given in boldface are limited to the sources with high confidence redshifts ($zconf>1$).}
\end{table*}

\begin{figure*}[htpb]
    \centering
    \includegraphics[width=7cm]{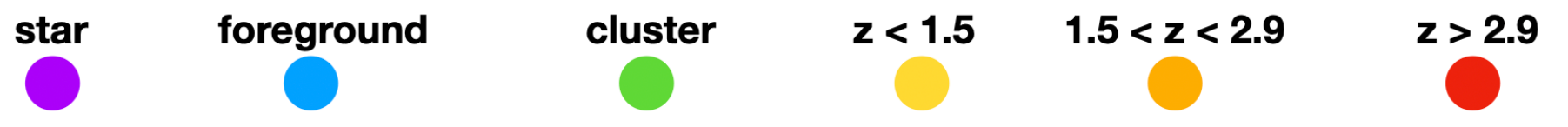}
    \includegraphics[width=11.7cm]{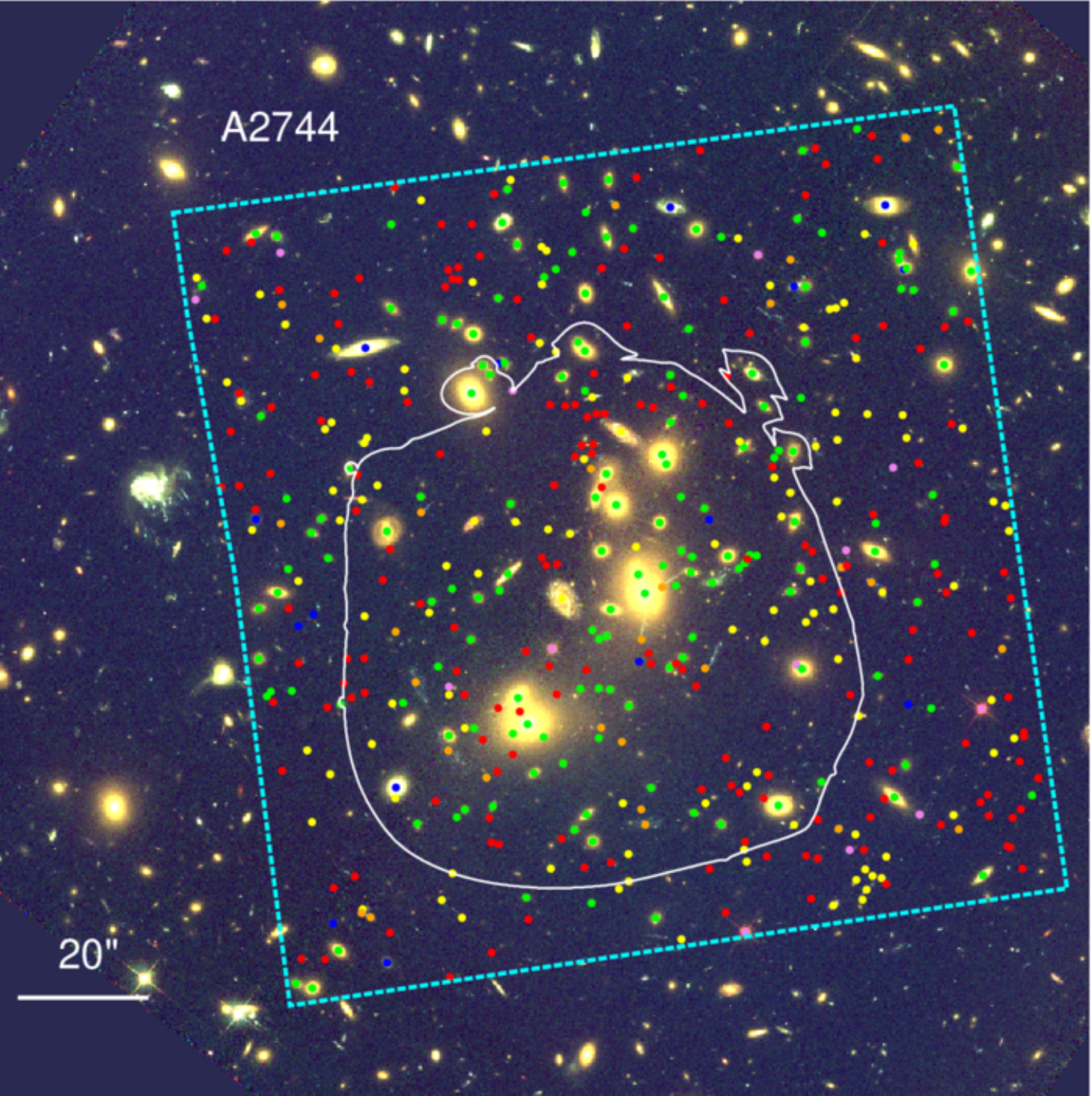}
    \includegraphics[width=11.7cm]{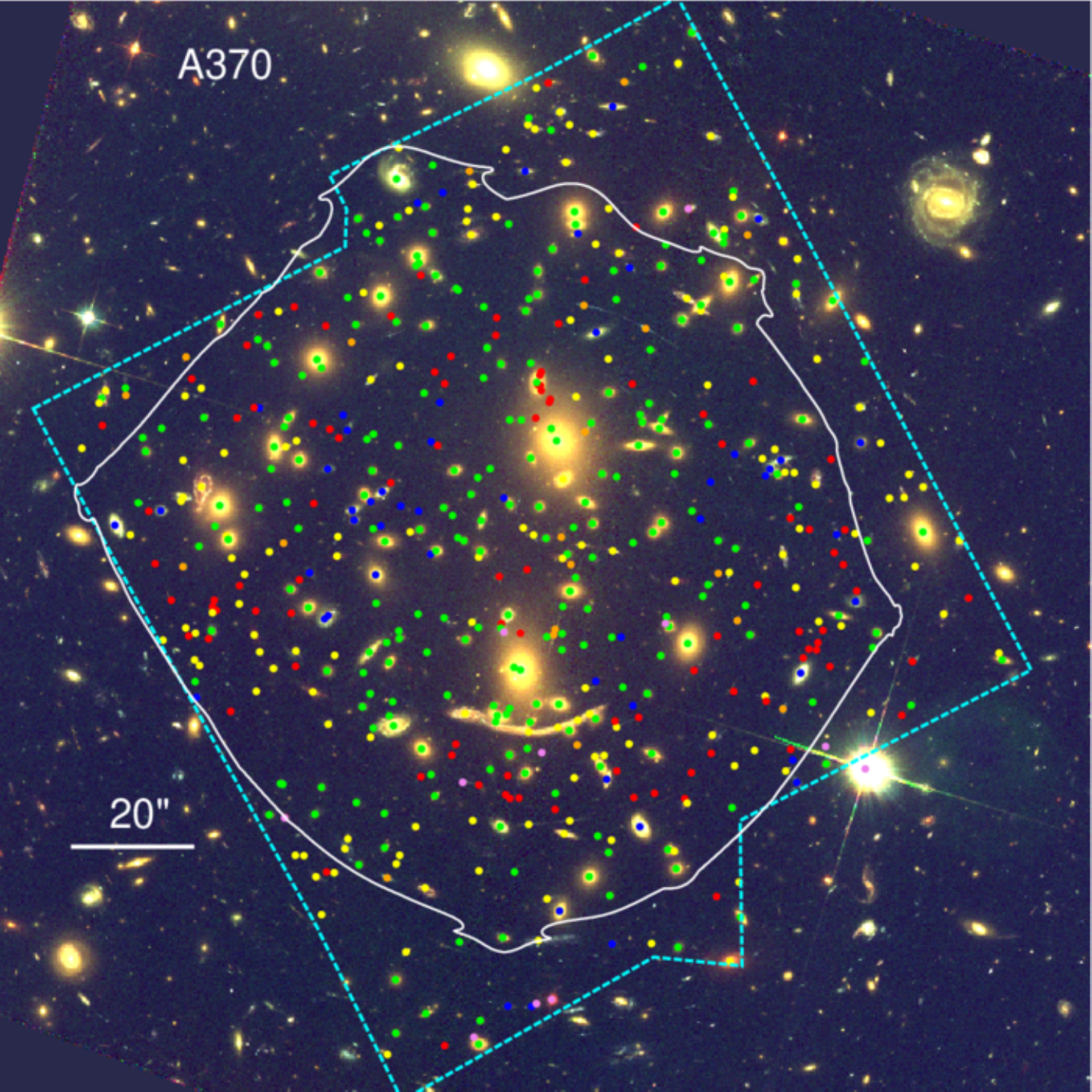}
    \caption{Location of MUSE spectroscopic redshifts over {\it HST} F606W-F814W-F160W colour composites. North is up and east is left.  The cyan dashed lines represent the limits of the MUSE datasets, while the solid white contour highlights the region where strongly-lensed multiple images are expected.}
    \label{fig:fullspec1}
\end{figure*}

\begin{figure*}[htpb]\ContinuedFloat
    \centering
    \includegraphics[width=12cm]{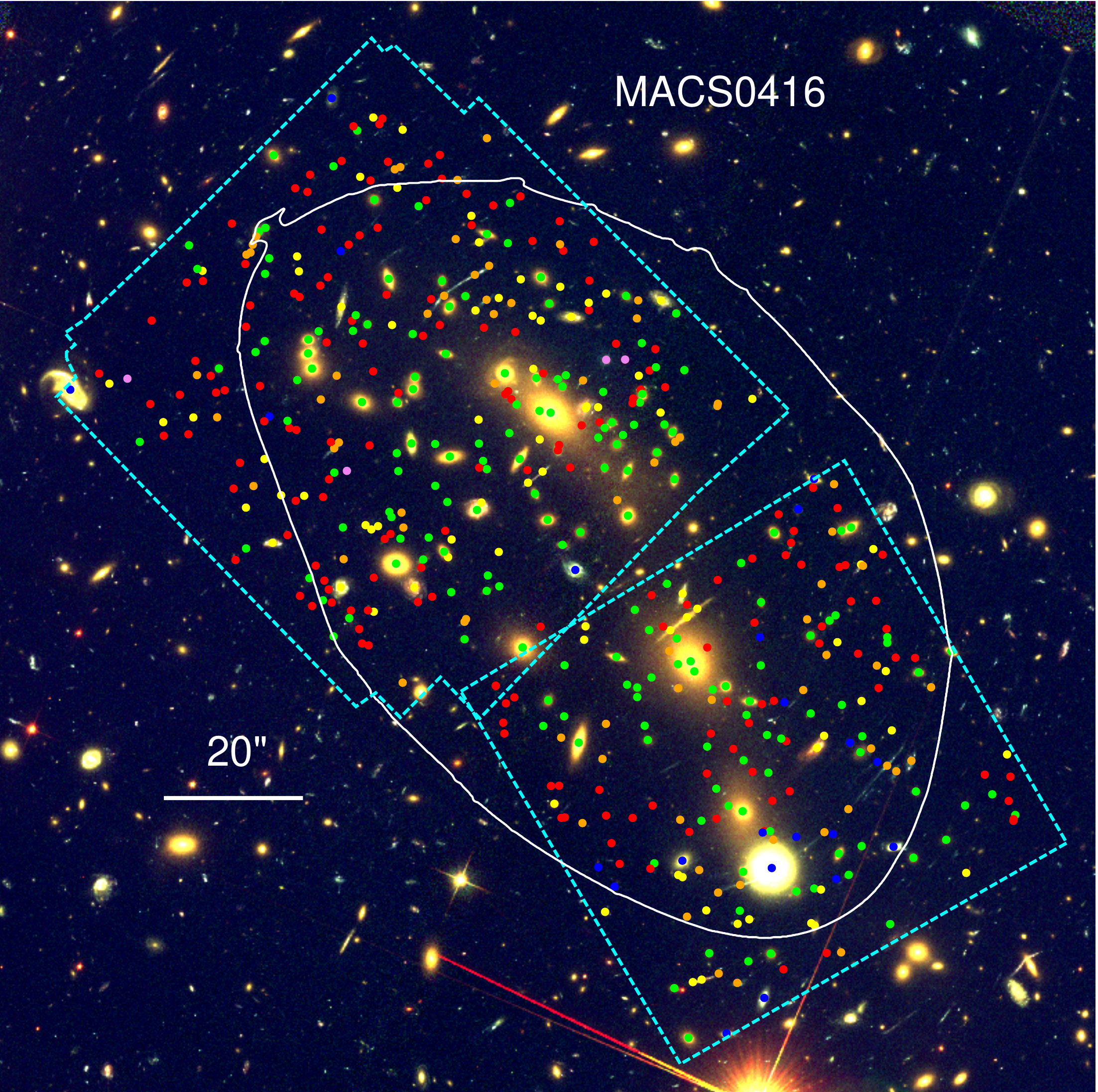}
    \includegraphics[width=12cm]{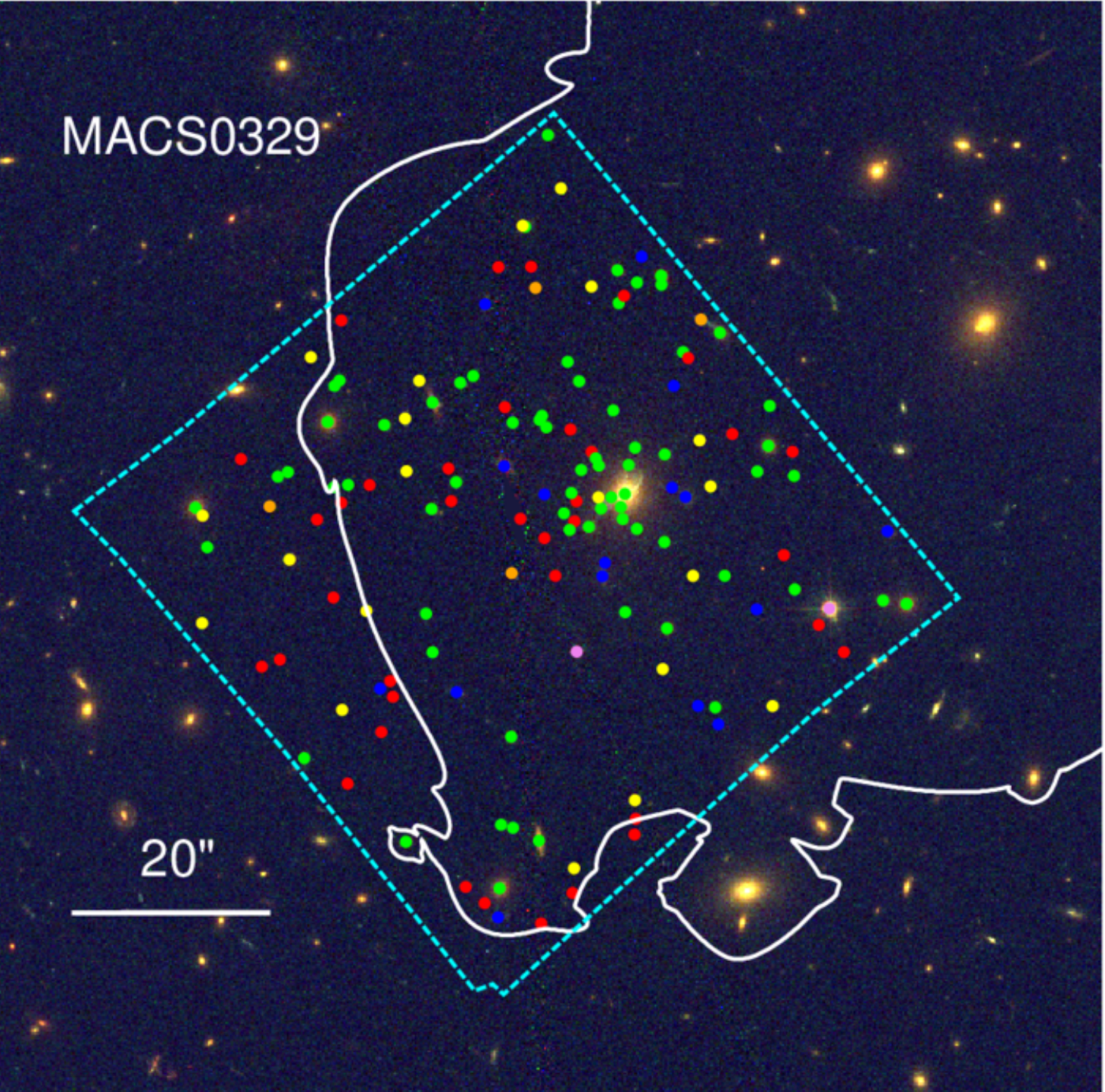}
%\caption{Continuation of Fig.\ref{fig:fullspec1}.}
\caption{(continued)}
\end{figure*}

\begin{figure*}[htpb]\ContinuedFloat
    \centering
\includegraphics[width=12cm]{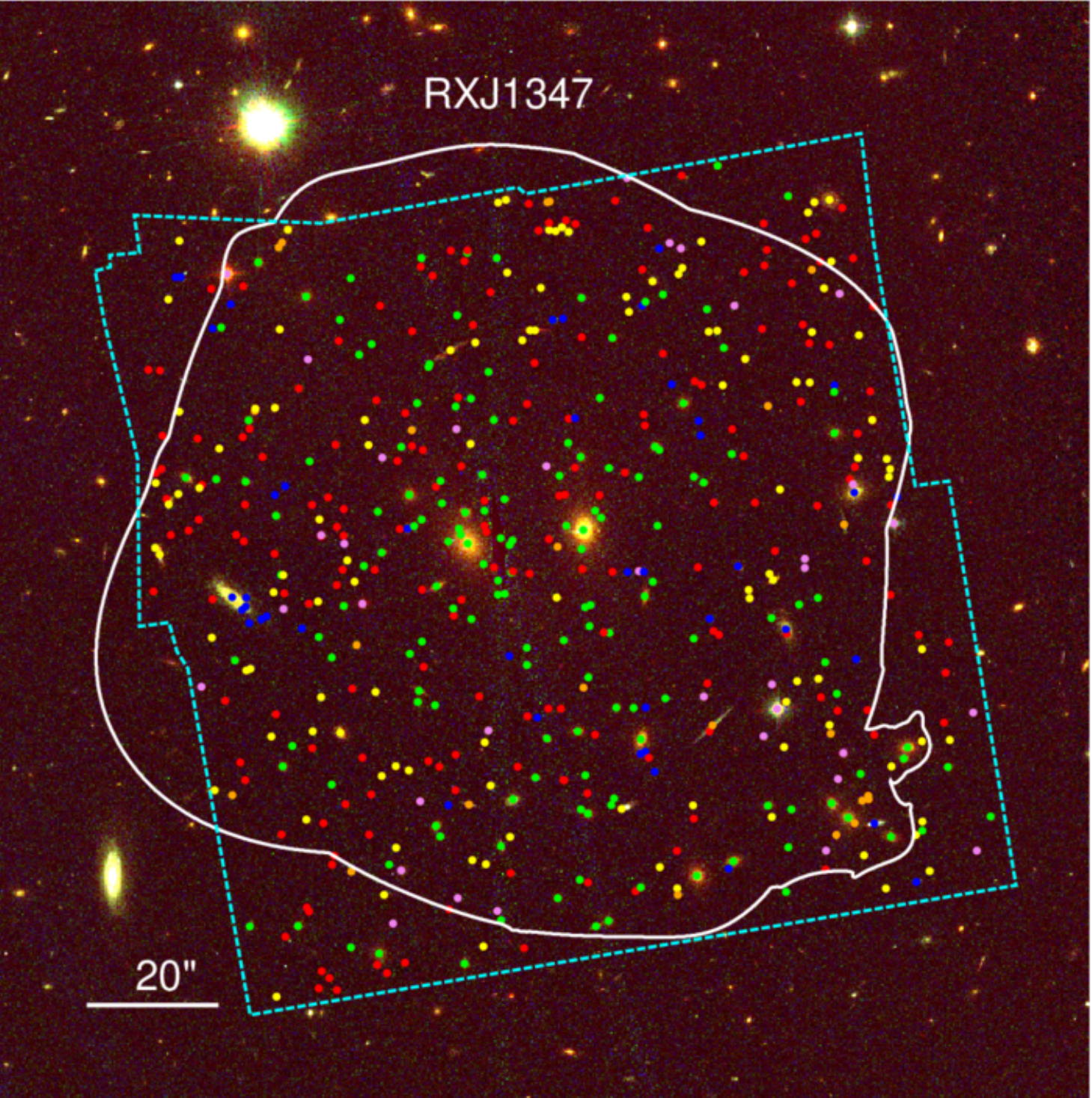}
\includegraphics[width=12cm]{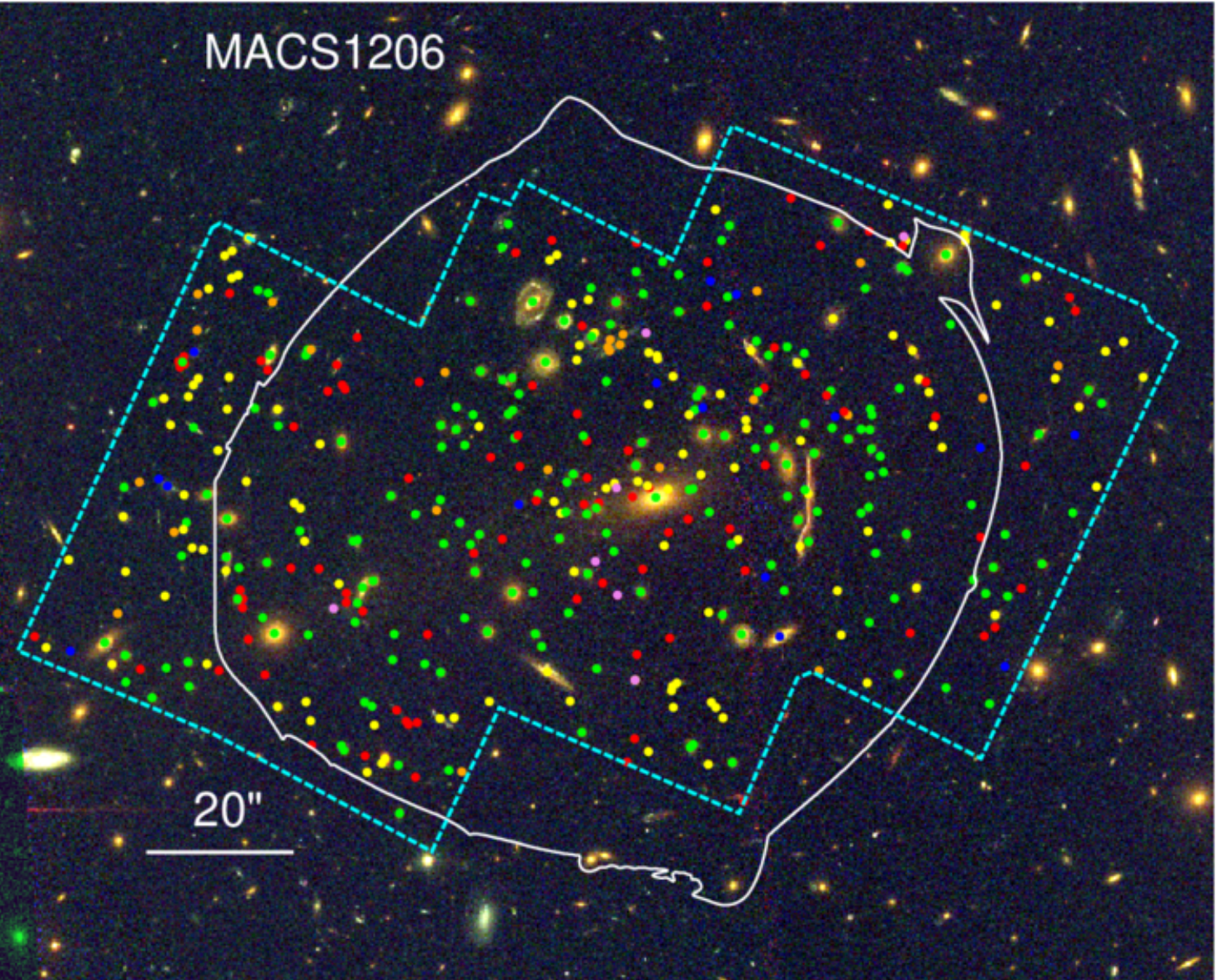}
     %\caption{Continuation of Fig.\ref{fig:fullspec1}.}
     \caption{(continued)}
\end{figure*}

\begin{figure*}[htpb]\ContinuedFloat
    \begin{minipage}{7.8cm}
    \includegraphics[width=7.8cm]{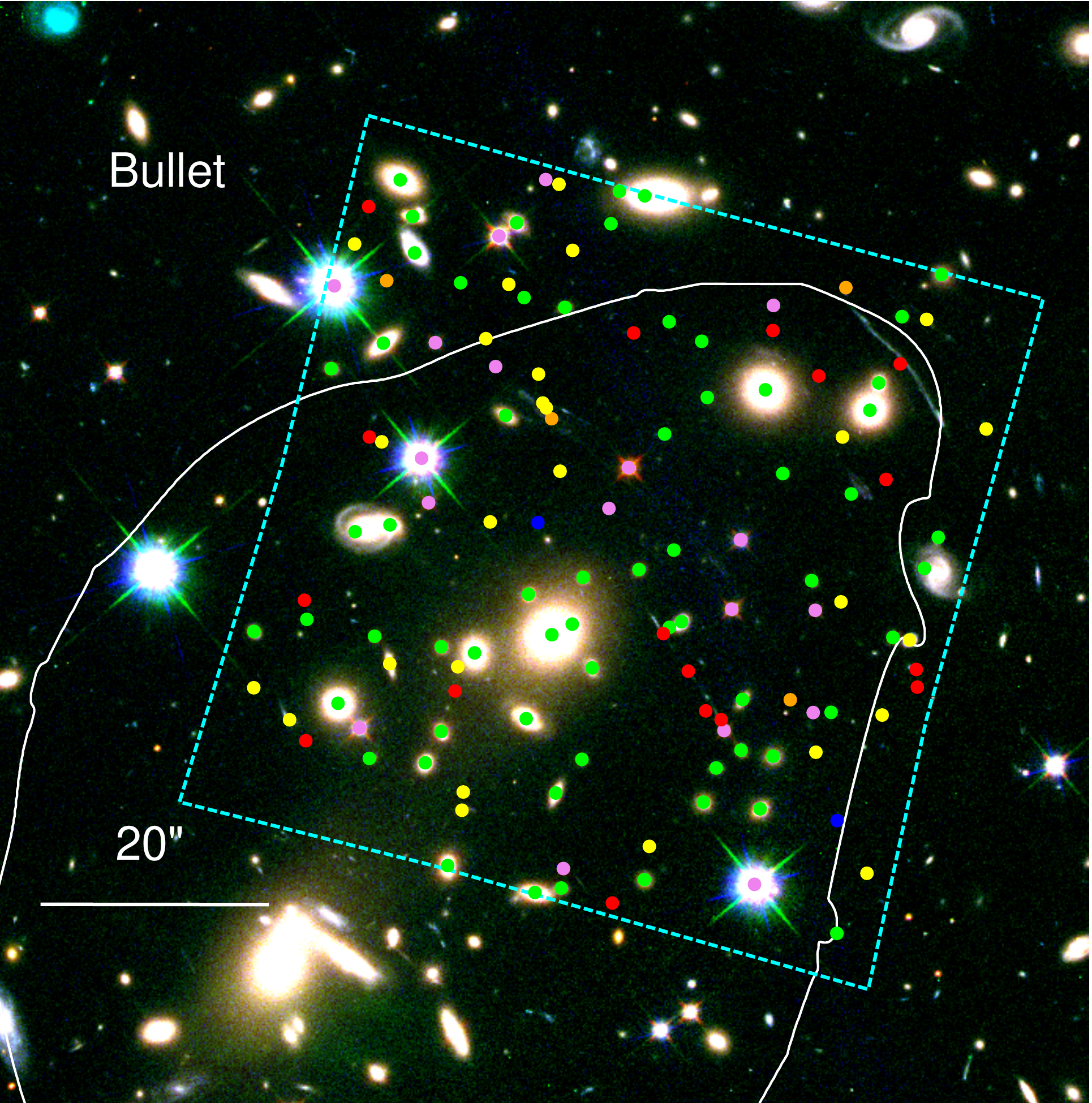}
    \includegraphics[width=7.8cm]{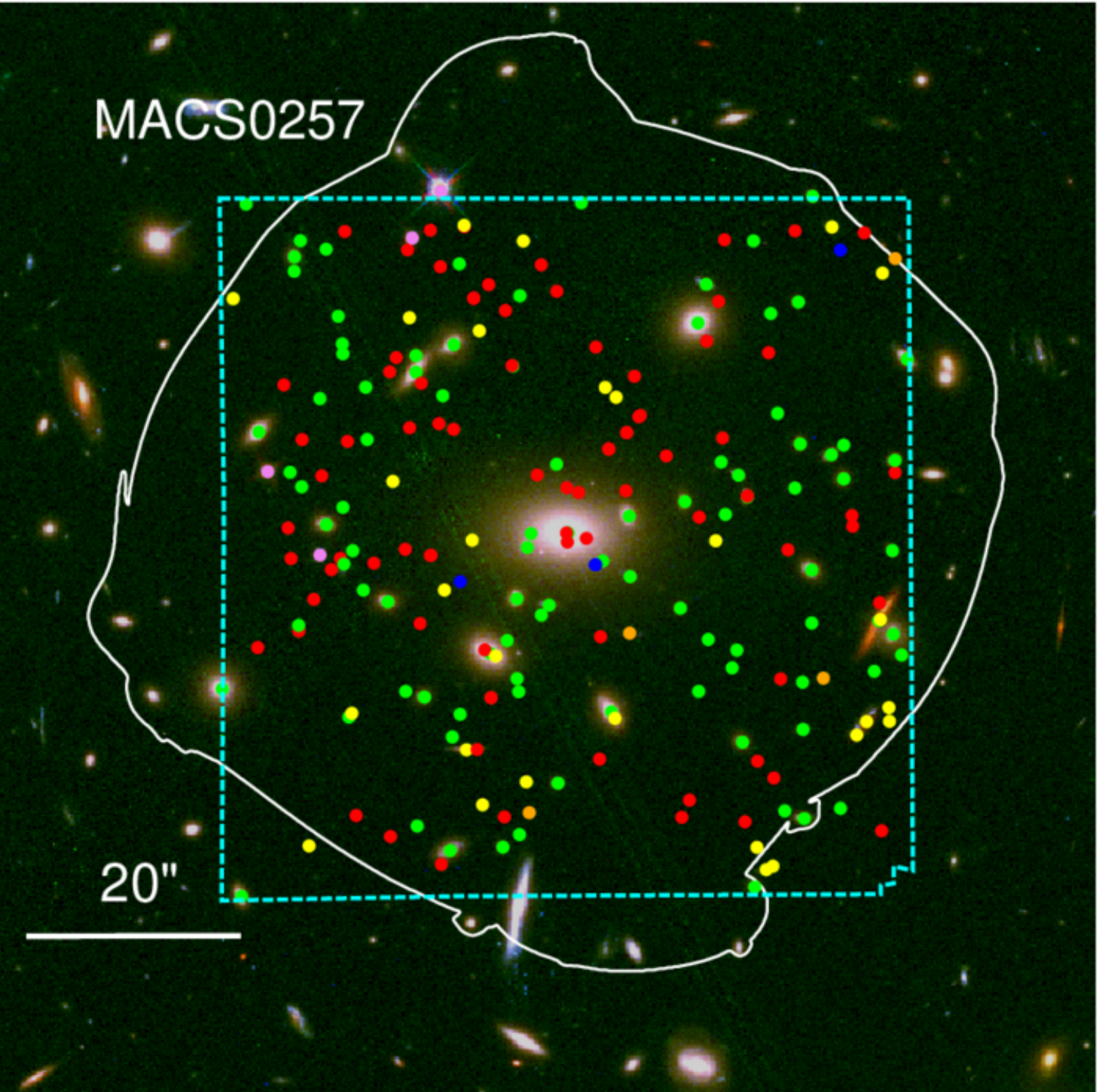}
    \includegraphics[width=7.8cm]{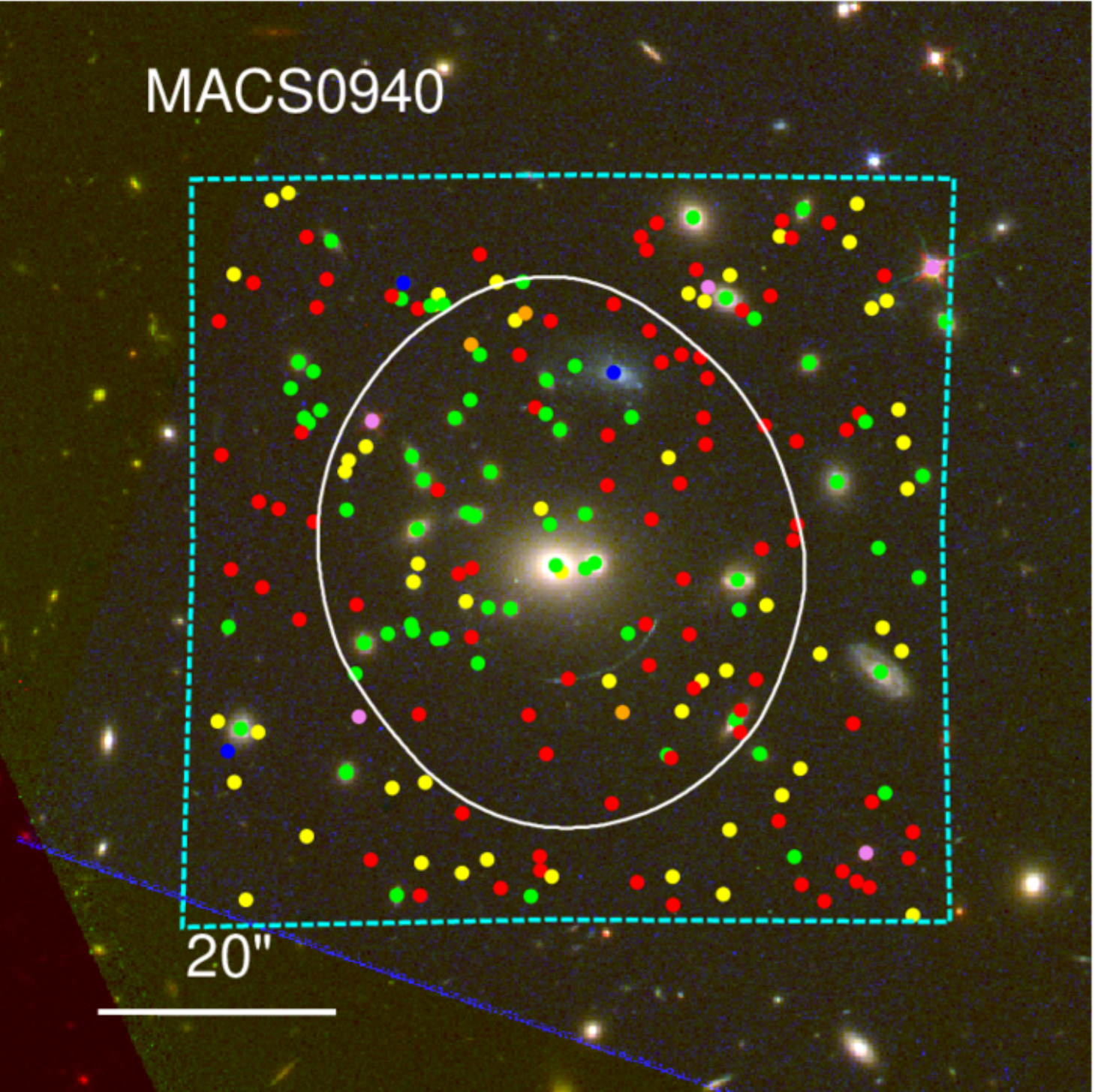}
    \end{minipage}
    \begin{minipage}{7.8cm}
    \includegraphics[width=7.8cm]{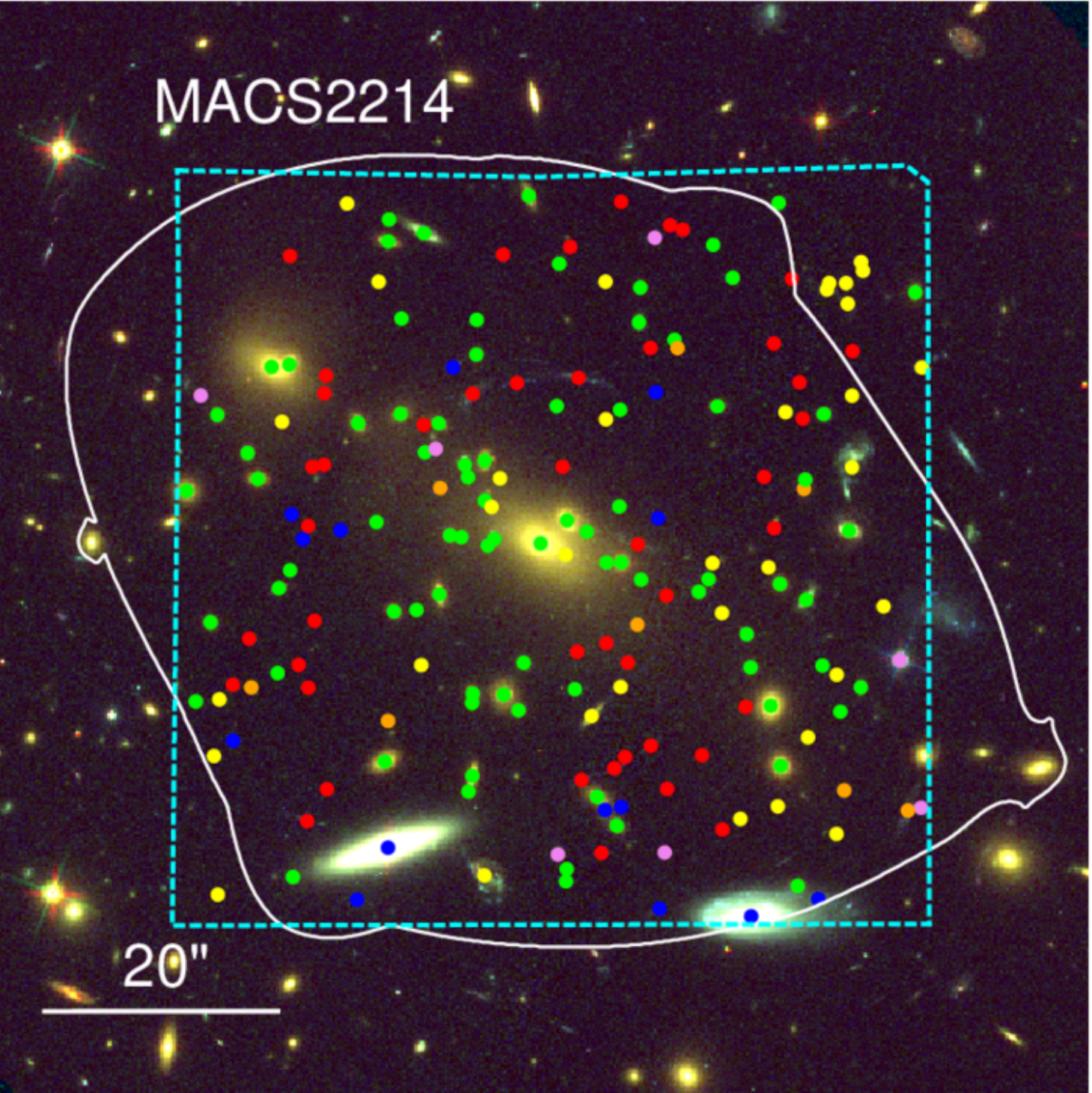}
    \includegraphics[width=7.8cm]{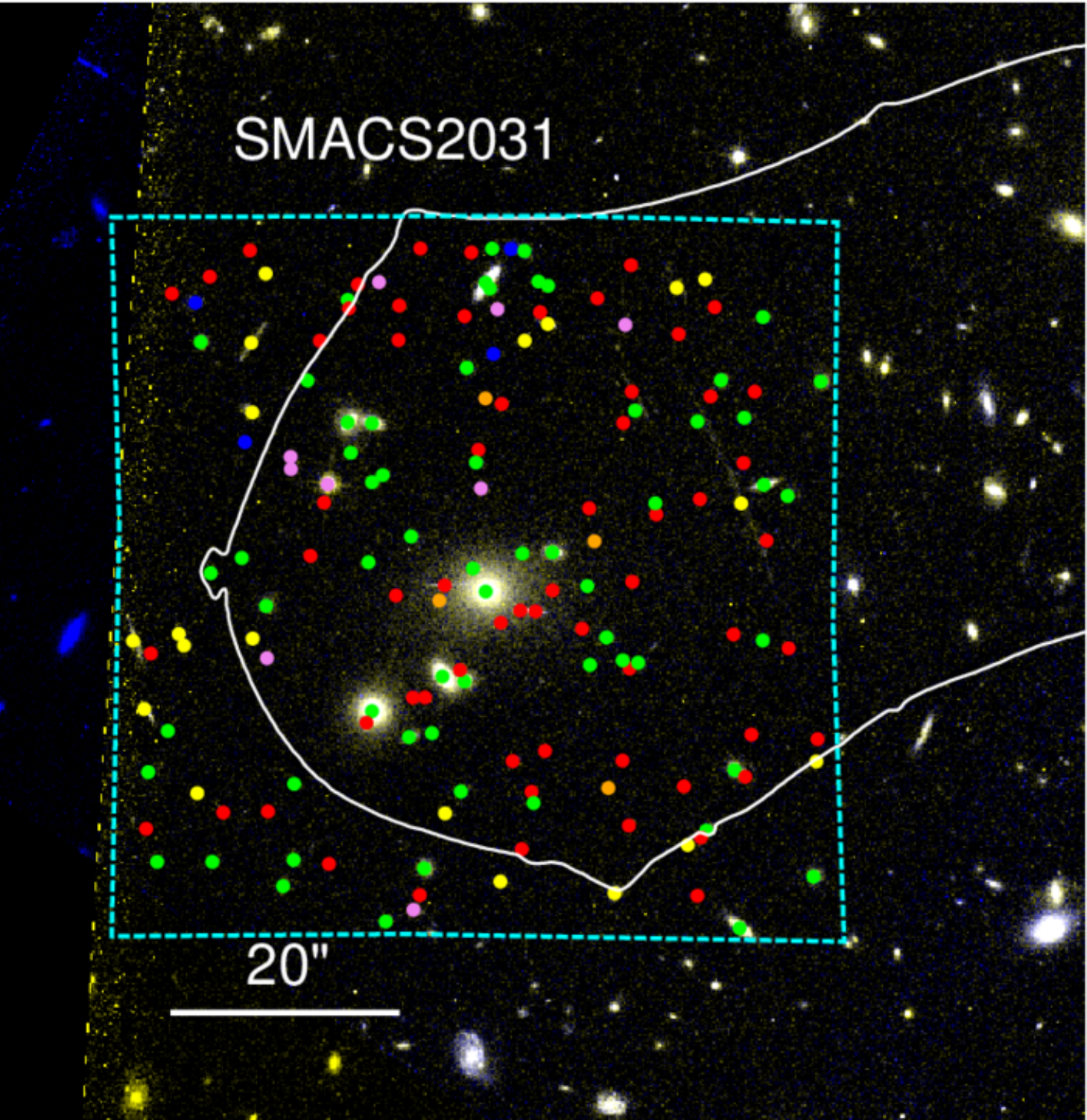}
    \includegraphics[width=7.8cm]{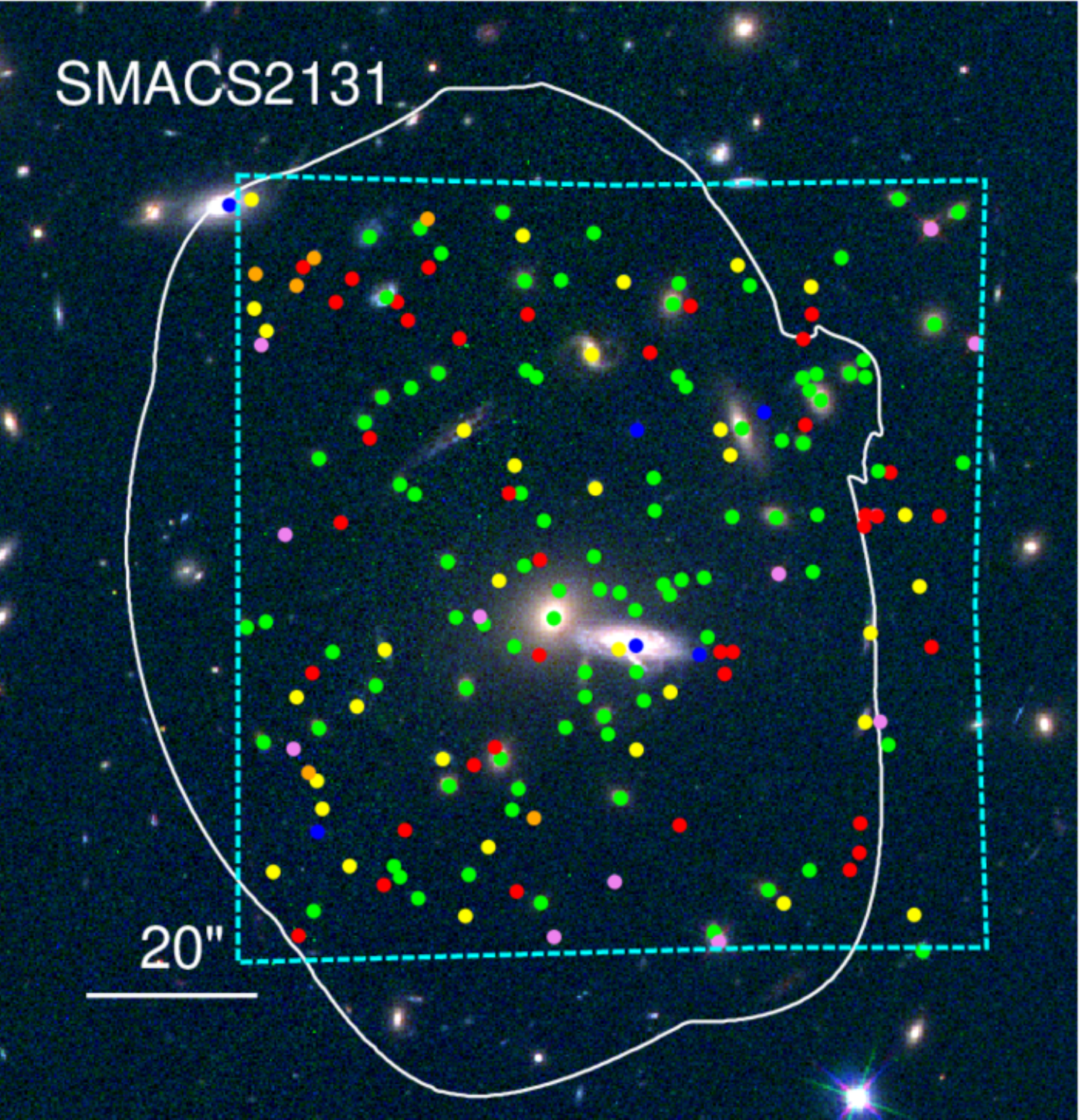}
    %\caption{Continuation of Fig.\ref{fig:fullspec1}.}
    \end{minipage}
    \caption{(continued)}
\end{figure*}
\clearpage

\section{Mass models}
\label{sec:massmodels}

%The physical interpretation of the measured spectrophotometric properties of background lensed sources requires a knowledge of the magnification factor, and more generally of the distortion induced by the massive cluster cores. We make use of the numerous multiple images identified in the MUSE catalogues to constrain a parametric model of the total mass distribution in the clusters. We describe in this Section the procedure we use for this analysis, which is similar to the mass models of the FF clusters published by the CATS (Clusters As Telescopes) team (e.g. \citealt{2014MNRAS.444..268R,2014MNRAS.443.1549J,2015MNRAS.452.1437J,2016MNRAS.457.2029J,2017MNRAS.469.3946L,2018MNRAS.473..663M,2019MNRAS.485.3738L}) and makes use of the latest version (v7.1) of the public {\sc Lenstool} software\footnote{Publicly available at:\\ \url{https://git-cral.univ-lyon1.fr/lenstool/lenstool},\\ more details at:\\ \url{https://projets.lam.fr/projects/lenstool/wiki/}} 
%\citep{1996ApJ...471..643K,2007NJPh....9..447J}. 
In order to interpret the measured spectrophotometric properties of lensed background sources in a more physical way, we must first account for (and correct) both the magnification factors and the general shape distortions that massive cluster cores impart to the observations.  To do this, we generate parametric models of each cluster’s total mass distribution, using the numerous multiple images identified in the MUSE catalogues as constraints. In this section, we describe the procedure we use for this analysis, which is similar to the process employed by the CATS (Clusters As Telescopes) team to generate mass models of the FF clusters (e.g. \citealt{2014MNRAS.444..268R,2014MNRAS.443.1549J,2015MNRAS.452.1437J,2016MNRAS.457.2029J,2017MNRAS.469.3946L,2018MNRAS.473..663M,2019MNRAS.485.3738L}) and makes use of the latest version (v7.1) of the public {\sc Lenstool} software\footnote{Publicly available at:\\ \url{https://git-cral.univ-lyon1.fr/lenstool/lenstool},\\ more details at:\\ \url{https://projets.lam.fr/projects/lenstool/wiki/}} 
\citep{1996ApJ...471..643K,2007NJPh....9..447J}. 

%The following assumptions were used: we restrict the constraints of the model to the spatial location of multiply-imaged systems, either confirmed spectroscopically (\textit{gold} systems, having at least 1 image with a spectroscopic redshift) or bright {\it HST} sources with an unambiguous lensing confirmation, as reported in previous works (\textit{silver} systems in the notation used in the FFs, e.g. \citealt{2017ApJ...837...97L}). Indeed, the large majority of known systems will have new spectroscopic redshifts with MUSE. This conservative approach allows us to reach a precision sufficient to assess additional multiple images in the spectroscopic catalogues, as discussed in Sect.~\ref{sec:mulimages}. We also make use of the preliminary mass models and known {\it HST} arcs from references listed in Table~\ref{tab:clusters_table} as a starting point for our strong-lensing analysis.
To reduce uncertainty, we use the spatial positions of only high-confidence multiple-image systems as constraints for a given model. Specifically, we include systems that are either confirmed spectroscopically (having at least one image with a spectroscopic redshift; called \textit{gold} systems), or bright {\it HST} sources without a spectroscopic redshift, but with an obvious, unambiguous lensing configuration, similar to the ones reported as \textit{silver} systems in previous works. Here, \textit{gold} and \textit{silver} are based on the notation used in previous FF studies (e.g. \citealt{2017ApJ...837...97L}). We note that the vast majority of known systems will have new spectroscopic redshifts thanks to MUSE, so this limitation does not strongly bias the final model. In fact, this conservative approach allows us to reach a sufficient precision such that we can assess additional multiple images in the spectroscopic catalogues, as discussed in Sect.~\ref{sec:mulimages}. We also make use of the preliminary mass models and known {\it HST} arcs from references listed in Table~\ref{tab:clusters_table} as a starting point for our strong-lensing analysis.

\subsection{Model parametrisation}

\textsc{Lenstool} uses a  Monte Carlo Markov Chain (MCMC) to sample the posterior probability distribution of the model, expressed as a function of the likelihood defined in \citet{2007NJPh....9..447J}. In practice, we minimise the distances in the image plane:

\begin{eqnarray}
\chi^{2} = \sum\limits_{i,j}  \frac{\norm {\Vec{\theta^{(i,j)}_{obs}} - \Vec{{\theta^{(i,j)}_{pred}}}}^{2}}{\sigma_{pos}^{2}}\ ,
\end{eqnarray}

\noindent with $\Vec{\theta^{(i,j)}_{obs}}$ and $\Vec{\theta^{(i,j)}_{pred}}$ representing the observed and predicted vector positions of multiple image $j$ in system $i$, respectively. Furthermore, $\sigma_{pos}$ is a global error on the position of all multiple images, which we fix at 0\farcs5. This value corresponds to the typical uncertainty in reproducing the strongly-lensed images, which is affected by the presence of mass substructures within the cluster or along the line of sight \citep{2010Sci...329..924J}. The best model found (which minimises the $\chi^2$ value) also minimises the global rms between ${\Vec{\theta^{(i,j)}_{obs}}}$ and ${\Vec{\theta^{(i,j)}_{pred}}}$ (herafter rms$_{\rm model}$). 

In order to estimate all the values of $\Vec{\theta^{(i,j)}_{pred}}$, {\sc Lenstool} inverts the lens equation with a parametric potential which we assume to be  a combination of double Pseudo Isothermal Elliptical mass profiles (dPIE, \citealt{2007arXiv0710.5636E,2010A&A...524A..94S}). dPIE potentials are isothermal profiles which are characterised by a central velocity dispersion $\sigma_0$ and include a core radius $r_{\rm core}$ (producing a flattening of the mass distribution at the centre), and a cut radius $r_{\rm cut}$  (producing a drop-off of the mass distribution on large scales). Together with the central position ($x_c$,$y_c$) and elliptical shape (ellipticity $e$, angle $\theta$), dPIE potentials are fully-defined by 7 parameters.

%For each cluster, we include a small number (up to 3) of cluster-scale dPIE profiles, which account for the smooth large-scale components of the mass distribution, as well as individual dPIE potentials assigned on each cluster member, which represent the galaxy-scale substructure. We fix the cut radius of cluster-scale mass components to 1 Mpc, as it is basically unconstrained with strong lensing, and let all other parameters free to vary.
For each cluster, we include a small number of cluster-scale dPIE profiles (up to three) to account for the smooth large-scale components of the mass distribution. Additionally, we assign individual dPIE potentials to each cluster member, which represent the galaxy-scale substructure. We fix the cut radius of cluster-scale mass components to 1 Mpc (e.g. \citealt{2007ApJ...668..643L}), as it is generally unconstrained by strong lensing, but allow the other parameters to vary freely in the fit.

Cluster members used as galaxy-scale potentials are selected through the red sequence of elliptical galaxies, which is well-identified in a colour-magnitude diagram based on the {\it HST} photometry. More details on this selection are provided in, for instance, \citet{2014MNRAS.444..268R} for the FF clusters. For the newest lens models and most recent {\it HST} observations, we make use of the F606W-F814W vs F160W colour-magnitude diagram to select cluster members down to 0.01 L$^*$, where L$^*$ is the characteristic luminosity at the cluster redshift based on the luminosity function from \citet{2006ApJ...650L..99L}. As a further refinement to this process, we cross-check the selected members with existing redshifts in our MUSE spectroscopic catalogue.

To reduce the number of free parameters in the model, we assume that the shape parameters ($x_c,y_c,e,\theta$) of %these galaxy
the galaxy-scale potentials follow the shape of their light distribution, and use the following scaling relations on the dPIE parameters with respect to their luminosity $L$:

\begin{equation}
r_{\rm core}=0.15 {\rm kpc},\ r_{\rm cut}=r_{\rm cut}^*\ \Big({\frac{L}{L^*}}\Big)^{1/2},\ \sigma_0=\sigma_0^*\ \Big({\frac{L}{L^*}}\Big)^{1/4}
\label{eq:scaling}
\end{equation}
which follow from the \citet{1976ApJ...204..668F} relation of elliptical galaxies and the assumption of a constant mass-to-light ratio. The fixed value for $r_{\rm core}$ is negligible and follows from the discussion in \citet{2007ApJ...668..643L}.

The mass models are constructed through an iterative process, using a single cluster-scale dPIE profile at first to fit the most confident sets of multiply imaged systems. Other systems are tested and included as additional constraints, and a second or third cluster-scale dPIE profile is added when it has a significant effect in reducing rms$_{\rm model}$. %Regarding galaxy-scale mass components, 
While most galaxy-scale mass components follow the general scaling relations presented in Equation \ref{eq:scaling}, we do fit the mass parameters of some individual galaxies separately, in special cases where we can constrain these values using additional information outside of the scaling relation.
%we fit the mass parameters of individual galaxies that we can constrain outside of the scaling relation. 
This is particularly the case for (a) BCGs, which are known not to follow the usual scaling relations of elliptical galaxies, (b) cluster members lying very close ($<\sim5$\arcsec) to multiple images, which locally influence their location  (and sometimes produce galaxy-galaxy lens systems), and (c) additional galaxies located slightly in the foreground or background of the cluster, which produce a local lensing effect on their surroundings but will not follow the same scaling relations.

Finally, for the most complex clusters (A370, MACS1206, RXJ1347) we have tested the addition of an external shear potential. Typically used when modelling galaxy-galaxy lensed systems (e.g. \citealt{2018MNRAS.479.2630D}), this potential accounts for an unknown effect of the large scale mass  environment surrounding the cluster, in the form of a constant shear $\gamma_{\rm ext}$ at an angle $\Phi_{\rm ext}$. However it does not bring any additional mass to the model (see also the discussions in \citealt{2018MNRAS.473..663M} and \citealt{2019MNRAS.485.3738L}).

\subsection{Results of the strong lensing analysis}

In Table~\ref{tab:modelsum}, we summarise the number of strong-lensing constraints (systems and images) which were confirmed by our strong lensing analysis for each cluster, along with the rms from the best fit models. The complete list of multiple images and the best-fit parameters of the mass models are detailed in Appendix \ref{app:mul}.

\begin{table*}
    \centering
    \begin{tabular}{c|c|c|c|c|c|c|c}
    \hline\hline             
    Cluster     & Nsys  & Nimg  & rms$_{\rm model}$  & $\theta_{\rm E}$ ($z=4$) & $M(\theta<\theta_{\rm E} (z=4))$ & $\sigma_{\rm model}$ & $\sigma_{\rm dynamics}$ \\
    & & & [\arcsec] & [\arcsec] & [10$^{14}$ M$_\odot$] & [km\,s$^{-1}$] & [km\,s$^{-1}$] \\
    \hline
A2744 &29 (\textbf{29}) &  83 (\textbf{83})& 0.67 & 23.9 & 0.62$\pm$0.04 & 1394$\pm$37 & 1357$\pm$138  \\ % 15.7$\pm$0.3
A370 & 45 (\textbf{39}) & 137 (\textbf{122}) & 0.78 & 39.5 & 2.53$\pm$0.03 & 1976$\pm$50 & 1789$\pm$109 \\ %33.7$\pm$0.1
MACS0257 & 25 (\textbf{25}) & 81 (\textbf{78}) & 0.78 & 28.1  & 1.10$\pm$0.02& 1487$\pm$30 & 1633$\pm$164\\ %27.5$\pm$0.1 0.322
MACS0329 & 9 (\textbf{8}) & 24 (\textbf{21}) & 0.53 & 29.9  & 1.56$\pm$0.08 & 1621$\pm$68 & 1231$\pm$130 \\ %24.2$\pm$0.1 0.450
MACS0416 & 71 (\textbf{71}) & 198 (\textbf{198}) & 0.58 & 29.9 & 1.05$\pm$0.04 & 1559$\pm$25 & 1277$\pm$93\\ %13.0$\pm$0.2 0.397
Bullet & 15  (\textbf{5}) & 40 (\textbf{15}) & 0.39 & 25.8  & 0.73$\pm$0.02 & 1601$\pm$90 & 1283$\pm$273\\ %12.4$\pm$0.8 0.296
MACS0940 & 7 (\textbf{7}) & 22 (\textbf{22}) & 0.23 & 10.9 & 0.17$\pm$0.01 & 859$\pm$291 & 856$\pm$192\\ %10.6$\pm$0.1 0.335
MACS1206 & 37 (\textbf{36}) & 113 (\textbf{110})& 0.52 & 31.5 & 1.76$\pm$0.06 & 1603$\pm$69 & 1842$\pm$184\\ %29.3$\pm$0.2 0.438
RXJ1347 & 36 (\textbf{35}) & 121 (\textbf{119})& 0.84& 38.6 & 2.75$\pm$0.02 & 1684$\pm$48 & 1097$\pm$121\\ %36.7$\pm$0.1 0.451 
SMACS2031 & 13 (\textbf{13}) & 46 (\textbf{45}) & 0.33 & 26.1 & 0.76$\pm$0.05 & 1683$\pm$24 & 1531$\pm$210\\ %15.0$\pm$0.1 0.331
SMACS2131 & 10 (\textbf{10}) & 28 (\textbf{29}) & 0.79 & 24.4 & 1.08$\pm$0.05 & 1270$\pm$40 & 1378$\pm$408\\ %22.3$\pm$0.1 0.442
MACS2214 & 15 (\textbf{15}) & 46 (\textbf{46}) & 0.41 & 22.5 & 0.99$\pm$0.02 & 1578$\pm$60 & 1359$\pm$224\\ %20.6$\pm$0.1 0.502
\hline
Total & 312 (\textbf{293}) & 939 (\textbf{888}) & -- & -- & --  & -- & -- \\ 
    \end{tabular}
    \caption{Global cluster properties measured from our strong lensing mass model. From left to right: number of multiply-imaged systems, total number of multiple images, global rms of the model, Einstein radius at $z=4$, total enclosed mass within the Einstein radius, global velocity dispersion measured from the lens model or the cluster dynamics. Numbers in boldface are confirmed with spectroscopy}
    \label{tab:modelsum}
\end{table*}

Overall, these models are among the most constrained (in terms of number of independent multiply imaged systems) strong-lensing cluster cores to date. We stress that this occurs not only for clusters with deep {\it HST} imaging, such as those available in the FFs, but even clusters in our sample observed with snapshot {\it HST} images are densely constrained; for example, MACS0257 %which 
has 25 systems with spectroscopic redshifts, producing 81 total multiple images over a single 1 arcmin$^2$ MUSE field. This is entirely dominated by the population of multiply-imaged Lyman-$\alpha$ emitters, which are very faint in {\it HST} but are easily revealed by the MUSE IFU.

Such a high density of constraints allows for a precise measurement of the mass profile, with a typical statistical error of $<1\%$ (e.g. \citealt{2014MNRAS.443.1549J,2019A&A...632A..36C}). The improvement is typically a factor of $\sim$5 with respect to the models prior to MUSE observations (e.g. \citealt{2015MNRAS.446L..16R}). This has many applications, allowing us for example to test different parametrisations of the mass models and line of sight effects (e.g. \citealt{2018A&A...614A...8C}) to improve the rms$_{\rm model}$ even further and reduce systematics. Some discussion on these effects were presented for the clusters Abell 2744 and Abell 370 in \citet{2018MNRAS.473..663M} and \citet{2019MNRAS.485.3738L}, respectively. One route to improve the models further is to combine the parametric mass models with a non-parametric (grid-like) mass distribution using a perturbative approach %(Beauchesne et al. 2020 submitted).
(\citealt{Beauchesnesubmitted} submitted).

Another application of these very-constrained models is to make use of the high density of constraints at the centre of the cluster, and probe the inner slope of the mass distribution to test predictions from $\Lambda$-CDM simulations \citep{2019AandA...632A..36C}. Several clusters in our sample show so-called 'hyperbolic-umbilic' lensing configurations \citep{2009MNRAS.399....2O}, which are rarely seen and provide us with very important constraints in the cluster core \citep{2009A&A...498...37R}. Finally, multiply-imaged systems appearing in the same regions of the cluster but originating from very different source plane redshifts are important for strong-lensing cosmography studies \citep{2010Sci...329..924J,2016A&A...587A..80C,2017MNRAS.470.1809A}.

A full discussion of the 2D mass distribution of individual clusters is beyond the scope of this paper. However, in Table \ref{tab:modelsum} we provide the main cluster parameters derived from our lens model which characterise their lensing power: namely the equivalent Einstein radius $\theta_{\rm E}$ at $z=4$ (defined from the area within the critical curve $\Sigma$, as $\theta_{\rm E}=\sqrt{\Sigma/\pi}$), and the enclosed projected mass within this radius $M<\theta_{\rm E}$. We selected $z=4$ as it is the average DLS/DS (the lensing efficiency factor, ratio of the distance between the lens and the source over the distance to the source) for Lyman-$\alpha$ emitters which contribute to the majority of constraints.

\section{Source properties}
\label{sec:sources}

\subsection{Redshift distribution}

\begin{figure*}
    \centering
    \includegraphics[width=9cm]{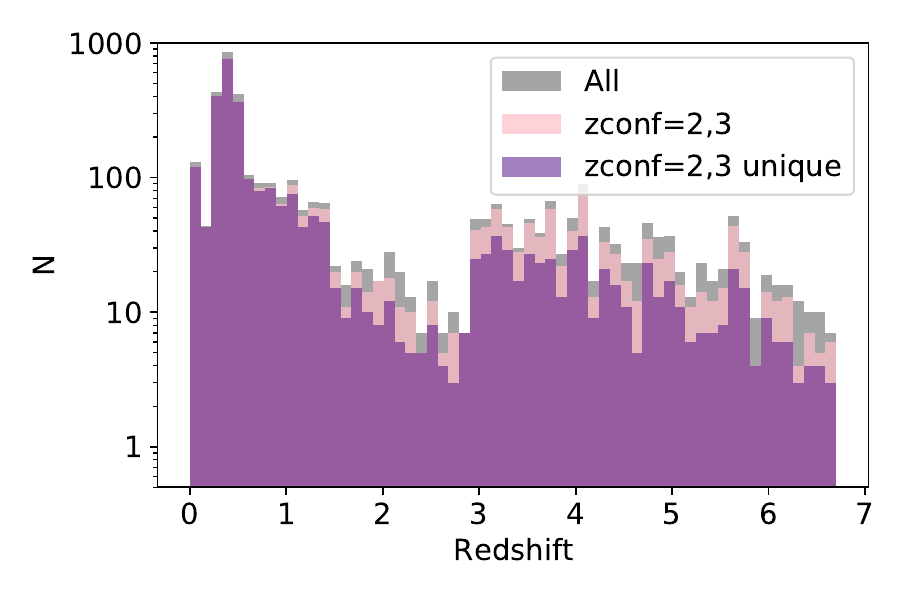}
    \includegraphics[width=9cm]{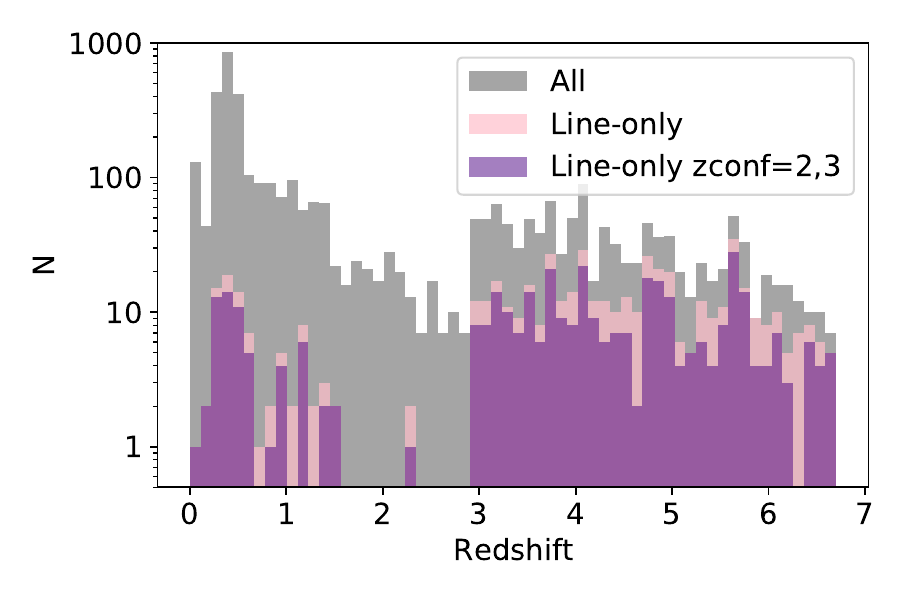}
    \caption{Global redshift distribution of MUSE sources in the lensing cluster fields. Left panel: comparison between the overall distribution and sources with high confidence redshifts ($zconf=2,3$), the latter before and after ('unique') accounting for image multiplicity. Right panel: comparison between the overall distribution and the line-only sources (pure line emitters without any {\it HST} counterpart, see Sect.\,\ref{sec:lineonly}).}
    \label{fig:histz}
\end{figure*}

The complete spectroscopic dataset of the 12 clusters amount to more than 3200 high-confidence redshifts, %which are coming from 
consisting of galaxies either in the foreground, in the cluster itself, or in the background. Overall, cluster members dominate the redshift distribution; this is clearly apparent in %the summary 
Table~\ref{tab:specsummary}, where cluster members account for $\sim$40\% of the total spectroscopic sample. We display a histogram of the global redshift distribution combining all clusters in Fig.~\ref{fig:histz}. The left panel shows the distribution of all and high confidence redshifts when limited to unique sources, that is to say removing the additional images of strongly lensed systems. The right panel shows the distribution of line-only sources (i.e. sources detected purely from line emission without any HST counterpart). 

%Since all clusters are located at a similar average redshift, there is a prominent peak around $z=0.4$ dominating the redshift distribution. Multiply imaged lensed sources start to represent a significant fraction (37\%) of all sources at $z>1.8$. The redshift desert (1.5$<z<$2.9) is also a clear feature of this distribution, when none of the strong emission lines ([\oii], Lyman-$\alpha$) are present in the MUSE wavelength range. The redshift histogram at $z>2.9$ is dominated by the population of Lyman$-\alpha$ emitters  which are more easily detected between sky lines, producing small gaps in the redshift distribution in particular at $z\sim4.6$ and $z>5.8$.
These histograms reveal a number of important features in the data, both intrinsic to our sample and unique to the MUSE instrument. Notably, since all of the clusters are located within a relatively narrow redshift range, there is a prominent peak around $z=0.4$ representing the cluster overdensities. At the same time, we see that multiply imaged lensed sources start to represent a significant fraction (37\%) of all galaxies at $z>1.8$, as evidenced by the increasing discrepancy between the ``all'' (grey) and ``unique'' (purple) distributions in the left panel. Furthermore, the MUSE redshift desert (1.5$<z<$2.9), a region where no strong emission emission lines ([\oii], Lyman-$\alpha$) are present in the MUSE wavelength range, is also clearly visible, with a significant deficit of redshifts (especially for line-only sources) in those bins. Finally, the redshift histogram at $z>2.9$ is dominated by the population of Lyman$-\alpha$ emitters, which are more easily detected between sky lines.  This produces small gaps in the redshift distribution at wavelengths where the sky is brighter, in particular at $z\sim4.6$ and $z>5.8$.

%+ few examples

Compared to the total redshift histograms, individual clusters tend to show more prominent peaks at specific redshifts. This is a known effect of galaxy clustering in the small area probed behind lensing clusters (e.g. \citealt{2004MNRAS.349.1211K}). Specific group-like structures are seen at $z\sim4$ behind Abell 2744 \citep{2018MNRAS.473..663M} and $z\sim1$ behind Abell 370 \citep{2019MNRAS.485.3738L}.

\subsection{Kinematics of the cluster cores}

\begin{figure*}
    \begin{minipage}{11cm}
    \includegraphics[width=11cm]{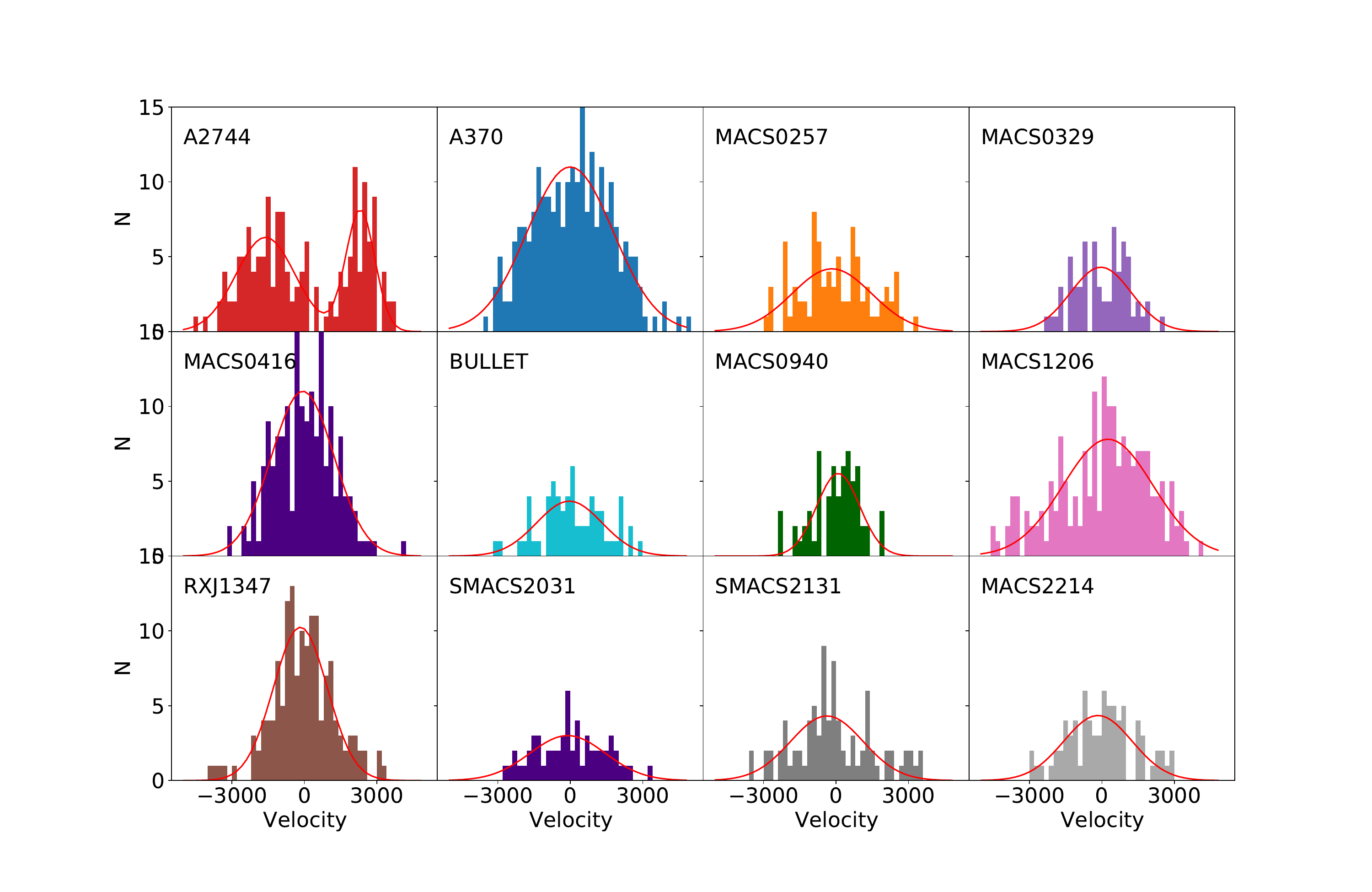}
    \end{minipage}
    \begin{minipage}{8cm}
    \includegraphics[width=8cm]{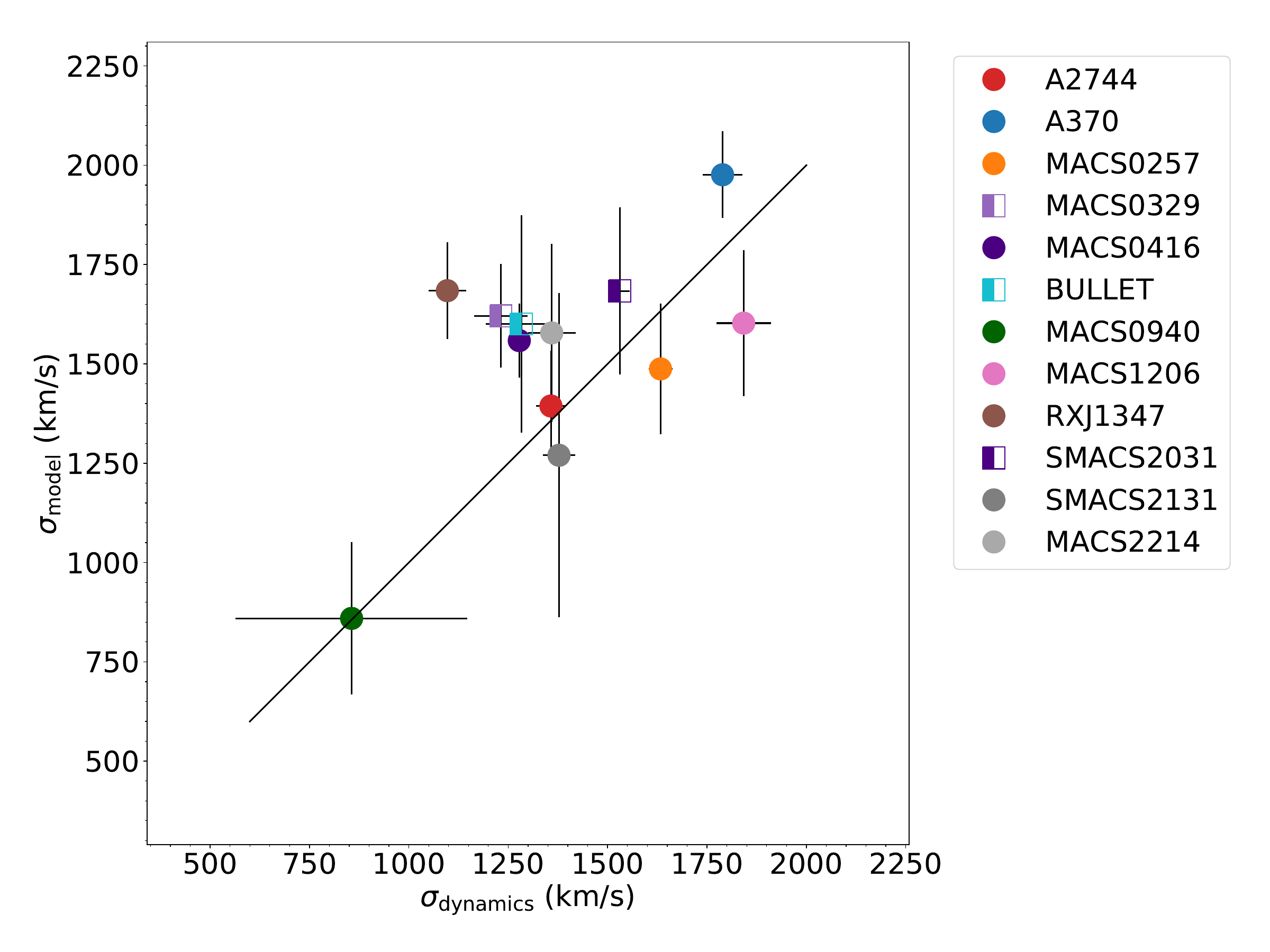}
    \end{minipage}
    \caption{ (Left panel) Velocity distribution of cluster members for each of the 12 fields. The red curves shows the best fit Gaussian model of the distribution (2 Gaussian components in Abell 2744).
    (Right panel) Comparison between the total velocity dispersion from the dPIE profiles in the mass model and the total velocity dispersion from the velocity distribution. The square symbols mark the three clusters with only partial MUSE coverage in the cluster cores.}
    \label{fig:clusterz}
\end{figure*}

The high velocities of galaxies in cluster cores appear clearly in the redshift distribution when zooming-in around the systemic cluster redshift (Fig.~\ref{fig:clusterz}). The shape of the distribution generally follows a single normal distribution, with the exception of Abell 2744  showing a clear bimodal distribution. %We have shown in \citet{2018MNRAS.473..663M} that we could associate the same velocity components seen for Abell 2744 in the core and in spectroscopy over Mpc scales \citep{2011ApJ...728...27O}.
For that particular cluster, \citet{2018MNRAS.473..663M} demonstrated that the two velocity components seen in the inner 400 kpc radius persist out to much larger distances, as they are associated (and in good agreement) with sparser spectroscopic data covering $>1$ Mpc scales \citep{2011ApJ...728...27O}.

Although a full analysis of the cluster kinematics is clearly limited by the spatial coverage of MUSE-only spectroscopy in the cluster core, the large number of cluster redshifts measured in the very central core ($R<300$ kpc) still provides insight into the overall velocity dispersion in the different cluster components seen in the distribution. We fit either 1 or 2 Gaussian components per cluster to the velocity distribution and estimate the global velocity dispersion $\sigma_{\rm dynamics}$ as the quadratic sum of their $\sigma$. We independently estimate a global velocity dispersion $\sigma_{\rm model}$ from our strong lensing model, by taking the quadratic sum of the $\sigma_0$ parameters in the cluster-scale dPIE components. We include a 1.3$\times$ correction factor between the line-of-sight velocity dispersion and the dPIE $\sigma_0$ parameters, as discussed in \citet{2007arXiv0710.5636E} for the typical values of core and cut radii in our cluster potentials. 

Figure~\ref{fig:clusterz} (right panel) directly compares the two velocity dispersion estimators for the 12 clusters. 
Taken at face value, we find that the clusters span a large range in $\sigma$, but there is a generally good agreement between the two estimators for most of the clusters. This shows that despite the strong differences in the dynamical state of the clusters, the mass components included in our mass models account for a similar amount of mass as seen in the cluster velocity distribution, although we cannot directly associate between the large-scale clumps of the lens model and the velocity components. 
We observe a large dispersion on $\sigma_{\rm dynamics}$, which is certainly due to the limited coverage of the velocity measurements and line-of-sight projection effects due to the overall geometry of the cluster. Moreover, clusters showing very high $\sigma_{\rm dynamics}$ (close to $\sim$2000 km/s, as for Abell 370 and MACS1206) are most likely formed of multiple lower $\sigma$ components which we cannot isolate individually. RXJ1347 is the most discrepant point in this diagram, with $\sigma_{\rm dynamics}\sim1100$ km/s despite it having the largest mass of the sample within its Einstein radius. This could be explained if multiple cluster structures are present with a low velocity difference, enhancing its lensing power while maintaining a low $\sigma_{\rm dynamics}$. 
Regardless, we find an average ratio $<\sigma_{\rm dynamics}/\sigma_{\rm model}>=$0.92$\pm$0.14. Even though the cluster dynamics only probe the inner regions ($<200$ kpc), we do not see a significant under- or overestimation of the global velocity dispersion in the velocity measurements.

\subsection{Magnification distribution and survey volume at high redshift}

One of the direct outputs of the mass models is an estimate of the magnification factor at a given source redshift and image position. These values are crucial in order to correct absolute physical measurements of lensed sources, such as the stellar mass, luminosity, and star-formation rate, etc. We include this estimated magnification $\mu$ at the central location of each source in the spectroscopic catalogue (Table \ref{tab:spectable}). These values are generally well-constrained within the multiply imaged regions, but tend to be overestimated in the vicinity of the critical lines (typically $\mu>25$), where the emergence of lensed pairs and the resolved sources limit very strong magnification factors. 

\begin{figure}
    \centering
    \includegraphics[width=9cm]{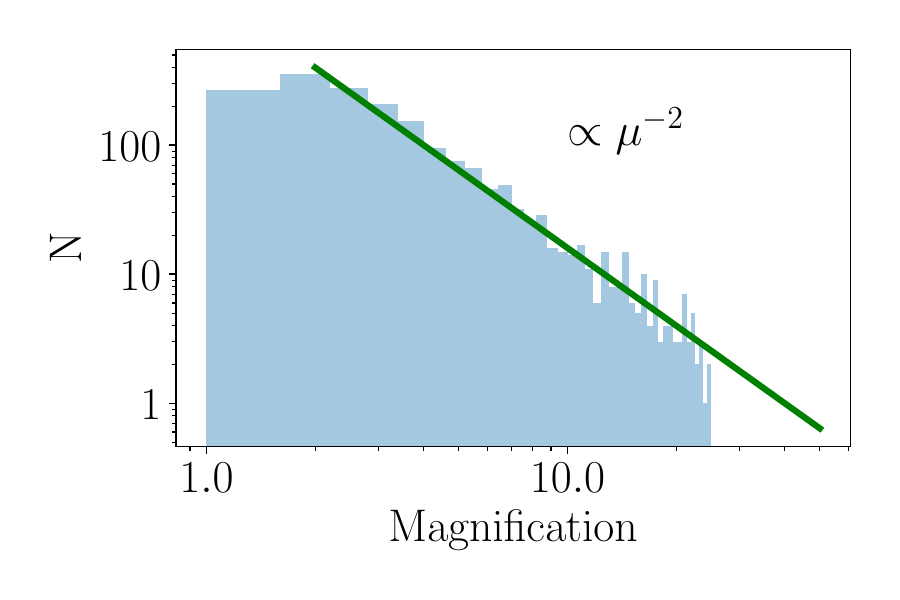}
    \caption{Magnification distribution for all background sources. We overlay a theoretical $\mu^{-2}$ relation to guide the eye. We observe a significant drop in the number of sources at low magnifications ($\mu<1.5$), due to the limits set by the FoV of the MUSE observations, and at very high magnifications ($\mu>25$) due to the very small statistics and possibly the resolved nature of Lyman-$\alpha$ emitters.}
    \label{fig:histmag}
\end{figure}

We present the magnification distribution of the full set of lensed images in Fig.\ref{fig:histmag}, where we count the magnification of each observed image individually (i.e., multiply-imaged systems are not combined into a single entry). The minimum magnification is set by the limits of the MUSE coverage in each cluster, but is typically $\mu\sim1.5-2$. The distribution $N(\mu)$ at $\mu>2$ follows a clear power-law, with an exponent $\mu^{-2.02\pm 0.09}$. This value is very close to the theoretical prediction of optical depth for strong lensing magnification, $\tau\propto\mu^{-2}$ \citep{1986ApJ...310..568B} in the case of smooth mass distributions like the ones used in our modelling. This shows that our detection process of MUSE lensed sources is not strongly affected and missing additional sources at very high magnifications.

\begin{figure*}
    \includegraphics[width=9cm]{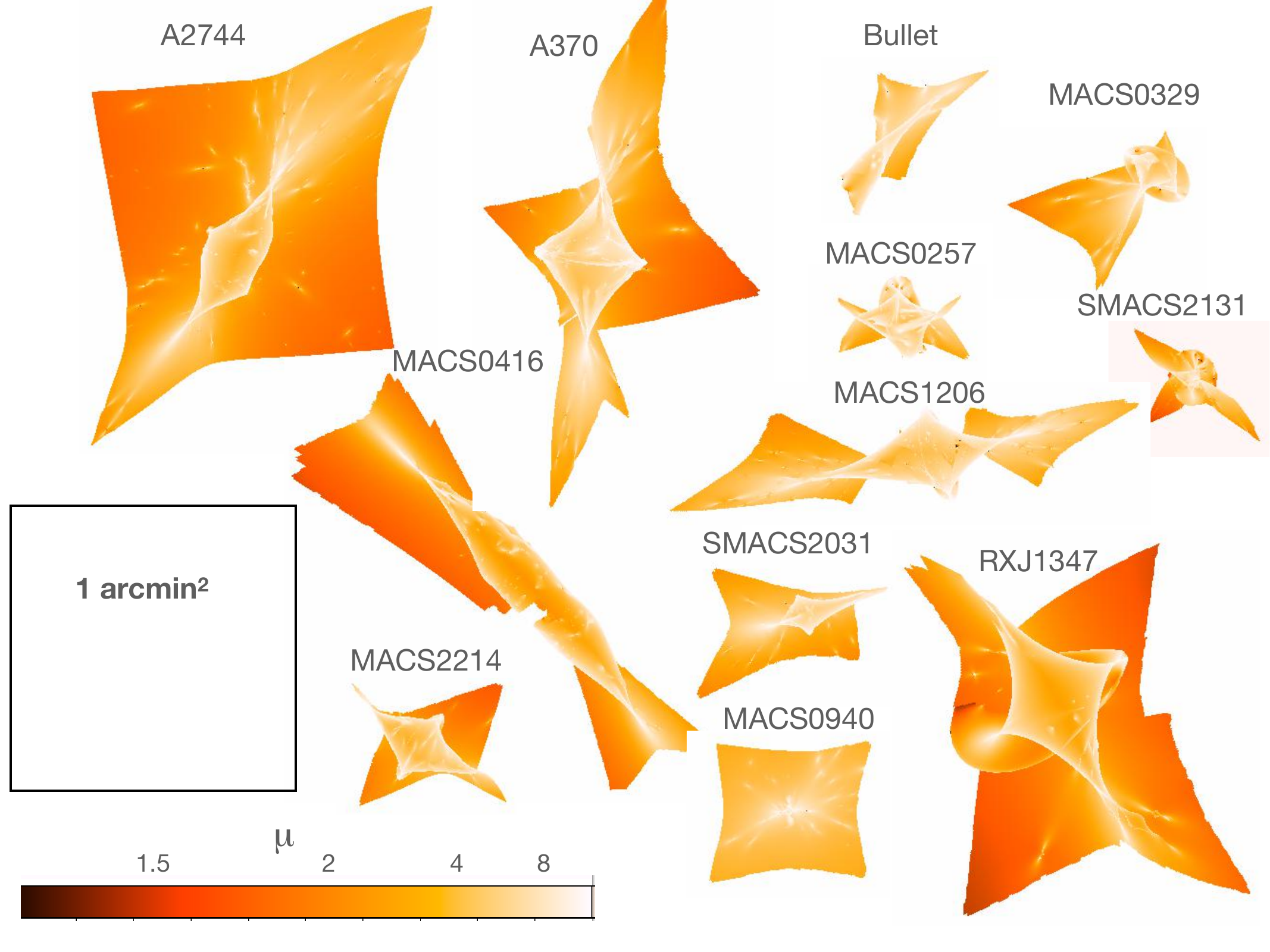}
    \includegraphics[width=9cm]{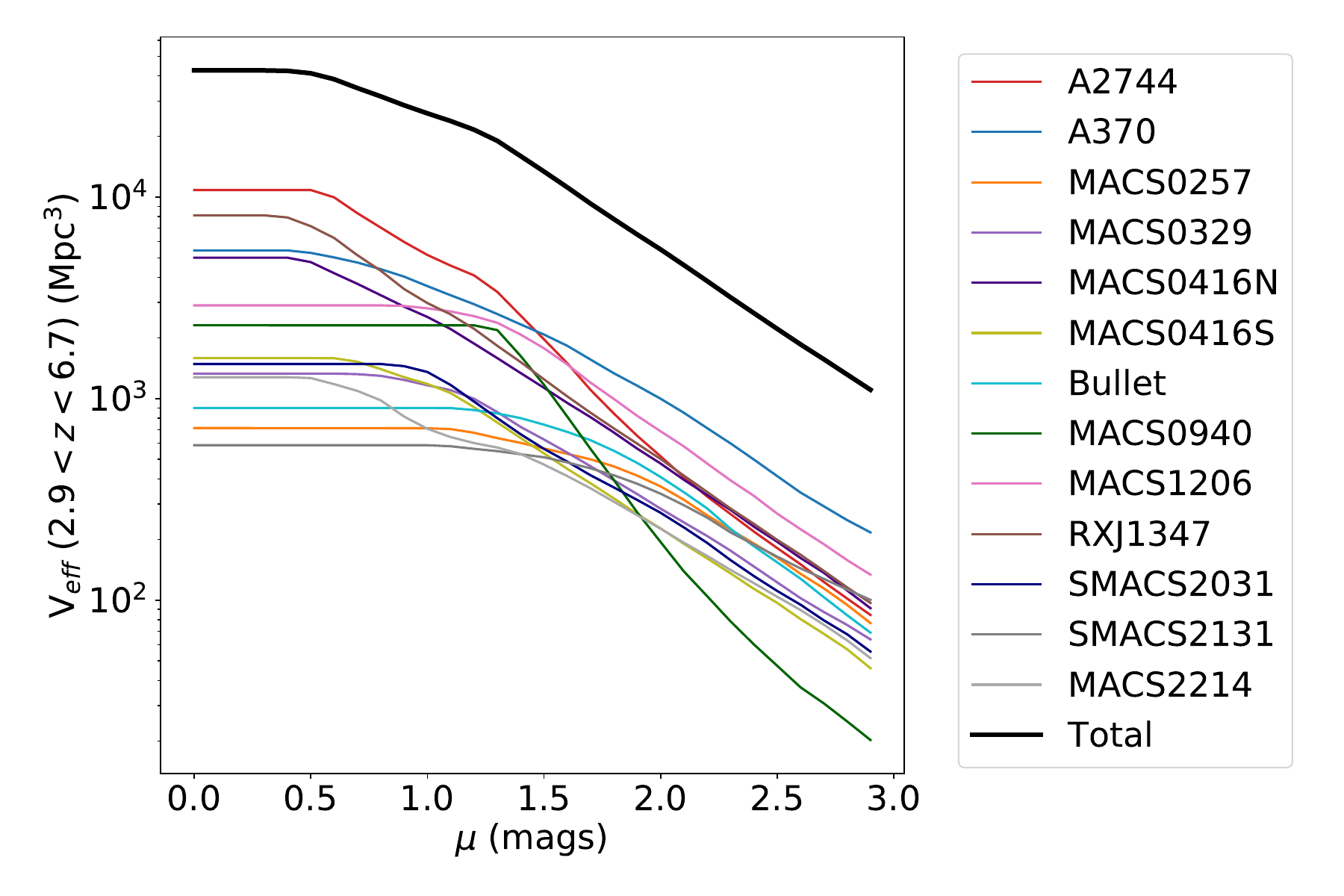}
    \caption{    \label{fig:splane}
(Left panel) Source plane coverage of each cluster at $z=4$, as a function of the strong lensing magnification of the most magnified image. The black square shows a 1 arcmin$^2$ for scale. (Right panel) Effective volume surveyed at $2.9<z<6.7$ with MUSE for Lyman-$\alpha$ emitters, as a function of the magnification of the brightest image. Presented are individual volumes behind each cluster and the total over all fields.}
\end{figure*}

%Another important measurement we can retrieve from the lens model is the equivalent source plane area covered by the MUSE observations. Indeed, the strong lensing effect will reduce the effective area covered in the source plane by the same amount as the magnification factor, and the associated survey volume in the background of the cluster. In addition, only a small fraction of this area will get very strongly magnified. Figure \ref{fig:splane} presents the source plane area for each cluster at a typical source redshift $z=4$, with the colorscale representing the  magnification factor for the most magnified image at a given source position. The strongly lensed region appears in the form of the caustic lines in the source plane.
Another important measurement we derive from the lens model is the equivalent source-plane area covered by the MUSE observations. This differs from the image-plane area (i.e., the MUSE footprint) due to the strong lensing effect, with the effective source-plane area reduced by the same amount as the magnification factor; unsurprisingly this also diminishes the associated survey volume behind the cluster. 
%In addition, only a small fraction of this area will get very strongly magnified. 
Figure \ref{fig:splane} presents the source-plane area for each cluster at a typical source redshift $z=4$, with the colourscale representing the greatest magnification factor at a given source position (recall that sources lying within the multiply-imaged region will have more than one magnification solution). In the figure, the strongly lensed region itself appears in the form of the caustic lines in the source plane.

Compared to the total observed area of $\sim23.5$ arcmin$^2$ on sky covered with MUSE, the total effective area covered in the source plane at high redshift is $4$ arcmin$^2$, a factor $6\times$ smaller. 
From these source plane areas we compute the effective volume covered at source redshifts $2.9<z<6.7$ when detecting lensed Lyman-$\alpha$ emitters, as a function of lensing magnification. This is important knowledge when probing the luminosity function of Lyman-$\alpha$ emitters behind clusters \citep{2016A&A...590A..14B,2019A&A...628A...3D} compared to blank fields \citep{2017A&A...608A...6D,2019A&A...621A.107H}. We stress that there are strong cluster-to-cluster variations in the volume surveyed depending on the geometry of the caustics and the MUSE coverage, with a maximum volume ranging between 600 and 10000 comoving Mpc$^3$ in individual clusters at any magnification. While a full volume computation would require more precise completeness measurements accounting for spatial variations of exposure time and the evolution of the noise as a function of wavelength, these variations can easily explain the cluster-to-cluster differences in the observed number counts of LAEs, as found by \citet{2019A&A...628A...3D}.

\subsection{Resolved properties of high redshift galaxies}

Despite the effective surface reduction in the source plane, one of the strong benefits from the magnification is to increase the apparent size of lensed galaxies, allowing us to reach much smaller intrinsic scales. Combined with MUSE integral field spectroscopy, this magnification gives access to resolved spectral properties in particular from the nebular line emission. Some of the brightest and most extended arcs at $0.5<z<1.5$ in the present sample have been studied in \citet{2018MNRAS.477...18P} to measure the detailed gas kinematics at sub-kpc scales, in relation with the presence of star-forming clumps and significant resolved metallicity gradients as a function of radius \citep{2019MNRAS.489..224P}. These bright extended arcs are rare cases of very strong magnification in typical L$^*$ galaxies. But the measurement of resolved kinematics can be pushed to fainter masses / luminosities and larger samples to study the Tully-Fisher relation and morpho-kinematics of lensed sources reconstructed in the source plane. This allows spatially resolved analyses which are  complementary to blank field studies, the latter being more strongly limited by the spatial resolution (\citealt{2016A&A...591A..49C,2020MNRAS.497..173G}, \citealt{Valentinasubmitted} submitted).

At higher redshifts ($z>2.9$) IFU observations have now established that the Lyman-$\alpha$ emission of distant galaxies is typically more spatially extended than the stellar UV continuum, illuminating the surrounding circumgalactic medium \citep{2016A&A...587A..98W,2017A&A...608A...8L}. The origin of this Lyman-$\alpha$ emission is complex, and only the brightest Lyman-$\alpha$ haloes are sufficiently extended to allow spatially resolved analyses in the absence of any lensing effect \citep{2018ApJ...862L..10E,2020A&A...635A..82L} and test predictions from simulations.  

\begin{figure}
    \centering
    \includegraphics[width=9cm]{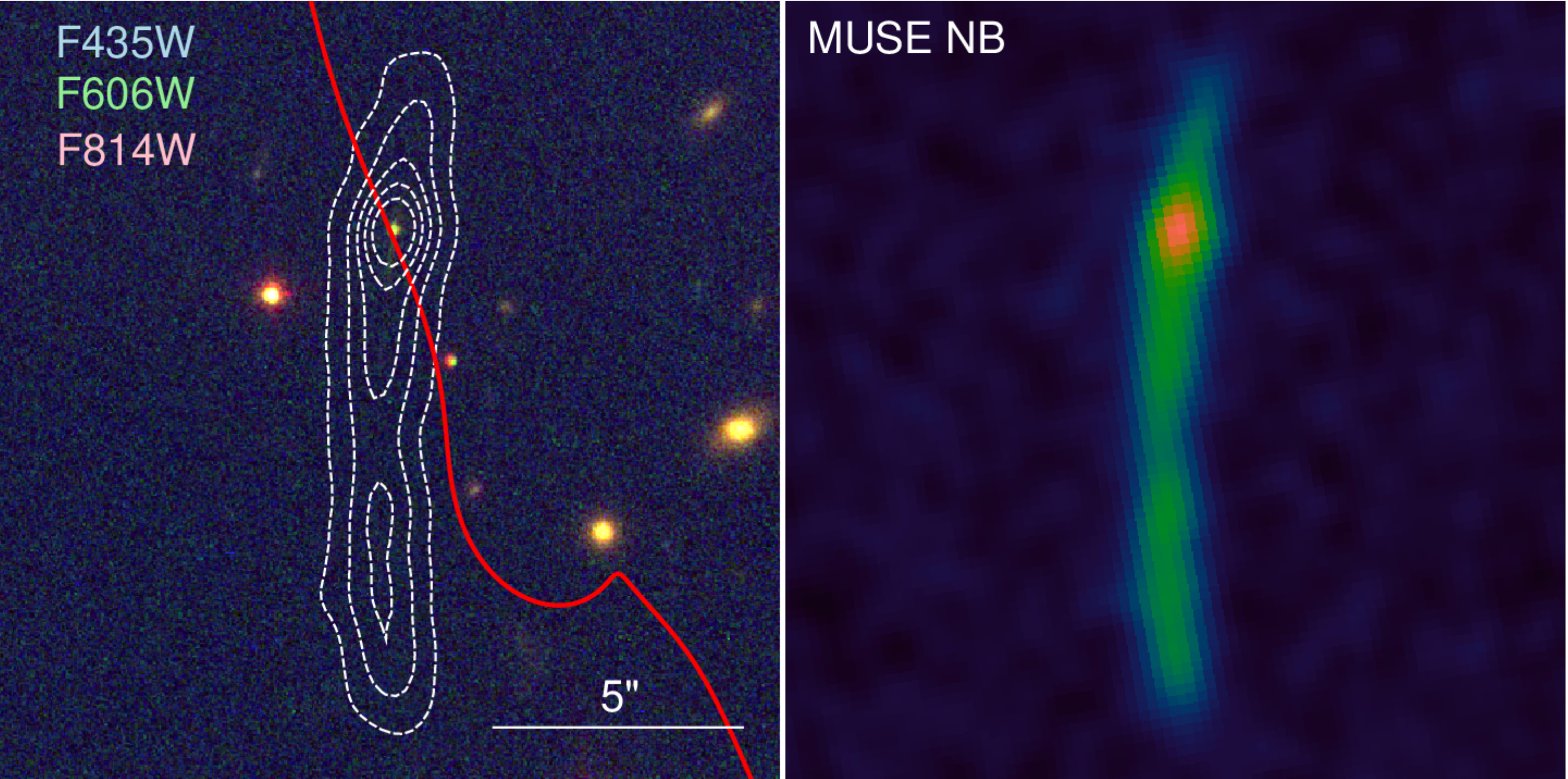}
    \caption{Example of very extended Lyman-$\alpha$ emitter in RXJ1347 (system 25), at $z=4.086$. Left panel shows the HST colour image and the right panel the MUSE NB image around the Lyman-$\alpha$ emission. White dashed contours of the NB are overlayed on HST. The red curve shows the critical line at this redshift, which crosses the arc-shaped Lyman-$\alpha$ emission at the location of a point-like source in HST.}
    \label{fig:llama}
\end{figure}

MUSE observations on lensing clusters offer the power to identify extended gas  around high redshift galaxies (once magnified), allowing detailed studies of the spectral line properties within multiple regions, in particular through bright Lyman-$\alpha$ extended haloes \citep{2016MNRAS.456.4191P,2017MNRAS.467.3306S,2016A&A...595A.100C,2017MNRAS.465.3803V,2019MNRAS.489.5022C}. Figure~\ref{fig:llama} shows a very clear example taken from our sample of extended emission detected over the $z=4.086$ source originally identified by \citet{2002ApJ...573..524C} in RXJ1347. This galaxy appears as a very compact source in the {\it HST} image (FWHM$<$0\farcs3, limited by the PSF), while the Lyman-$\alpha$ emission reaches an extension of 13\arcsec\ in the form of an arc. The critical line at $z=4.086$ straddles the arc at the location of the UV source, which appears as a bright \hii\ knot coincidentally located over the caustic line in the source plane. This example illustrates the capability of MUSE observations to clearly disentangle between stellar and nebular emission at these high redshifts. We identified 2-3 such magnified (spanning $>$5\arcsec) Lyman-$\alpha$ emitters per cluster, for which we will be able to retrieve detailed physical properties at scales better than typically $\sim$1 kpc (Claeyssens et al. in prep.).
%     \centering
%     \includegraphics[width=5cm]{vfield.jpg}
%     \caption{Draft: example of velocity field?}
%     \label{fig:vfield}
% \end{figure}

\section{Summary and conclusion}
\label{sec:conclusion}

We have presented a spectroscopic survey of MUSE/VLT observations towards 12 massive clusters selected from the MACS, CLASH, and FF samples. The data represent $\sim$23 arcmin$^2$ on sky as single pointings or mosaics, with an effective exposure time ranging between 2 and 15 hrs and a grand total of $>125$ hrs.

Thanks to improvements in the data reduction and the tools developed for source detection and inspection in the datacubes, our final spectroscopic catalogue presents $>3200$ spectroscopic redshifts with high confidence ($zconf=2$ or 3). Improvements include careful self-calibration and specific masking to remove instrumental systematics and other artefacts in the datacube, a better estimate of the noise properties, and the establishment of a clear process for spectral analysis.

Deep ($>8$ hrs per pointing) MUSE observations reveal a very high density of faint lensed LAEs, which dominate the population of background galaxies with spectroscopic redshifts. A very significant fraction of high-redshift galaxies are multiply-imaged systems, which we use to precisely constrain mass models in the cluster cores. The number of strong lensing constraints in a single cluster, reaching 70 systems producing 189 multiple images in the case of MACS0416, and the fraction of spectroscopically confirmed images ($>90\%$ in the models used in our analysis) are among the highest in any known cluster including the FFs. These parametric mass models are fully included in our redshift assessment process to help confirm or reject some multiple images.

Based on our cluster mass models, we assess the source plane survey area at high redshift ($z>3$) and find that the effective surface area is $\sim6$ arcmin$^2$, a factor of 6 smaller than in the image plane due to the lensing magnification. This shows that the number density of LAEs is very high at the low luminosities probed through lensing magnification. The  magnification distribution $N(\mu)$ is compatible with the theoretical predictions in the very strong magnification regime (2$<\mu<25$), showing that our detection process for lensed sources is not missing additional sources in the high magnification regime. We note that, in the area covered by MUSE observations, multiple images dominate the detected sources at $z>1.8$, while the fraction of multiple images is much lower at $z<1.5$ (Fig.\,\ref{fig:histz}). This difference is related to the magnification bias \citep{1995ApJ...438...49B}, which affects the source populations differently depending on the redshift and therefore intrinsic luminosity. The redshift range $1.8<z<6.7$ is therefore ideal to identify a large number of multiply imaged sources and makes MUSE very efficient for this purpose at $3<z<6.7$. The future BlueMUSE instrument \citep{2019arXiv190601657R} will complement this redshift range by targetting LAEs at $1.9<z<4$, where the density of multiply-imaged LAEs is expected to be very high.

We make the results of this analysis publicly available in the form of a first data release\footnote{\url{https://cral-perso.univ-lyon1.fr/labo/perso/johan.richard/MUSE_data_release/}}, including the reduced datacubes, spectrophotometric catalogues, extracted spectra, and mass models. Additional MUSE observations of lensing clusters taken under the GTO programme and archival observations (including lower redshift clusters such as Abell 1689, 2390 and 2667) will be inspected in a similar manner and their data released in subsequent catalogues. Overall, this very rich dataset has a strong legacy value, allowing for a large variety of statistical studies from cluster physics to very high redshift galaxies. The sample of magnified Lyman-$\alpha$ emitters will be particularly suitable for follow-ups at longer wavelengths with the James Webb Space Telescope (JWST) and the Atacama Large Millimeter/submillimeter Array (ALMA) to better constrain the resolved star formation and clump properties through ISM diagnostics.

%%%%%%%%%%%%%%%%%%%%%%%%%%%%%%%%%%%%%%%%%%%%%%%%%%


\begin{acknowledgements} 
We would like to thank the anonymous referee for their helpful comments on the manuscript.
JR, AC, DL, DC, BC, GM, VP, and JM acknowledge support from the ERC starting grant 336736-CALENDS. 
We acknowledge support from  the Programa de Cooperacion Cientifica - ECOS SUD/CONICYT Program C16U02 (JR, AC, FEB, RP, LG, DL, GV), ANID grants CATA-Basal AFB-170002 (FEB, LG, GPL), FONDECYT Regular 1190818 (FEB) and 1200495 (FEB, GPL) and Millennium Science Initiative ICN12\_009 (FEB, GPL). JPK acknowledges support from the ERC advanced grant 290975-LIDA. This work has been carried out thanks to the support of the OCEVU Labex (ANR-11-LABX-0060) and the A*MIDEX project (ANR11-IDEX-0001-02) funded by the "Investissements d’Avenir" French government programme managed by the ANR. The authors gratefully acknowledge financial support from STScI grant 15696. 
HE gratefully acknowledges support from STScI grants GO-9722, -12166, and -12884. 
Based on observations made with ESO Telescopes at the La Silla Paranal Observatory under programme IDs listed in the second column of Table \ref{tab:museobs}. We thank all the staff at Paranal Observatory for their valuable support during the MUSE-GTO observations. 

Also based on observations made with the NASA/ESA Hubble Space Telescope, which is operated by the Association of Universities for Research in Astronomy, Inc., under NASA 
contract NAS 5-26555.

This research made use several open source Python libraries: {\sc numpy} \citep{2011CSE....13b..22V}, {\sc scipy} \citep{2020SciPy-NMeth}, {\sc matplotlib} \citep{2007CSE.....9...90H}, and {\sc astropy}, a community-developed core Python package for Astronomy \citep{2013A&A...558A..33A}. This research has made use of the VizieR catalogue access tool, CDS, Strasbourg, France. The original description of the VizieR service was published in A\&AS 143, 23.  
This research made use of {\sc Montage}. It is funded by the National Science Foundation under Grant Number ACI-1440620, and was previously funded by the National Aeronautics and Space Administration's Earth Science Technology Office, Computation Technologies Project, under Cooperative Agreement Number NCC5-626 between NASA and the California Institute of Technology.

\end{acknowledgements}

\bibliographystyle{aa}
\bibliography{references.bib}

\begin{thebibliography}{119}
\expandafter\ifx\csname natexlab\endcsname\relax\def\natexlab#1{#1}\fi

\bibitem[{{Abril-Melgarejo} {et~al.}(2020){Abril-Melgarejo}, {Epinat},
  {Mercier}, {Contini}, {Boogaard}, {Brinchmann}, \&
  {Finley}}]{Valentinasubmitted}
{Abril-Melgarejo}, V., {Epinat}, B., {Mercier}, W., {et~al.} 2020, A\&A
  submitted

\bibitem[{{Acebron} {et~al.}(2017){Acebron}, {Jullo}, {Limousin}, {Tilquin},
  {Giocoli}, {Jauzac}, {Mahler}, \& {Richard}}]{2017MNRAS.470.1809A}
{Acebron}, A., {Jullo}, E., {Limousin}, M., {et~al.} 2017, \mnras, 470, 1809

\bibitem[{{Astropy Collaboration} {et~al.}(2013){Astropy Collaboration},
  {Robitaille}, {Tollerud}, {Greenfield}, {Droettboom}, {Bray}, {Aldcroft},
  {Davis}, {Ginsburg}, {Price-Whelan}, {Kerzendorf}, {Conley}, {Crighton},
  {Barbary}, {Muna}, {Ferguson}, {Grollier}, {Parikh}, {Nair}, {Unther},
  {Deil}, {Woillez}, {Conseil}, {Kramer}, {Turner}, {Singer}, {Fox}, {Weaver},
  {Zabalza}, {Edwards}, {Azalee Bostroem}, {Burke}, {Casey}, {Crawford},
  {Dencheva}, {Ely}, {Jenness}, {Labrie}, {Lim}, {Pierfederici}, {Pontzen},
  {Ptak}, {Refsdal}, {Servillat}, \& {Streicher}}]{2013A&A...558A..33A}
{Astropy Collaboration}, {Robitaille}, T.~P., {Tollerud}, E.~J., {et~al.} 2013,
  \aap, 558, A33

\bibitem[{{Atek} {et~al.}(2015){Atek}, {Richard}, {Kneib}, {Jauzac},
  {Schaerer}, {Clement}, {Limousin}, {Jullo}, {Natarajan}, {Egami}, \&
  {Ebeling}}]{2015ApJ...800...18A}
{Atek}, H., {Richard}, J., {Kneib}, J.-P., {et~al.} 2015, \apj, 800, 18

\bibitem[{{Atek} {et~al.}(2018){Atek}, {Richard}, {Kneib}, \&
  {Schaerer}}]{2018MNRAS.479.5184A}
{Atek}, H., {Richard}, J., {Kneib}, J.-P., \& {Schaerer}, D. 2018, \mnras, 479,
  5184

\bibitem[{{Bacon} {et~al.}(2010){Bacon}, {Accardo}, {Adjali}, {Anwand},
  {Bauer}, {Biswas}, {Blaizot}, {Boudon}, {Brau-Nogue}, {Brinchmann},
  {Caillier}, {Capoani}, {Carollo}, {Contini}, {Couderc}, {{Daguis{\'e}}},
  {Deiries}, {Delabre}, {Dreizler}, {Dubois}, {Dupieux}, {Dupuy}, {Emsellem},
  {Fechner}, {Fleischmann}, {{Fran{\c{c}}ois}}, {Gallou}, {Gharsa},
  {Glindemann}, {Gojak}, {Guiderdoni}, {Hansali}, {Hahn}, {Jarno}, {Kelz},
  {Koehler}, {Kosmalski}, {Laurent}, {{Le Floch}}, {Lilly}, {Lizon}, {Loupias},
  {Manescau}, {Monstein}, {Nicklas}, {Olaya}, {Pares}, {Pasquini},
  {{P{\'e}contal-Rousset}}, {{Pell{\'o}}}, {Petit}, {Popow}, {Reiss},
  {Remillieux}, {Renault}, {Roth}, {Rupprecht}, {Serre}, {Schaye}, {Soucail},
  {Steinmetz}, {Streicher}, {Stuik}, {Valentin}, {Vernet}, {Weilbacher},
  {Wisotzki}, \& {Yerle}}]{2010SPIE.7735E..08B}
{Bacon}, R., {Accardo}, M., {Adjali}, L., {et~al.} 2010, in Society of
  Photo-Optical Instrumentation Engineers (SPIE) Conference Series, Vol. 7735,
  Ground-based and Airborne Instrumentation for Astronomy III, 773508

\bibitem[{{Bacon} {et~al.}(2015){Bacon}, {Brinchmann}, {Richard}, {Contini},
  {Drake}, {Franx}, {Tacchella}, {Vernet}, {Wisotzki}, {Blaizot}, {Bouch{\'e}},
  {Bouwens}, {Cantalupo}, {Carollo}, {Carton}, {Caruana}, {Cl{\'e}ment},
  {Dreizler}, {Epinat}, {Guiderdoni}, {Herenz}, {Husser}, {Kamann}, {Kerutt},
  {Kollatschny}, {Krajnovic}, {Lilly}, {Martinsson}, {Michel-Dansac},
  {Patricio}, {Schaye}, {Shirazi}, {Soto}, {Soucail}, {Steinmetz}, {Urrutia},
  {Weilbacher}, \& {de Zeeuw}}]{2015A&A...575A..75B}
{Bacon}, R., {Brinchmann}, J., {Richard}, J., {et~al.} 2015, \aap, 575, A75

\bibitem[{{Bacon} {et~al.}(2017){Bacon}, {Conseil}, {Mary}, {Brinchmann},
  {Shepherd}, {Akhlaghi}, {Weilbacher}, {Piqueras}, {Wisotzki}, {Lagattuta},
  {Epinat}, {Guerou}, {Inami}, {Cantalupo}, {Courbot}, {Contini}, {Richard},
  {Maseda}, {Bouwens}, {Bouch{\'e}}, {Kollatschny}, {Schaye}, {Marino},
  {Pello}, {Herenz}, {Guiderdoni}, \& {Carollo}}]{2017A&A...608A...1B}
{Bacon}, R., {Conseil}, S., {Mary}, D., {et~al.} 2017, \aap, 608, A1

\bibitem[{{Baldry} {et~al.}(2014){Baldry}, {Alpaslan}, {Bauer}, {Bland
  -Hawthorn}, {Brough}, {Cluver}, {Croom}, {Davies}, {Driver}, {Gunawardhana},
  {Holwerda}, {Hopkins}, {Kelvin}, {Liske}, {L{\'o}pez-S{\'a}nchez}, {Loveday},
  {Norberg}, {Peacock}, {Robotham}, \& {Taylor}}]{2014MNRAS.441.2440B}
{Baldry}, I.~K., {Alpaslan}, M., {Bauer}, A.~E., {et~al.} 2014, \mnras, 441,
  2440

\bibitem[{{Balestra} {et~al.}(2016){Balestra}, {Mercurio}, {Sartoris},
  {Girardi}, {Grillo}, {Nonino}, {Rosati}, {Biviano}, {Ettori}, {Forman},
  {Jones}, {Koekemoer}, {Medezinski}, {Merten}, {Ogrean}, {Tozzi}, {Umetsu},
  {Vanzella}, {van Weeren}, {Zitrin}, {Annunziatella}, {Caminha}, {Broadhurst},
  {Coe}, {Donahue}, {Fritz}, {Frye}, {Kelson}, {Lombardi}, {Maier},
  {Meneghetti}, {Monna}, {Postman}, {Scodeggio}, {Seitz}, \&
  {Ziegler}}]{2016ApJS..224...33B}
{Balestra}, I., {Mercurio}, A., {Sartoris}, B., {et~al.} 2016, \apjs, 224, 33

\bibitem[{{Bayliss} {et~al.}(2011){Bayliss}, {Hennawi}, {Gladders}, {Koester},
  {Sharon}, {Dahle}, \& {Oguri}}]{2011ApJS..193....8B}
{Bayliss}, M.~B., {Hennawi}, J.~F., {Gladders}, M.~D., {et~al.} 2011, \apjs,
  193, 8

\bibitem[{{Beauchesne} {et~al.}(2020){Beauchesne}, {Cl\'ement}, {Richard}, \&
  {Kneib}}]{Beauchesnesubmitted}
{Beauchesne}, B., {Cl\'ement}, B., {Richard}, J., \& {Kneib}, J.-P. 2020, MNRAS
  submitted

\bibitem[{{Bertin} \& {Arnouts}(1996)}]{1996A&AS..117..393B}
{Bertin}, E. \& {Arnouts}, S. 1996, \aaps, 117, 393

\bibitem[{{Bertin} {et~al.}(2002){Bertin}, {Mellier}, {Radovich}, {Missonnier},
  {Didelon}, \& {Morin}}]{2002ASPC..281..228B}
{Bertin}, E., {Mellier}, Y., {Radovich}, M., {et~al.} 2002, in Astronomical
  Society of the Pacific Conference Series, Vol. 281, Astronomical Data
  Analysis Software and Systems XI, ed. D.~A. {Bohlender}, D.~{Durand}, \&
  T.~H. {Handley}, 228

\bibitem[{{Bina} {et~al.}(2016){Bina}, {Pell{\'o}}, {Richard}, {Lewis},
  {Patr{\'\i}cio}, {Cantalupo}, {Herenz}, {Soto}, {Weilbacher}, {Bacon},
  {Vernet}, {Wisotzki}, {Cl{\'e}ment}, {Cuby}, {Lagattuta}, {Soucail}, \&
  {Verhamme}}]{2016A&A...590A..14B}
{Bina}, D., {Pell{\'o}}, R., {Richard}, J., {et~al.} 2016, \aap, 590, A14

\bibitem[{{Blandford} \& {Narayan}(1986)}]{1986ApJ...310..568B}
{Blandford}, R. \& {Narayan}, R. 1986, \apj, 310, 568

\bibitem[{{Brinchmann} {et~al.}(2004){Brinchmann}, {Charlot}, {White},
  {Tremonti}, {Kauffmann}, {Heckman}, \& {Brinkmann}}]{2004MNRAS.351.1151B}
{Brinchmann}, J., {Charlot}, S., {White}, S.~D.~M., {et~al.} 2004, \mnras, 351,
  1151

\bibitem[{{Broadhurst} {et~al.}(2005){Broadhurst}, {Ben{\'\i}tez}, {Coe},
  {Sharon}, {Zekser}, {White}, {Ford}, {Bouwens}, {Blakeslee}, {Clampin},
  {Cross}, {Franx}, {Frye}, {Hartig}, {Illingworth}, {Infante}, {Menanteau},
  {Meurer}, {Postman}, {Ardila}, {Bartko}, {Brown}, {Burrows}, {Cheng},
  {Feldman}, {Golimowski}, {Goto}, {Gronwall}, {Herranz}, {Holden}, {Homeier},
  {Krist}, {Lesser}, {Martel}, {Miley}, {Rosati}, {Sirianni}, {Sparks},
  {Steindling}, {Tran}, {Tsvetanov}, \& {Zheng}}]{2005ApJ...621...53B}
{Broadhurst}, T., {Ben{\'\i}tez}, N., {Coe}, D., {et~al.} 2005, \apj, 621, 53

\bibitem[{{Broadhurst} {et~al.}(1995){Broadhurst}, {Taylor}, \&
  {Peacock}}]{1995ApJ...438...49B}
{Broadhurst}, T.~J., {Taylor}, A.~N., \& {Peacock}, J.~A. 1995, \apj, 438, 49

\bibitem[{{Caminha} {et~al.}(2016{\natexlab{a}}){Caminha}, {Grillo}, {Rosati},
  {Balestra}, {Karman}, {Lombardi}, {Mercurio}, {Nonino}, {Tozzi}, {Zitrin},
  {Biviano}, {Girardi}, {Koekemoer}, {Melchior}, {Meneghetti}, {Munari},
  {Suyu}, {Umetsu}, {Annunziatella}, {Borgani}, {Broadhurst}, {Caputi}, {Coe},
  {Delgado-Correal}, {Ettori}, {Fritz}, {Frye}, {Gobat}, {Maier}, {Monna},
  {Postman}, {Sartoris}, {Seitz}, {Vanzella}, \&
  {Ziegler}}]{2016A&A...587A..80C}
{Caminha}, G.~B., {Grillo}, C., {Rosati}, P., {et~al.} 2016{\natexlab{a}},
  \aap, 587, A80

\bibitem[{{Caminha} {et~al.}(2017{\natexlab{a}}){Caminha}, {Grillo}, {Rosati},
  {Balestra}, {Mercurio}, {Vanzella}, {Biviano}, {Caputi}, {Delgado-Correal},
  {Karman}, {Lombardi}, {Meneghetti}, {Sartoris}, \&
  {Tozzi}}]{2017A&A...600A..90C}
{Caminha}, G.~B., {Grillo}, C., {Rosati}, P., {et~al.} 2017{\natexlab{a}},
  \aap, 600, A90

\bibitem[{{Caminha} {et~al.}(2017{\natexlab{b}}){Caminha}, {Grillo}, {Rosati},
  {Meneghetti}, {Mercurio}, {Ettori}, {Balestra}, {Biviano}, {Umetsu},
  {Vanzella}, {Annunziatella}, {Bonamigo}, {Delgado-Correal}, {Girardi},
  {Lombardi}, {Nonino}, {Sartoris}, {Tozzi}, {Bartelmann}, {Bradley}, {Caputi},
  {Coe}, {Ford}, {Fritz}, {Gobat}, {Postman}, {Seitz}, \&
  {Zitrin}}]{2017A&A...607A..93C}
{Caminha}, G.~B., {Grillo}, C., {Rosati}, P., {et~al.} 2017{\natexlab{b}},
  \aap, 607, A93

\bibitem[{{Caminha} {et~al.}(2016{\natexlab{b}}){Caminha}, {Karman}, {Rosati},
  {Caputi}, {Arrigoni Battaia}, {Balestra}, {Grillo}, {Mercurio}, {Nonino}, \&
  {Vanzella}}]{2016A&A...595A.100C}
{Caminha}, G.~B., {Karman}, W., {Rosati}, P., {et~al.} 2016{\natexlab{b}},
  \aap, 595, A100

\bibitem[{{Caminha} {et~al.}(2019{\natexlab{a}}){Caminha}, {Rosati}, {Grillo},
  {Rosani}, {Caputi}, {Meneghetti}, {Mercurio}, {Balestra}, {Bergamini},
  {Biviano}, {Nonino}, {Umetsu}, {Vanzella}, {Annunziatella}, {Broadhurst},
  {Delgado-Correal}, {Demarco}, {Koekemoer}, {Lombardi}, {Maier}, {Verdugo}, \&
  {Zitrin}}]{2019AandA...632A..36C}
{Caminha}, G.~B., {Rosati}, P., {Grillo}, C., {et~al.} 2019{\natexlab{a}},
  \aap, 632, A36

\bibitem[{{Caminha} {et~al.}(2019{\natexlab{b}}){Caminha}, {Rosati}, {Grillo},
  {Rosani}, {Caputi}, {Meneghetti}, {Mercurio}, {Balestra}, {Bergamini},
  {Biviano}, {Nonino}, {Umetsu}, {Vanzella}, {Annunziatella}, {Broadhurst},
  {Delgado-Correal}, {Demarco}, {Koekemoer}, {Lombardi}, {Maier}, {Verdugo}, \&
  {Zitrin}}]{2019A&A...632A..36C}
{Caminha}, G.~B., {Rosati}, P., {Grillo}, C., {et~al.} 2019{\natexlab{b}},
  \aap, 632, A36

\bibitem[{{Campusano} {et~al.}(2001){Campusano}, {Pell{\'o}}, {Kneib}, {Le
  Borgne}, {Fort}, {Ellis}, {Mellier}, \& {Smail}}]{2001A&A...378..394C}
{Campusano}, L.~E., {Pell{\'o}}, R., {Kneib}, J.~P., {et~al.} 2001, \aap, 378,
  394

\bibitem[{{Chiriv{\`\i}} {et~al.}(2018){Chiriv{\`\i}}, {Suyu}, {Grillo},
  {Halkola}, {Balestra}, {Caminha}, {Mercurio}, \&
  {Rosati}}]{2018A&A...614A...8C}
{Chiriv{\`\i}}, G., {Suyu}, S.~H., {Grillo}, C., {et~al.} 2018, \aap, 614, A8

\bibitem[{{Christensen} {et~al.}(2012){Christensen}, {Richard}, {Hjorth},
  {Milvang-Jensen}, {Laursen}, {Limousin}, {Dessauges-Zavadsky}, {Grillo}, \&
  {Ebeling}}]{2012MNRAS.427.1953C}
{Christensen}, L., {Richard}, J., {Hjorth}, J., {et~al.} 2012, \mnras, 427,
  1953

\bibitem[{{Claeyssens} {et~al.}(2019){Claeyssens}, {Richard}, {Blaizot},
  {Garel}, {Leclercq}, {Patr{\'\i}cio}, {Verhamme}, {Wisotzki}, {Bacon},
  {Carton}, {Cl{\'e}ment}, {Herenz}, {Marino}, {Muzahid}, {Saust}, \&
  {Schaye}}]{2019MNRAS.489.5022C}
{Claeyssens}, A., {Richard}, J., {Blaizot}, J., {et~al.} 2019, \mnras, 489,
  5022

\bibitem[{Cohen \& Kneib(2002)}]{2002ApJ...573..524C}
Cohen, J.~G. \& Kneib, J.-P. 2002, The Astrophysical Journal, 573, 524

\bibitem[{{Contini} {et~al.}(2016){Contini}, {Epinat}, {Bouch{\'e}},
  {Brinchmann}, {Boogaard}, {Ventou}, {Bacon}, {Richard}, {Weilbacher},
  {Wisotzki}, {Krajnovi{\'c}}, {Vielfaure}, {Emsellem}, {Finley}, {Inami},
  {Schaye}, {Swinbank}, {Gu{\'e}rou}, {Martinsson}, {Michel-Dansac},
  {Schroetter}, {Shirazi}, \& {Soucail}}]{2016A&A...591A..49C}
{Contini}, T., {Epinat}, B., {Bouch{\'e}}, N., {et~al.} 2016, \aap, 591, A49

\bibitem[{{de La Vieuville} {et~al.}(2019){de La Vieuville}, {Bina}, {Pello},
  {Mahler}, {Richard}, {Drake}, {Herenz}, {Bauer}, {Cl{\'e}ment}, {Lagattuta},
  {Laporte}, {Martinez}, {Patr{\'\i}cio}, {Wisotzki}, {Zabl}, {Bouwens},
  {Contini}, {Garel}, {Guiderdoni}, {Marino}, {Maseda}, {Matthee}, {Schaye}, \&
  {Soucail}}]{2019A&A...628A...3D}
{de La Vieuville}, G., {Bina}, D., {Pello}, R., {et~al.} 2019, \aap, 628, A3

\bibitem[{{Desprez} {et~al.}(2018){Desprez}, {Richard}, {Jauzac}, {Martinez},
  {Siana}, \& {Cl{\'e}ment}}]{2018MNRAS.479.2630D}
{Desprez}, G., {Richard}, J., {Jauzac}, M., {et~al.} 2018, \mnras, 479, 2630

\bibitem[{{Drake} {et~al.}(2017){Drake}, {Garel}, {Wisotzki}, {Leclercq},
  {Hashimoto}, {Richard}, {Bacon}, {Blaizot}, {Caruana}, {Conseil}, {Contini},
  {Guiderdoni}, {Herenz}, {Inami}, {Lewis}, {Mahler}, {Marino}, {Pello},
  {Schaye}, {Verhamme}, {Ventou}, \& {Weilbacher}}]{2017A&A...608A...6D}
{Drake}, A.~B., {Garel}, T., {Wisotzki}, L., {et~al.} 2017, \aap, 608, A6

\bibitem[{{Ebeling} {et~al.}(2007){Ebeling}, {Barrett}, {Donovan}, {Ma},
  {Edge}, \& {van Speybroeck}}]{2007ApJ...661L..33E}
{Ebeling}, H., {Barrett}, E., {Donovan}, D., {et~al.} 2007, \apjl, 661, L33

\bibitem[{{Ebeling} {et~al.}(2000){Ebeling}, {Edge}, {Allen}, {Crawford},
  {Fabian}, \& {Huchra}}]{2000MNRAS.318..333E}
{Ebeling}, H., {Edge}, A.~C., {Allen}, S.~W., {et~al.} 2000, \mnras, 318, 333

\bibitem[{{Ebeling} {et~al.}(1998){Ebeling}, {Edge}, {Bohringer}, {Allen},
  {Crawford}, {Fabian}, {Voges}, \& {Huchra}}]{1998MNRAS.301..881E}
{Ebeling}, H., {Edge}, A.~C., {Bohringer}, H., {et~al.} 1998, \mnras, 301, 881

\bibitem[{{Ebeling} {et~al.}(2001){Ebeling}, {Edge}, \&
  {Henry}}]{2001ApJ...553..668E}
{Ebeling}, H., {Edge}, A.~C., \& {Henry}, J.~P. 2001, \apj, 553, 668

\bibitem[{{Ebeling} {et~al.}(2009){Ebeling}, {Ma}, {Kneib}, {Jullo},
  {Courtney}, {Barrett}, {Edge}, \& {Le Borgne}}]{2009MNRAS.395.1213E}
{Ebeling}, H., {Ma}, C.~J., {Kneib}, J.~P., {et~al.} 2009, \mnras, 395, 1213

\bibitem[{{El{\'\i}asd{\'o}ttir} {et~al.}(2007){El{\'\i}asd{\'o}ttir},
  {Limousin}, {Richard}, {Hjorth}, {Kneib}, {Natarajan}, {Pedersen}, {Jullo},
  \& {Paraficz}}]{2007arXiv0710.5636E}
{El{\'\i}asd{\'o}ttir}, {\'A}., {Limousin}, M., {Richard}, J., {et~al.} 2007,
  ApJ submitted, arXiv:0710.5636

\bibitem[{{Erb} {et~al.}(2018){Erb}, {Steidel}, \&
  {Chen}}]{2018ApJ...862L..10E}
{Erb}, D.~K., {Steidel}, C.~C., \& {Chen}, Y. 2018, \apjl, 862, L10

\bibitem[{{Faber} \& {Jackson}(1976)}]{1976ApJ...204..668F}
{Faber}, S.~M. \& {Jackson}, R.~E. 1976, \apj, 204, 668

\bibitem[{{Fujimoto} {et~al.}(2020){Fujimoto}, {Oguri}, {Brammer}, {Yoshimura},
  Nicolas, {González-López}, {Caminha}, {Kohno}, {Zitrin}, \&
  {Richard}}]{Fujimotosubmitted}
{Fujimoto}, S., {Oguri}, M., {Brammer}, G., {et~al.} 2020, ApJ submitted

\bibitem[{{Gaia Collaboration} {et~al.}(2018){Gaia Collaboration}, {Brown},
  {Vallenari}, {Prusti}, {de Bruijne}, {Babusiaux}, {Bailer-Jones}, {Biermann},
  {Evans}, {Eyer}, {Jansen}, {Jordi}, {Klioner}, {Lammers}, {Lindegren},
  {Luri}, {Mignard}, {Panem}, {Pourbaix}, {Randich}, {Sartoretti}, {Siddiqui},
  {Soubiran}, {van Leeuwen}, {Walton}, {Arenou}, {Bastian}, {Cropper},
  {Drimmel}, {Katz}, {Lattanzi}, {Bakker}, {Cacciari}, {Casta{\~n}eda},
  {Chaoul}, {Cheek}, {De Angeli}, {Fabricius}, {Guerra}, {Holl}, {Masana},
  {Messineo}, {Mowlavi}, {Nienartowicz}, {Panuzzo}, {Portell}, {Riello},
  {Seabroke}, {Tanga}, {Th{\'e}venin}, {Gracia-Abril}, {Comoretto},
  {Garcia-Reinaldos}, {Teyssier}, {Altmann}, {Andrae}, {Audard},
  {Bellas-Velidis}, {Benson}, {Berthier}, {Blomme}, {Burgess}, {Busso},
  {Carry}, {Cellino}, {Clementini}, {Clotet}, {Creevey}, {Davidson}, {De
  Ridder}, {Delchambre}, {Dell'Oro}, {Ducourant},
  {Fern{\'a}ndez-Hern{\'a}ndez}, {Fouesneau}, {Fr{\'e}mat}, {Galluccio},
  {Garc{\'\i}a-Torres}, {Gonz{\'a}lez-N{\'u}{\~n}ez}, {Gonz{\'a}lez-Vidal},
  {Gosset}, {Guy}, {Halbwachs}, {Hambly}, {Harrison}, {Hern{\'a}ndez},
  {Hestroffer}, {Hodgkin}, {Hutton}, {Jasniewicz}, {Jean-Antoine-Piccolo},
  {Jordan}, {Korn}, {Krone-Martins}, {Lanzafame}, {Lebzelter}, {L{\"o}ffler},
  {Manteiga}, {Marrese}, {Mart{\'\i}n-Fleitas}, {Moitinho}, {Mora}, {Muinonen},
  {Osinde}, {Pancino}, {Pauwels}, {Petit}, {Recio-Blanco}, {Richards},
  {Rimoldini}, {Robin}, {Sarro}, {Siopis}, {Smith}, {Sozzetti}, {S{\"u}veges},
  {Torra}, {van Reeven}, {Abbas}, {Abreu Aramburu}, {Accart}, {Aerts},
  {Altavilla}, {{\'A}lvarez}, {Alvarez}, {Alves}, {Anderson}, {Andrei},
  {Anglada Varela}, {Antiche}, {Antoja}, {Arcay}, {Astraatmadja}, {Bach},
  {Baker}, {Balaguer-N{\'u}{\~n}ez}, {Balm}, {Barache}, {Barata}, {Barbato},
  {Barblan}, {Barklem}, {Barrado}, {Barros}, {Barstow}, {Bartholom{\'e}
  Mu{\~n}oz}, {Bassilana}, {Becciani}, {Bellazzini}, {Berihuete}, {Bertone},
  {Bianchi}, {Bienaym{\'e}}, {Blanco-Cuaresma}, {Boch}, {Boeche}, {Bombrun},
  {Borrachero}, {Bossini}, {Bouquillon}, {Bourda}, {Bragaglia}, {Bramante},
  {Breddels}, {Bressan}, {Brouillet}, {Br{\"u}semeister}, {Brugaletta},
  {Bucciarelli}, {Burlacu}, {Busonero}, {Butkevich}, {Buzzi}, {Caffau},
  {Cancelliere}, {Cannizzaro}, {Cantat-Gaudin}, {Carballo}, {Carlucci},
  {Carrasco}, {Casamiquela}, {Castellani}, {Castro-Ginard}, {Charlot},
  {Chemin}, {Chiavassa}, {Cocozza}, {Costigan}, {Cowell}, {Crifo}, {Crosta},
  {Crowley}, {Cuypers}, {Dafonte}, {Damerdji}, {Dapergolas}, {David}, {David},
  {de Laverny}, {De Luise}, {De March}, {de Martino}, {de Souza}, {de Torres},
  {Debosscher}, {del Pozo}, {Delbo}, {Delgado}, {Delgado}, {Di Matteo},
  {Diakite}, {Diener}, {Distefano}, {Dolding}, {Drazinos}, {Dur{\'a}n},
  {Edvardsson}, {Enke}, {Eriksson}, {Esquej}, {Eynard Bontemps}, {Fabre},
  {Fabrizio}, {Faigler}, {Falc{\~a}o}, {Farr{\`a}s Casas}, {Federici},
  {Fedorets}, {Fernique}, {Figueras}, {Filippi}, {Findeisen}, {Fonti},
  {Fraile}, {Fraser}, {Fr{\'e}zouls}, {Gai}, {Galleti}, {Garabato},
  {Garc{\'\i}a-Sedano}, {Garofalo}, {Garralda}, {Gavel}, {Gavras}, {Gerssen},
  {Geyer}, {Giacobbe}, {Gilmore}, {Girona}, {Giuffrida}, {Glass}, {Gomes},
  {Granvik}, {Gueguen}, {Guerrier}, {Guiraud}, {Guti{\'e}rrez-S{\'a}nchez},
  {Haigron}, {Hatzidimitriou}, {Hauser}, {Haywood}, {Heiter}, {Helmi}, {Heu},
  {Hilger}, {Hobbs}, {Hofmann}, {Holland}, {Huckle}, {Hypki}, {Icardi},
  {Jan{\ss}en}, {Jevardat de Fombelle}, {Jonker}, {Juh{\'a}sz}, {Julbe},
  {Karampelas}, {Kewley}, {Klar}, {Kochoska}, {Kohley}, {Kolenberg},
  {Kontizas}, {Kontizas}, {Koposov}, {Kordopatis}, {Kostrzewa-Rutkowska},
  {Koubsky}, {Lambert}, {Lanza}, {Lasne}, {Lavigne}, {Le Fustec}, {Le
  Poncin-Lafitte}, {Lebreton}, {Leccia}, {Leclerc}, {Lecoeur-Taibi},
  {Lenhardt}, {Leroux}, {Liao}, {Licata}, {Lindstr{\o}m}, {Lister}, {Livanou},
  {Lobel}, {L{\'o}pez}, {Managau}, {Mann}, {Mantelet}, {Marchal}, {Marchant},
  {Marconi}, {Marinoni}, {Marschalk{\'o}}, {Marshall}, {Martino}, {Marton},
  {Mary}, {Massari}, {Matijevi{\v{c}}}, {Mazeh}, {McMillan}, {Messina},
  {Michalik}, {Millar}, {Molina}, {Molinaro}, {Moln{\'a}r}, {Montegriffo},
  {Mor}, {Morbidelli}, {Morel}, {Morris}, {Mulone}, {Muraveva}, {Musella},
  {Nelemans}, {Nicastro}, {Noval}, {O'Mullane}, {Ord{\'e}novic},
  {Ord{\'o}{\~n}ez-Blanco}, {Osborne}, {Pagani}, {Pagano}, {Pailler},
  {Palacin}, {Palaversa}, {Panahi}, {Pawlak}, {Piersimoni}, {Pineau}, {Plachy},
  {Plum}, {Poggio}, {Poujoulet}, {Pr{\v{s}}a}, {Pulone}, {Racero}, {Ragaini},
  {Rambaux}, {Ramos-Lerate}, {Regibo}, {Reyl{\'e}}, {Riclet}, {Ripepi}, {Riva},
  {Rivard}, {Rixon}, {Roegiers}, {Roelens}, {Romero-G{\'o}mez}, {Rowell},
  {Royer}, {Ruiz-Dern}, {Sadowski}, {Sagrist{\`a} Sell{\'e}s}, {Sahlmann},
  {Salgado}, {Salguero}, {Sanna}, {Santana-Ros}, {Sarasso}, {Savietto},
  {Schultheis}, {Sciacca}, {Segol}, {Segovia}, {S{\'e}gransan}, {Shih},
  {Siltala}, {Silva}, {Smart}, {Smith}, {Solano}, {Solitro}, {Sordo}, {Soria
  Nieto}, {Souchay}, {Spagna}, {Spoto}, {Stampa}, {Steele},
  {Steidelm{\"u}ller}, {Stephenson}, {Stoev}, {Suess}, {Surdej}, {Szabados},
  {Szegedi-Elek}, {Tapiador}, {Taris}, {Tauran}, {Taylor}, {Teixeira},
  {Terrett}, {Teyssand ier}, {Thuillot}, {Titarenko}, {Torra Clotet}, {Turon},
  {Ulla}, {Utrilla}, {Uzzi}, {Vaillant}, {Valentini}, {Valette}, {van Elteren},
  {Van Hemelryck}, {van Leeuwen}, {Vaschetto}, {Vecchiato}, {Veljanoski},
  {Viala}, {Vicente}, {Vogt}, {von Essen}, {Voss}, {Votruba}, {Voutsinas},
  {Walmsley}, {Weiler}, {Wertz}, {Wevers}, {Wyrzykowski}, {Yoldas},
  {{\v{Z}}erjal}, {Ziaeepour}, {Zorec}, {Zschocke}, {Zucker}, {Zurbach}, \&
  {Zwitter}}]{2018A&A...616A...1G}
{Gaia Collaboration}, {Brown}, A.~G.~A., {Vallenari}, A., {et~al.} 2018, \aap,
  616, A1

\bibitem[{{Girard} {et~al.}(2020){Girard}, {Mason}, {Fontana},
  {Dessauges-Zavadsky}, {Morishita}, {Amor{\'\i}n}, {Fisher}, {Jones},
  {Schaerer}, {Schmidt}, {Treu}, \& {Vulcani}}]{2020MNRAS.497..173G}
{Girard}, M., {Mason}, C.~A., {Fontana}, A., {et~al.} 2020, \mnras, 497, 173

\bibitem[{{Grillo} {et~al.}(2016){Grillo}, {Karman}, {Suyu}, {Rosati},
  {Balestra}, {Mercurio}, {Lombardi}, {Treu}, {Caminha}, {Halkola}, {Rodney},
  {Gavazzi}, \& {Caputi}}]{2016ApJ...822...78G}
{Grillo}, C., {Karman}, W., {Suyu}, S.~H., {et~al.} 2016, \apj, 822, 78

\bibitem[{{Halkola} {et~al.}(2008){Halkola}, {Hildebrandt}, {Schrabback},
  {Lombardi}, {Brada{\v{c}}}, {Erben}, {Schneider}, \&
  {Wuttke}}]{2008AandA...481...65H}
{Halkola}, A., {Hildebrandt}, H., {Schrabback}, T., {et~al.} 2008, \aap, 481,
  65

\bibitem[{{Hashimoto} {et~al.}(2017){Hashimoto}, {Garel}, {Guiderdoni},
  {Drake}, {Bacon}, {Blaizot}, {Richard}, {Leclercq}, {Inami}, {Verhamme},
  {Bouwens}, {Brinchmann}, {Cantalupo}, {Carollo}, {Caruana}, {Herenz},
  {Kerutt}, {Marino}, {Mitchell}, \& {Schaye}}]{2017A&A...608A..10H}
{Hashimoto}, T., {Garel}, T., {Guiderdoni}, B., {et~al.} 2017, \aap, 608, A10

\bibitem[{{Herenz} {et~al.}(2017){Herenz}, {Urrutia}, {Wisotzki}, {Kerutt},
  {Saust}, {Werhahn}, {Schmidt}, {Caruana}, {Diener}, {Bacon}, {Brinchmann},
  {Schaye}, {Maseda}, \& {Weilbacher}}]{2017A&A...606A..12H}
{Herenz}, E.~C., {Urrutia}, T., {Wisotzki}, L., {et~al.} 2017, \aap, 606, A12

\bibitem[{{Herenz} {et~al.}(2019){Herenz}, {Wisotzki}, {Saust}, {Kerutt},
  {Urrutia}, {Diener}, {Schmidt}, {Marino}, {de la Vieuville}, {Boogaard},
  {Schaye}, {Guiderdoni}, {Richard}, \& {Bacon}}]{2019A&A...621A.107H}
{Herenz}, E.~C., {Wisotzki}, L., {Saust}, R., {et~al.} 2019, \aap, 621, A107

\bibitem[{{Hinton}(2016)}]{2016ascl.soft05001H}
{Hinton}, S. 2016, {MARZ: Redshifting Program}

\bibitem[{{Hoag} {et~al.}(2015){Hoag}, {Brada{\v{c}}}, {Huang}, {Ryan},
  {Sharon}, {Schrabback}, {Schmidt}, {Cain}, {Gonzalez}, {Hildebrand t},
  {Hinz}, {Lemaux}, {von der Linden}, {Lubin}, {Treu}, \&
  {Zaritsky}}]{2015ApJ...813...37H}
{Hoag}, A., {Brada{\v{c}}}, M., {Huang}, K.~H., {et~al.} 2015, \apj, 813, 37

\bibitem[{{Hoag} {et~al.}(2016){Hoag}, {Huang}, {Treu}, {Brada{\v{c}}},
  {Schmidt}, {Wang}, {Brammer}, {Broussard}, {Amorin}, {Castellano}, {Fontana},
  {Merlin}, {Schrabback}, {Trenti}, \& {Vulcani}}]{2016ApJ...831..182H}
{Hoag}, A., {Huang}, K.~H., {Treu}, T., {et~al.} 2016, \apj, 831, 182

\bibitem[{{Horne}(1986)}]{1986PASP...98..609H}
{Horne}, K. 1986, \pasp, 98, 609

\bibitem[{{Hunter}(2007)}]{2007CSE.....9...90H}
{Hunter}, J.~D. 2007, Computing in Science and Engineering, 9, 90

\bibitem[{{Inami} {et~al.}(2017){Inami}, {Bacon}, {Brinchmann}, {Richard},
  {Contini}, {Conseil}, {Hamer}, {Akhlaghi}, {Bouch{\'e}}, {Cl{\'e}ment},
  {Desprez}, {Drake}, {Hashimoto}, {Leclercq}, {Maseda}, {Michel-Dansac},
  {Paalvast}, {Tresse}, {Ventou}, {Kollatschny}, {Boogaard}, {Finley},
  {Marino}, {Schaye}, \& {Wisotzki}}]{2017A&A...608A...2I}
{Inami}, H., {Bacon}, R., {Brinchmann}, J., {et~al.} 2017, \aap, 608, A2

\bibitem[{{Jauzac} {et~al.}(2014){Jauzac}, {Cl{\'e}ment}, {Limousin},
  {Richard}, {Jullo}, {Ebeling}, {Atek}, {Kneib}, {Knowles}, {Natarajan},
  {Eckert}, {Egami}, {Massey}, \& {Rexroth}}]{2014MNRAS.443.1549J}
{Jauzac}, M., {Cl{\'e}ment}, B., {Limousin}, M., {et~al.} 2014, \mnras, 443,
  1549

\bibitem[{{Jauzac} {et~al.}(2020){Jauzac}, {Klein}, {Kneib}, {Richard},
  {Rexroth}, {Sch{\"a}fer}, \& {Verdier}}]{2020arXiv200610700J}
{Jauzac}, M., {Klein}, B., {Kneib}, J.-P., {et~al.} 2020, MNRAS submitted,
  arXiv:2006.10700

\bibitem[{{Jauzac} {et~al.}(2019){Jauzac}, {Mahler}, {Edge}, {Sharon},
  {Gillman}, {Ebeling}, {Harvey}, {Richard}, {Hamer}, {Fumagalli}, {Mark
  Swinbank}, {Kneib}, {Massey}, \& {Salom{\'e}}}]{2019MNRAS.483.3082J}
{Jauzac}, M., {Mahler}, G., {Edge}, A.~C., {et~al.} 2019, \mnras, 483, 3082

\bibitem[{{Jauzac} {et~al.}(2015){Jauzac}, {Richard}, {Jullo}, {Cl{\'e}ment},
  {Limousin}, {Kneib}, {Ebeling}, {Natarajan}, {Rodney}, {Atek}, {Massey},
  {Eckert}, {Egami}, \& {Rexroth}}]{2015MNRAS.452.1437J}
{Jauzac}, M., {Richard}, J., {Jullo}, E., {et~al.} 2015, \mnras, 452, 1437

\bibitem[{{Jauzac} {et~al.}(2016){Jauzac}, {Richard}, {Limousin}, {Knowles},
  {Mahler}, {Smith}, {Kneib}, {Jullo}, {Natarajan}, {Ebeling}, {Atek},
  {Cl{\'e}ment}, {Eckert}, {Egami}, {Massey}, \&
  {Rexroth}}]{2016MNRAS.457.2029J}
{Jauzac}, M., {Richard}, J., {Limousin}, M., {et~al.} 2016, \mnras, 457, 2029

\bibitem[{{Jullo} {et~al.}(2007){Jullo}, {Kneib}, {Limousin},
  {El{\'\i}asd{\'o}ttir}, {Marshall}, \& {Verdugo}}]{2007NJPh....9..447J}
{Jullo}, E., {Kneib}, J.~P., {Limousin}, M., {et~al.} 2007, New Journal of
  Physics, 9, 447

\bibitem[{{Jullo} {et~al.}(2010){Jullo}, {Natarajan}, {Kneib}, {D'Aloisio},
  {Limousin}, {Richard}, \& {Schimd}}]{2010Sci...329..924J}
{Jullo}, E., {Natarajan}, P., {Kneib}, J.~P., {et~al.} 2010, Science, 329, 924

\bibitem[{{Karman} {et~al.}(2015){Karman}, {Caputi}, {Grillo}, {Balestra},
  {Rosati}, {Vanzella}, {Coe}, {Christensen}, {Koekemoer}, {Kr{\"u}hler},
  {Lombardi}, {Mercurio}, {Nonino}, \& {van der Wel}}]{2015A&A...574A..11K}
{Karman}, W., {Caputi}, K.~I., {Grillo}, C., {et~al.} 2015, \aap, 574, A11

\bibitem[{{Kneib} {et~al.}(1996){Kneib}, {Ellis}, {Smail}, {Couch}, \&
  {Sharples}}]{1996ApJ...471..643K}
{Kneib}, J.~P., {Ellis}, R.~S., {Smail}, I., {Couch}, W.~J., \& {Sharples},
  R.~M. 1996, \apj, 471, 643

\bibitem[{{Kneib} \& {Natarajan}(2011)}]{2011A&ARv..19...47K}
{Kneib}, J.-P. \& {Natarajan}, P. 2011, \aapr, 19, 47

\bibitem[{{Kneib} {et~al.}(2004){Kneib}, {van der Werf}, {Kraiberg Knudsen},
  {Smail}, {Blain}, {Frayer}, {Barnard}, \& {Ivison}}]{2004MNRAS.349.1211K}
{Kneib}, J.-P., {van der Werf}, P.~P., {Kraiberg Knudsen}, K., {et~al.} 2004,
  \mnras, 349, 1211

\bibitem[{{Koekemoer} {et~al.}(2011){Koekemoer}, {Faber}, {Ferguson}, {Grogin},
  {Kocevski}, {Koo}, {Lai}, {Lotz}, {Lucas}, {McGrath}, {Ogaz}, {Rajan},
  {Riess}, {Rodney}, {Strolger}, {Casertano}, {Castellano}, {Dahlen},
  {Dickinson}, {Dolch}, {Fontana}, {Giavalisco}, {Grazian}, {Guo}, {Hathi},
  {Huang}, {van der Wel}, {Yan}, {Acquaviva}, {Alexander}, {Almaini}, {Ashby},
  {Barden}, {Bell}, {Bournaud}, {Brown}, {Caputi}, {Cassata}, {Challis},
  {Chary}, {Cheung}, {Cirasuolo}, {Conselice}, {Roshan Cooray}, {Croton},
  {Daddi}, {Dav{\'e}}, {de Mello}, {de Ravel}, {Dekel}, {Donley}, {Dunlop},
  {Dutton}, {Elbaz}, {Fazio}, {Filippenko}, {Finkelstein}, {Frazer}, {Gardner},
  {Garnavich}, {Gawiser}, {Gruetzbauch}, {Hartley}, {H{\"a}ussler},
  {Herrington}, {Hopkins}, {Huang}, {Jha}, {Johnson}, {Kartaltepe},
  {Khostovan}, {Kirshner}, {Lani}, {Lee}, {Li}, {Madau}, {McCarthy},
  {McIntosh}, {McLure}, {McPartland}, {Mobasher}, {Moreira}, {Mortlock},
  {Moustakas}, {Mozena}, {Nandra}, {Newman}, {Nielsen}, {Niemi}, {Noeske},
  {Papovich}, {Pentericci}, {Pope}, {Primack}, {Ravindranath}, {Reddy},
  {Renzini}, {Rix}, {Robaina}, {Rosario}, {Rosati}, {Salimbeni}, {Scarlata},
  {Siana}, {Simard}, {Smidt}, {Snyder}, {Somerville}, {Spinrad}, {Straughn},
  {Telford}, {Teplitz}, {Trump}, {Vargas}, {Villforth}, {Wagner}, {Wand ro},
  {Wechsler}, {Weiner}, {Wiklind}, {Wild}, {Wilson}, {Wuyts}, \&
  {Yun}}]{2011ApJS..197...36K}
{Koekemoer}, A.~M., {Faber}, S.~M., {Ferguson}, H.~C., {et~al.} 2011, \apjs,
  197, 36

\bibitem[{{Lagattuta} {et~al.}(2019){Lagattuta}, {Richard}, {Bauer},
  {Cl{\'e}ment}, {Mahler}, {Soucail}, {Carton}, {Kneib}, {Laporte}, {Martinez},
  {Patr{\'\i}cio}, {Payne}, {Pell{\'o}}, {Schmidt}, \& {de la
  Vieuville}}]{2019MNRAS.485.3738L}
{Lagattuta}, D.~J., {Richard}, J., {Bauer}, F.~E., {et~al.} 2019, \mnras, 485,
  3738

\bibitem[{{Lagattuta} {et~al.}(2017){Lagattuta}, {Richard}, {Cl{\'e}ment},
  {Mahler}, {Patr{\'\i}cio}, {Pell{\'o}}, {Soucail}, {Schmidt}, {Wisotzki},
  {Martinez}, \& {Bina}}]{2017MNRAS.469.3946L}
{Lagattuta}, D.~J., {Richard}, J., {Cl{\'e}ment}, B., {et~al.} 2017, \mnras,
  469, 3946

\bibitem[{{Laporte} {et~al.}(2017){Laporte}, {Bauer}, {Troncoso-Iribarren},
  {Huang}, {Gonz{\'a}lez-L{\'o}pez}, {Kim}, {Anguita}, {Aravena}, {Barrientos},
  {Bouwens}, {Bradley}, {Brammer}, {Carrasco}, {Carvajal}, {Coe}, {Demarco},
  {Ellis}, {Ford}, {Francke}, {Ibar}, {Infante}, {Kneissl}, {Koekemoer},
  {Messias}, {Mu{\~n}oz Arancibia}, {Nagar}, {Padilla}, {Pell{\'o}}, {Postman},
  {Qu{\'e}nard}, {Romero-Ca{\~n}izales}, {Treister}, {Villard}, {Zheng}, \&
  {Zitrin}}]{2017A&A...604A.132L}
{Laporte}, N., {Bauer}, F.~E., {Troncoso-Iribarren}, P., {et~al.} 2017, \aap,
  604, A132

\bibitem[{{Leclercq} {et~al.}(2020){Leclercq}, {Bacon}, {Verhamme}, {Garel},
  {Blaizot}, {Brinchmann}, {Cantalupo}, {Claeyssens}, {Conseil}, {Contini},
  {Hashimoto}, {Herenz}, {Kusakabe}, {Marino}, {Maseda}, {Matthee}, {Mitchell},
  {Pezzulli}, {Richard}, {Schmidt}, \& {Wisotzki}}]{2020A&A...635A..82L}
{Leclercq}, F., {Bacon}, R., {Verhamme}, A., {et~al.} 2020, \aap, 635, A82

\bibitem[{{Leclercq} {et~al.}(2017){Leclercq}, {Bacon}, {Wisotzki}, {Mitchell},
  {Garel}, {Verhamme}, {Blaizot}, {Hashimoto}, {Herenz}, {Conseil},
  {Cantalupo}, {Inami}, {Contini}, {Richard}, {Maseda}, {Schaye}, {Marino},
  {Akhlaghi}, {Brinchmann}, \& {Carollo}}]{2017A&A...608A...8L}
{Leclercq}, F., {Bacon}, R., {Wisotzki}, L., {et~al.} 2017, \aap, 608, A8

\bibitem[{{Leethochawalit} {et~al.}(2016){Leethochawalit}, {Jones}, {Ellis},
  {Stark}, \& {Zitrin}}]{2016ApJ...831..152L}
{Leethochawalit}, N., {Jones}, T.~A., {Ellis}, R.~S., {Stark}, D.~P., \&
  {Zitrin}, A. 2016, \apj, 831, 152

\bibitem[{{Leibundgut} {et~al.}(2017){Leibundgut}, {Bacon}, {Jaff{\'e}},
  {Johnston}, {Kuntschner}, {Selman}, {Valenti}, {Vernet}, \&
  {Vogt}}]{2017Msngr.170...20L}
{Leibundgut}, B., {Bacon}, R., {Jaff{\'e}}, Y.~L., {et~al.} 2017, The
  Messenger, 170, 20

\bibitem[{{Limousin} {et~al.}(2007){Limousin}, {Richard}, {Jullo}, {Kneib},
  {Fort}, {Soucail}, {El{\'\i}asd{\'o}ttir}, {Natarajan}, {Ellis}, {Smail},
  {Czoske}, {Smith}, {Hudelot}, {Bardeau}, {Ebeling}, {Egami}, \&
  {Knudsen}}]{2007ApJ...668..643L}
{Limousin}, M., {Richard}, J., {Jullo}, E., {et~al.} 2007, \apj, 668, 643

\bibitem[{{Lin} {et~al.}(2006){Lin}, {Mohr}, {Gonzalez}, \&
  {Stanford}}]{2006ApJ...650L..99L}
{Lin}, Y.-T., {Mohr}, J.~J., {Gonzalez}, A.~H., \& {Stanford}, S.~A. 2006,
  \apjl, 650, L99

\bibitem[{{Lotz} {et~al.}(2017){Lotz}, {Koekemoer}, {Coe}, {Grogin}, {Capak},
  {Mack}, {Anderson}, {Avila}, {Barker}, {Borncamp}, {Brammer}, {Durbin},
  {Gunning}, {Hilbert}, {Jenkner}, {Khandrika}, {Levay}, {Lucas}, {MacKenty},
  {Ogaz}, {Porterfield}, {Reid}, {Robberto}, {Royle}, {Smith},
  {Storrie-Lombardi}, {Sunnquist}, {Surace}, {Taylor}, {Williams}, {Bullock},
  {Dickinson}, {Finkelstein}, {Natarajan}, {Richard}, {Robertson}, {Tumlinson},
  {Zitrin}, {Flanagan}, {Sembach}, {Soifer}, \&
  {Mountain}}]{2017ApJ...837...97L}
{Lotz}, J.~M., {Koekemoer}, A., {Coe}, D., {et~al.} 2017, \apj, 837, 97

\bibitem[{{Mahler} {et~al.}(2018){Mahler}, {Richard}, {Cl{\'e}ment},
  {Lagattuta}, {Schmidt}, {Patr{\'\i}cio}, {Soucail}, {Bacon}, {Pello},
  {Bouwens}, {Maseda}, {Martinez}, {Carollo}, {Inami}, {Leclercq}, \&
  {Wisotzki}}]{2018MNRAS.473..663M}
{Mahler}, G., {Richard}, J., {Cl{\'e}ment}, B., {et~al.} 2018, \mnras, 473, 663

\bibitem[{{Mahler} {et~al.}(2019){Mahler}, {Sharon}, {Fox}, {Coe}, {Jauzac},
  {Strait}, {Edge}, {Acebron}, {Andrade-Santos}, {Avila}, {Brada{\v{c}}},
  {Bradley}, {Carrasco}, {Cerny}, {Cibirka}, {Czakon}, {Dawson}, {Frye},
  {Hoag}, {Huang}, {Johnson}, {Jones}, {Kikuchihara}, {Lam}, {Livermore},
  {Lovisari}, {Mainali}, {Ogaz}, {Ouchi}, {Paterno-Mahler}, {Roederer}, {Ryan},
  {Salmon}, {Sendra-Server}, {Stark}, {Toft}, {Trenti}, {Umetsu}, {Vulcani}, \&
  {Zitrin}}]{2019ApJ...873...96M}
{Mahler}, G., {Sharon}, K., {Fox}, C., {et~al.} 2019, \apj, 873, 96

\bibitem[{{Maseda} {et~al.}(2018){Maseda}, {Bacon}, {Franx}, {Brinchmann},
  {Schaye}, {Boogaard}, {Bouch{\'e}}, {Bouwens}, {Cantalupo}, {Contini},
  {Hashimoto}, {Inami}, {Marino}, {Muzahid}, {Nanayakkara}, {Richard},
  {Schmidt}, {Verhamme}, \& {Wisotzki}}]{2018ApJ...865L...1M}
{Maseda}, M.~V., {Bacon}, R., {Franx}, M., {et~al.} 2018, \apjl, 865, L1

\bibitem[{{Meneghetti} {et~al.}(2020){Meneghetti}, {Davoli}, {Bergamini},
  {Rosati}, {Natarajan}, {Giocoli}, {Caminha}, {Metcalf}, {Rasia}, {Borgani},
  {Calura}, {Grillo}, {Mercurio}, \& {Vanzella}}]{2020Sci...369.1347M}
{Meneghetti}, M., {Davoli}, G., {Bergamini}, P., {et~al.} 2020, Science, 369,
  1347

\bibitem[{{Moffat}(1969)}]{1969A&A.....3..455M}
{Moffat}, A.~F.~J. 1969, \aap, 3, 455

\bibitem[{{Orban de Xivry} \& {Marshall}(2009)}]{2009MNRAS.399....2O}
{Orban de Xivry}, G. \& {Marshall}, P. 2009, \mnras, 399, 2

\bibitem[{{Owers} {et~al.}(2011){Owers}, {Randall}, {Nulsen}, {Couch}, {David},
  \& {Kempner}}]{2011ApJ...728...27O}
{Owers}, M.~S., {Randall}, S.~W., {Nulsen}, P. E.~J., {et~al.} 2011, \apj, 728,
  27

\bibitem[{{Paraficz} {et~al.}(2016){Paraficz}, {Kneib}, {Richard}, {Morandi},
  {Limousin}, {Jullo}, \& {Martinez}}]{2016AandA...594A.121P}
{Paraficz}, D., {Kneib}, J.~P., {Richard}, J., {et~al.} 2016, \aap, 594, A121

\bibitem[{{Patr{\'\i}cio} {et~al.}(2018){Patr{\'\i}cio}, {Richard}, {Carton},
  {Contini}, {Epinat}, {Brinchmann}, {Schmidt}, {Krajnovi{\'c}}, {Bouch{\'e}},
  {Weilbacher}, {Pell{\'o}}, {Caruana}, {Maseda}, {Finley}, {Bauer},
  {Martinez}, {Mahler}, {Lagattuta}, {Cl{\'e}ment}, {Soucail}, \&
  {Wisotzki}}]{2018MNRAS.477...18P}
{Patr{\'\i}cio}, V., {Richard}, J., {Carton}, D., {et~al.} 2018, \mnras, 477,
  18

\bibitem[{{Patr{\'\i}cio} {et~al.}(2019){Patr{\'\i}cio}, {Richard}, {Carton},
  {P{\'e}roux}, {Contini}, {Brinchmann}, {Schaye}, {Weilbacher}, {Nanayakkara},
  {Maseda}, {Mahler}, \& {Wisotzki}}]{2019MNRAS.489..224P}
{Patr{\'\i}cio}, V., {Richard}, J., {Carton}, D., {et~al.} 2019, \mnras, 489,
  224

\bibitem[{{Patr{\'\i}cio} {et~al.}(2016){Patr{\'\i}cio}, {Richard}, {Verhamme},
  {Wisotzki}, {Brinchmann}, {Turner}, {Christensen}, {Weilbacher}, {Blaizot},
  {Bacon}, {Contini}, {Lagattuta}, {Cantalupo}, {Cl{\'e}ment}, \&
  {Soucail}}]{2016MNRAS.456.4191P}
{Patr{\'\i}cio}, V., {Richard}, J., {Verhamme}, A., {et~al.} 2016, \mnras, 456,
  4191

\bibitem[{{Piqueras} {et~al.}(2019){Piqueras}, {Conseil}, {Shepherd}, {Bacon},
  {Leclercq}, \& {Richard}}]{2019ASPC..521..545P}
{Piqueras}, L., {Conseil}, S., {Shepherd}, M., {et~al.} 2019, in Astronomical
  Society of the Pacific Conference Series, Vol. 521, Astronomical Data
  Analysis Software and Systems XXVI, ed. M.~{Molinaro}, K.~{Shortridge}, \&
  F.~{Pasian}, 545

\bibitem[{{Postman} {et~al.}(2012){Postman}, {Coe}, {Ben{\'\i}tez}, {Bradley},
  {Broadhurst}, {Donahue}, {Ford}, {Graur}, {Graves}, {Jouvel}, {Koekemoer},
  {Lemze}, {Medezinski}, {Molino}, {Moustakas}, {Ogaz}, {Riess}, {Rodney},
  {Rosati}, {Umetsu}, {Zheng}, {Zitrin}, {Bartelmann}, {Bouwens}, {Czakon},
  {Golwala}, {Host}, {Infante}, {Jha}, {Jimenez-Teja}, {Kelson}, {Lahav},
  {Lazkoz}, {Maoz}, {McCully}, {Melchior}, {Meneghetti}, {Merten}, {Moustakas},
  {Nonino}, {Patel}, {Reg{\"o}s}, {Sayers}, {Seitz}, \& {Van der
  Wel}}]{2012ApJS..199...25P}
{Postman}, M., {Coe}, D., {Ben{\'\i}tez}, N., {et~al.} 2012, \apjs, 199, 25

\bibitem[{{Repp} \& {Ebeling}(2018)}]{2018MNRAS.479..844R}
{Repp}, A. \& {Ebeling}, H. 2018, \mnras, 479, 844

\bibitem[{{Repp} {et~al.}(2016){Repp}, {Ebeling}, \&
  {Richard}}]{2016MNRAS.457.1399R}
{Repp}, A., {Ebeling}, H., \& {Richard}, J. 2016, \mnras, 457, 1399

\bibitem[{{Rescigno} {et~al.}(2020){Rescigno}, {Grillo}, {Lombardi}, {Rosati},
  {Caminha}, {Meneghetti}, {Mercurio}, {Bergamini}, \&
  {Coe}}]{2020A&A...635A..98R}
{Rescigno}, U., {Grillo}, C., {Lombardi}, M., {et~al.} 2020, \aap, 635, A98

\bibitem[{{Rexroth} {et~al.}(2017){Rexroth}, {Kneib}, {Joseph}, {Richard}, \&
  {Her}}]{2017arXiv170309239R}
{Rexroth}, M., {Kneib}, J.-P., {Joseph}, R., {Richard}, J., \& {Her}, R. 2017,
  arXiv e-prints, arXiv:1703.09239

\bibitem[{{Richard} {et~al.}(2019){Richard}, {Bacon}, {Blaizot}, {Boissier},
  {Boselli}, {NicolasBouch{\'e}}, {Brinchmann}, {Castro}, {Ciesla}, {Crowther},
  {Daddi}, {Dreizler}, {Duc}, {Elbaz}, {Epinat}, {Evans}, {Fossati},
  {Fumagalli}, {Garcia}, {Garel}, {Hayes}, {Herrero}, {Hugot}, {Humphrey},
  {Jablonka}, {Kamann}, {Kaper}, {Kelz}, {Kneib}, {de Koter}, {Krajnovi{\'c}},
  {Kudritzki}, {Langer}, {Lardo}, {Leclercq}, {Lennon}, {Mahler}, {Martins},
  {Massey}, {Mitchell}, {Monreal-Ibero}, {Najarro}, {Opitom}, {Papaderos},
  {P{\'e}roux}, {Revaz}, {Roth}, {Rousselot}, {Sander}, {Simmonds Wagemann},
  {Smail}, {Swinbank}, {Tramper}, {Urrutia}, {Verhamme}, {Vink}, {Walsh},
  {Weilbacher}, {Wendt}, {Wisotzki}, \& {Yang}}]{2019arXiv190601657R}
{Richard}, J., {Bacon}, R., {Blaizot}, J., {et~al.} 2019, arXiv e-prints,
  arXiv:1906.01657

\bibitem[{{Richard} {et~al.}(2014){Richard}, {Jauzac}, {Limousin}, {Jullo},
  {Cl{\'e}ment}, {Ebeling}, {Kneib}, {Atek}, {Natarajan}, {Egami}, {Livermore},
  \& {Bower}}]{2014MNRAS.444..268R}
{Richard}, J., {Jauzac}, M., {Limousin}, M., {et~al.} 2014, \mnras, 444, 268

\bibitem[{{Richard} {et~al.}(2015){Richard}, {Patricio}, {Martinez}, {Bacon},
  {Clement}, {Weilbacher}, {Soto}, {Wisotzki}, {Vernet}, {Pello}, {Schaye},
  {Turner}, \& {Martinsson}}]{2015MNRAS.446L..16R}
{Richard}, J., {Patricio}, V., {Martinez}, J., {et~al.} 2015, \mnras, 446, L16

\bibitem[{{Richard} {et~al.}(2009){Richard}, {Pei}, {Limousin}, {Jullo}, \&
  {Kneib}}]{2009A&A...498...37R}
{Richard}, J., {Pei}, L., {Limousin}, M., {Jullo}, E., \& {Kneib}, J.~P. 2009,
  \aap, 498, 37

\bibitem[{{Richard} {et~al.}(2010){Richard}, {Smith}, {Kneib}, {Ellis},
  {Sanderson}, {Pei}, {Targett}, {Sand}, {Swinbank}, {Dannerbauer}, {Mazzotta},
  {Limousin}, {Egami}, {Jullo}, {Hamilton-Morris}, \&
  {Moran}}]{2010MNRAS.404..325R}
{Richard}, J., {Smith}, G.~P., {Kneib}, J.-P., {et~al.} 2010, \mnras, 404, 325

\bibitem[{{Roth} {et~al.}(2018){Roth}, {Sandin}, {Kamann}, {Husser},
  {Weilbacher}, {Monreal-Ibero}, {Bacon}, {den Brok}, {Dreizler}, {Kelz},
  {Marino}, \& {Steinmetz}}]{2018A&A...618A...3R}
{Roth}, M.~M., {Sandin}, C., {Kamann}, S., {et~al.} 2018, \aap, 618, A3

\bibitem[{{Schmidt} {et~al.}(2014){Schmidt}, {Treu}, {Brammer}, {Brada{\v{c}}},
  {Wang}, {Dijkstra}, {Dressler}, {Fontana}, {Gavazzi}, {Henry}, {Hoag},
  {Jones}, {Kelly}, {Malkan}, {Mason}, {Pentericci}, {Poggianti}, {Stiavelli},
  {Trenti}, {von der Linden}, \& {Vulcani}}]{2014ApJ...782L..36S}
{Schmidt}, K.~B., {Treu}, T., {Brammer}, G.~B., {et~al.} 2014, \apjl, 782, L36

\bibitem[{{Seitz} {et~al.}(1998){Seitz}, {Saglia}, {Bender}, {Hopp}, {Belloni},
  \& {Ziegler}}]{1998MNRAS.298..945S}
{Seitz}, S., {Saglia}, R.~P., {Bender}, R., {et~al.} 1998, \mnras, 298, 945

\bibitem[{{Smit} {et~al.}(2017){Smit}, {Swinbank}, {Massey}, {Richard},
  {Smail}, \& {Kneib}}]{2017MNRAS.467.3306S}
{Smit}, R., {Swinbank}, A.~M., {Massey}, R., {et~al.} 2017, \mnras, 467, 3306

\bibitem[{{Soto} {et~al.}(2016){Soto}, {Lilly}, {Bacon}, {Richard}, \&
  {Conseil}}]{2016MNRAS.458.3210S}
{Soto}, K.~T., {Lilly}, S.~J., {Bacon}, R., {Richard}, J., \& {Conseil}, S.
  2016, \mnras, 458, 3210

\bibitem[{{Soucail} {et~al.}(1988){Soucail}, {Mellier}, {Fort}, {Mathez}, \&
  {Cailloux}}]{1988A&A...191L..19S}
{Soucail}, G., {Mellier}, Y., {Fort}, B., {Mathez}, G., \& {Cailloux}, M. 1988,
  \aap, 191, L19

\bibitem[{{Suyu} \& {Halkola}(2010)}]{2010A&A...524A..94S}
{Suyu}, S.~H. \& {Halkola}, A. 2010, \aap, 524, A94

\bibitem[{{Tremonti} {et~al.}(2004){Tremonti}, {Heckman}, {Kauffmann},
  {Brinchmann}, {Charlot}, {White}, {Seibert}, {Peng}, {Schlegel}, {Uomoto},
  {Fukugita}, \& {Brinkmann}}]{2004ApJ...613..898T}
{Tremonti}, C.~A., {Heckman}, T.~M., {Kauffmann}, G., {et~al.} 2004, \apj, 613,
  898

\bibitem[{{Treu} {et~al.}(2015){Treu}, {Schmidt}, {Brammer}, {Vulcani}, {Wang},
  {Brada{\v{c}}}, {Dijkstra}, {Dressler}, {Fontana}, {Gavazzi}, {Henry},
  {Hoag}, {Huang}, {Jones}, {Kelly}, {Malkan}, {Mason}, {Pentericci},
  {Poggianti}, {Stiavelli}, {Trenti}, \& {von der
  Linden}}]{2015ApJ...812..114T}
{Treu}, T., {Schmidt}, K.~B., {Brammer}, G.~B., {et~al.} 2015, \apj, 812, 114

\bibitem[{{Urrutia} {et~al.}(2019){Urrutia}, {Wisotzki}, {Kerutt}, {Schmidt},
  {Herenz}, {Klar}, {Saust}, {Werhahn}, {Diener}, {Caruana}, {Krajnovi{\'c}},
  {Bacon}, {Boogaard}, {Brinchmann}, {Enke}, {Maseda}, {Nanayakkara},
  {Richard}, {Steinmetz}, \& {Weilbacher}}]{2019A&A...624A.141U}
{Urrutia}, T., {Wisotzki}, L., {Kerutt}, J., {et~al.} 2019, \aap, 624, A141

\bibitem[{{van der Walt} {et~al.}(2011){van der Walt}, {Colbert}, \&
  {Varoquaux}}]{2011CSE....13b..22V}
{van der Walt}, S., {Colbert}, S.~C., \& {Varoquaux}, G. 2011, Computing in
  Science and Engineering, 13, 22

\bibitem[{{Vanzella} {et~al.}(2017){Vanzella}, {Balestra}, {Gronke}, {Karman},
  {Caminha}, {Dijkstra}, {Rosati}, {De Barros}, {Caputi}, {Grillo}, {Tozzi},
  {Meneghetti}, {Mercurio}, \& {Gilli}}]{2017MNRAS.465.3803V}
{Vanzella}, E., {Balestra}, I., {Gronke}, M., {et~al.} 2017, \mnras, 465, 3803

\bibitem[{{Vanzella} {et~al.}(2020{\natexlab{a}}){Vanzella}, {Caminha},
  {Rosati}, {Mercurio}, {Castellano}, {Meneghetti}, {Grillo}, {Sani},
  {Bergamini}, {Calura}, {Caputi}, {Cristiani}, {Cupani}, {Fontana}, {Gilli},
  {Grazian}, {Gronke}, {Mignoli}, {Nonino}, {Pentericci}, {Tozzi}, {Treu},
  {Balestra}, \& {Dijkstra}}]{2020arXiv200908458V}
{Vanzella}, E., {Caminha}, G.~B., {Rosati}, P., {et~al.} 2020{\natexlab{a}},
  A\&A in press, arXiv:2009.08458

\bibitem[{{Vanzella} {et~al.}(2020{\natexlab{b}}){Vanzella}, {Meneghetti},
  {Caminha}, {Castellano}, {Calura}, {Rosati}, {Grillo}, {Dijkstra}, {Gronke},
  {Sani}, {Mercurio}, {Tozzi}, {Nonino}, {Cristiani}, {Mignoli}, {Pentericci},
  {Gilli}, {Treu}, {Caputi}, {Cupani}, {Fontana}, {Grazian}, \&
  {Balestra}}]{2020MNRAS.494L..81V}
{Vanzella}, E., {Meneghetti}, M., {Caminha}, G.~B., {et~al.}
  2020{\natexlab{b}}, \mnras, 494, L81

\bibitem[{{Virtanen} {et~al.}(2020){Virtanen}, {Gommers}, {Oliphant},
  {Haberland}, {Reddy}, {Cournapeau}, {Burovski}, {Peterson}, {Weckesser},
  {Bright}, {van der Walt}, {Brett}, {Wilson}, {Jarrod Millman}, {Mayorov},
  {Nelson}, {Jones}, {Kern}, {Larson}, {Carey}, {Polat}, {Feng}, {Moore}, {Vand
  erPlas}, {Laxalde}, {Perktold}, {Cimrman}, {Henriksen}, {Quintero}, {Harris},
  {Archibald}, {Ribeiro}, {Pedregosa}, {van Mulbregt}, \&
  {Contributors}}]{2020SciPy-NMeth}
{Virtanen}, P., {Gommers}, R., {Oliphant}, T.~E., {et~al.} 2020, Nature
  Methods, 17, 261

\bibitem[{{Weilbacher} {et~al.}(2020){Weilbacher}, {Palsa}, {Streicher},
  {Bacon}, {Urrutia}, {Wisotzki}, {Conseil}, {Husemann}, {Jarno}, {Kelz},
  {P{\'e}contal-Rousset}, {Richard}, {Roth}, {Selman}, \&
  {Vernet}}]{2020A&A...641A..28W}
{Weilbacher}, P.~M., {Palsa}, R., {Streicher}, O., {et~al.} 2020, \aap, 641,
  A28

\bibitem[{{Wisotzki} {et~al.}(2016){Wisotzki}, {Bacon}, {Blaizot},
  {Brinchmann}, {Herenz}, {Schaye}, {Bouch{\'e}}, {Cantalupo}, {Contini},
  {Carollo}, {Caruana}, {Courbot}, {Emsellem}, {Kamann}, {Kerutt}, {Leclercq},
  {Lilly}, {Patr{\'\i}cio}, {Sandin}, {Steinmetz}, {Straka}, {Urrutia},
  {Verhamme}, {Weilbacher}, \& {Wendt}}]{2016A&A...587A..98W}
{Wisotzki}, L., {Bacon}, R., {Blaizot}, J., {et~al.} 2016, \aap, 587, A98

\bibitem[{{Zitrin} {et~al.}(2011){Zitrin}, {Broadhurst}, {Barkana}, {Rephaeli},
  \& {Ben{\'\i}tez}}]{2011MNRAS.410.1939Z}
{Zitrin}, A., {Broadhurst}, T., {Barkana}, R., {Rephaeli}, Y., \&
  {Ben{\'\i}tez}, N. 2011, \mnras, 410, 1939

\bibitem[{{Zitrin} {et~al.}(2012){Zitrin}, {Moustakas}, {Bradley}, {Coe},
  {Moustakas}, {Postman}, {Shu}, {Zheng}, {Ben{\'\i}tez}, {Bouwens},
  {Broadhurst}, {Ford}, {Host}, {Jouvel}, {Koekemoer}, {Meneghetti}, {Rosati},
  {Donahue}, {Grillo}, {Kelson}, {Lemze}, {Medezinski}, {Molino}, {Nonino}, \&
  {Ogaz}}]{2012ApJ...747L...9Z}
{Zitrin}, A., {Moustakas}, J., {Bradley}, L., {et~al.} 2012, \apjl, 747, L9

\end{thebibliography}
\begin{appendix} %First appendix
%
% \longtab[1]{
% \begin{landscape}
% \begin{longtable}{lrcrrrrrrrrl}
% ...
% \end{longtable}
% \end{landscape}
% }% End longtab

\section{Description of spectroscopic catalogues}

\begin{sidewaystable*}
\centering
\begin{tabular}{c|c| c|c|c|c|c|c|c|c|c|c|c|c|c|c|c}
\hline\hline
 ID & IDFROM & RA & DEC & $z$ & $zconf$ & m$_{F606W}$  & [...] & $z_{\rm forbid}$ & $z_{\rm forbid_{err}}$ &  $z_{\rm balmer}$ & $z_{\rm balmer_{err}}$ &  $z_{\rm Ly \alpha}$ & $z_{\rm Ly \alpha _{err}}$ & $\mu$ & $\mu_{\rm err}$ & MUL \\
  & & [deg] & [deg] & & & [mag] & & & & & & & & & \\
 \hline
     183& PRIOR & 145.22577  & 7.73872 & 4.03679 & 3 & 22.62 & [...] & 4.03354 & 0.00029 & -- & -- &4.03658 & 0.00002 &  12.8 & 2.6 & 1.2 \\
     364 & MUSELET & 145.22850  & 7.73786 & 0.33751 & 2  & --  & [...] & 0.33769& 0.00002& 0.33774&0.00007 & -- & -- & 1.0 & 0.0 & -- \\
      ... & ...  &  ...  & ... & ... & ... & ...  & [...] & ... & ... & ... &... & ... & ... & ...& ...& ...\\
\end{tabular}
\caption{\label{tab:spectable}Example entries of the spectroscopic table for the two sources presented in Fig.~\ref{fig:pyplatefitl}.
From left to right: source identification and origin of catalog (PRIOR, MUSELET or EXTERN), sky coordinates, redshift and confidence level, HST photometry in all bands, redshift measurements and errors from {\tt pyplatefit} for each family of spectral lines, magnification and error estimate, multiple image identification.}
\end{sidewaystable*}
\begin{sidewaystable*}
\centering
\begin{tabular}{c| c|c|c|c|c|c|c|c|c|c|c|c|c|c|c}
\hline\hline             
 ID & IDFROM & RA & DEC  & $zconf$ &z$_{\rm init}$ & Family   & Line & $\lambda _{\rm rf}$ & z  & z$_{\rm err}$ & [...] & Flux & Flux$_{\rm err}$ & SNR & $\lambda _{\rm obs}$  \\
 \hline
     %P183 & 145.22577  & 7.73872 & 3 & 4.03679 & O$_{\rm VI}$ 1032 & O$_{\rm VI}$ 1032 & 1031.91 & 4.03606 & 0.00016  & [...] & 11.2 & 5195.31\\
    % &   & &   & & Abs & O$_{\rm VI}$ 1032 & 1031.91 & 4.03647 & 0.00061  & [...] & 1.9 & 5195.74\\
    % &   & &   & & Abs & O$_{\rm VI}$ 1038 & 1037.61 & 4.03647 & 0.00061  & [...] & 0.13 & 5224.44\\
    % &   & &   & & O$_{\rm VI}$ 1032 & O$_{\rm VI}$ 1038 & 1037.61& 4.03606 & 0.00016  & [...] & 9.8 & 5224.01\\
     183 & PRIOR & 145.22577  & 7.73872 & 3 & 4.03679 & Ly$\alpha$ & Ly$\alpha$ & 1215.67& 4.03658 & 0.00002  & [...] & 5616 & 32 & 173.0 & 6121.13\\
    % &   & &   & & Abs  & Ly$\alpha$ & 1215.67& 4.03647 & 0.00061  & [...] & 1.1 & 6120.99\\
    % &   & &   & & N$_{\rm V}$ 1238  & N$_{\rm V}$ 1238 & 1238.82& 4.03737 & 0.00049  & [...] & 0.14 & 6238.66\\
    % &   & &   & & Abs  & N$_{\rm V}$ 1238 & 1238.82& 4.03647 & 0.00061  & [...] & 0.07 & 6237.55\\
     & &   & &   & & \nv~$\lambda$1238  & N$_{\rm V}$~$\lambda$1243 & 1242.80& 4.03737 & 0.00049  & [...] & 121 & 30 & 4.01 & 6258.41\\
     %&   & &   & & Abs  & N$_{\rm V}$ 1243 & 1242.80& 4.03647 & 0.00061  & [...] & 0.20 & 6257.60\\
    % &   & &   & & Abs  & Si$_{\rm II}$ 1260 & 1260.42 & 4.03647 & 0.00061  & [...] & 0.12 & 6346.32\\
    % &   & &   & & Abs  & O$_{\rm I}$ 1302 & 1302.17 & 4.03647 & 0.00061  & [...] & 1.58 & 6556.53\\
    % &   & &   & & Abs  & Si$_{\rm II}$ 1304 & 1304.37 & 4.03647 & 0.00061  & [...] & 0.12 & 6567.61\\
    % &   & &   & & Abs  & C$_{\rm II}$ 1334 & 1334.53 & 4.03647 & 0.00061  & [...] & 2.77 & 6719.47\\
    % &   & &   & & Abs  & Si$_{\rm IV}$ 1394 & 1393.76 & 4.03647 & 0.00061  & [...] & 0.17 & 7017.69\\
     %&   & &   & & Abs  & Si$_{\rm II}$ 1403 & 1402.77 & 4.03647 & 0.00061  & [...] & 0.58 & 7063.06\\
    % &   & &   & & Forbidden & N$_{\rm IV}$ 1483 & 1483.32 & 4.03354 & 0.00029  & [...] & 2.18 & 7464.29\\
     %&   & &   & & Forbidden  & N$_{\rm IV}$ 1487 & 1486.50 & 4.03354 & 0.00029  & [...] & 2.67 & 7480.29\\
    % &   & &   & & Abs  & C$_{\rm IV}$ 1548 & 1548.19 & 4.03647 & 0.00061  & [...] & 2.68 & 7795.32\\
     & &   & &   & & \civ  & \civ~$\lambda$1548 & 1548.19 & 4.03466 & 0.00013  & [...] & 220 & 26 & 8.41 & 7792.50\\
    % &   & &   & & Abs  & C$_{\rm IV}$ 1551 & 1550.77 & 4.03647 & 0.00061  & [...] & 0.33 & 7808.26\\
     & &   & &   & & \civ  & \civ$\lambda$1551 & 1550.77 & 4.03466 & 0.00013  & [...] & 128 & 16 & 7.90 & 7805.44\\
    % &   & &   & & Abs  & Fe$_{\rm II}$ 1608 & 1608.45 & 4.03647 & 0.00061  & [...] & 0.87 & 8098.68\\
    % &   & &   & & Abs  & Fe$_{\rm II}$ 1611 & 1611.20 & 4.03647 & 0.00061  & [...] & 0.28 & 8112.53\\
    % &   & &   & & Forbidden  & He$_{\rm II} $1640 & 1640.42 & 4.03354 & 0.00029  & [...] & 0.14 & 8254.95\\
     &   & &   & & &  Forbidden & \oiii]~$\lambda$1660 & 1660.81 & 4.03354 & 0.00029  & [...] & 120 & 23 & 5.14 & 8357.45\\
     &   & &   & & & Forbidden  & [\oiii]~$\lambda$1666 & 1666.15 & 4.03354 & 0.00029  & [...] & 346 & 60 & 5.78 & 8384.32\\
    % &   & &   & & Abs  & Al$_{\rm II}$ 1671 & 1670.79 & 4.03647 & 0.00061  & [...] & 1.14 & 8412.57\\
     %&   & &   & & Forbidden  & N$_{\rm III}$ 1750 & 1749.67 & 4.03354 & 0.00029  & [...] & 0.85 & 8884.62\\
    % &   & &   & & Abs  & Al$_{\rm II}$ 1854 & 1854.10 & 4.03647 & 0.00061  &[...] & 0.55 & 9335.56\\
     364 & MUSELET & 145.22850  & 7.73786 & 2 & 0.33751 & Forbidden  & [\oii]~$\lambda$3727 & 3727.09 & 0.33769 & 0.00002  & [...] & 51 & 10 & 4.9 & 4984.31\\
      & &   &  &  &  & Forbidden  & [\oii]~$\lambda$3729 & 3729.88 & 0.33769 & 0.00002  & [...] & 31 & 9 & 3.3 & 4988.04\\
       & &   &  &  &  & Balmer  &H$\beta$ &  4862.68  & 0.33774 & 0.00007  & [...] & 52 & 10 & 5.49 & 6503.17\\
      % &   &  &  &  & Forbidden  &O$_{\rm III}$ 4960 & 4960.29 & 0.33769 & 0.00002  & [...] & 1.39 & 6633.53\\
       & &   &  &  &  & Forbidden  & [\oiii] & 5008.24 & 0.33769 & 0.00002  & [...] & 125 & 9 & 14.06 & 6697.64\\
      % &   &  &  &  & Abs  &MgB & 5175.44 & 0.33753 & 0.00015  & [...] & 0.75 & 6920.38\\
       %&   &  &  &  & Forbidden  &He$_{\rm I}$ 5877 & 5877.25 & 0.33769 & 0.00002  & [...] & 0.23 & 7859.79\\
       %&   &  &  &  & Abs  &NaD & 5891.94 & 0.33753 & 0.00015  & [...] & 0.18 & 7878.46\\
       %&   &  &  &  & Forbidden  &O$_{\rm I}$ 6302 & 6302.05 & 0.33769 & 0.00002  & [...] & 0.29 & 8427.89\\
      % &   &  &  &  & Forbidden  &N$_{\rm II}$ 6550 & 6549.85 & 0.33769 & 0.00002  & [...] & 0.31 & 8759.28\\
       & &   &  &  &  & Balmer  &H$\alpha$ &  6564.61  & 0.33774 & 0.00007  & [...] & 112 & 34 & 3.29 & 8779.29\\
       %&   &  &  &  & Abs  &H$_{\alpha}$ & 6564.61  & 0.33753 & 0.00015  & [...] & 0.76 & 8777.93\\
       %&   &  &  &  & Forbidden  &N$_{\rm II}$ 6585 & 6585.28 & 0.33769 & 0.00002  & [...] & 2.32 & 8806.66\\
      % &   &  &  &  & Forbidden  & [\sii]~\lambda6718 & 6718.29 & 0.33769 & 0.00002  & [...] & 3.83 & 8984.54\\
      % &   &  &  &  & Forbidden  & [\sii]~\lambda6733 & 6732.66 & 0.33769 & 0.00002  & [...] & 1.60 & 9003.77\\
      & ...  & ...  & ... & ... & ... & ... &... & ... & ... & ...  & [...] & ... & ... & ... & ...\\

\end{tabular}
\caption{\label{tab:speclinestable}Example entries of the spectroscopic lines table, for the same sources as Table~\ref{tab:spectable}. Each row gives the best parameters of a {\sc pyplatefit} line fit for the most significant lines detected. Fluxes are in units of 10$^{-20}$ erg\,s$^{-1}$\,cm$^{-2}$.}
\end{sidewaystable*}

\clearpage

\section{Mass model parameters and multiple images}

We present for all new mass models the best fit parameters from Lenstool in Table \ref{tab:massmodel} and the list of constraints used in Tables \ref{tab:multiples} to \ref{tab:multiples2}. Figure \ref{fig:MACS0257mul} mark these multiply imaged systems over the HST colour image in each cluster. A full description of the Abell 2744 and Abell 370 mass models has been provided in \citet{2018MNRAS.473..663M} and \citet{2019MNRAS.485.3738L} respectively, and they were released as v4 of the Clusters As Telescopes (CATs) in the Frontier Field mass models\footnote{Available at https://archive.stsci.edu/prepds/frontier/lensmodels/}. The new MACS0416 model is very similar to v4 of the CATS model available on the same page.

\label{app:mul}
\begin{table*}[ht]
    \centering
    \begin{tabular}{c|c|c|c|c|c|c|c}
\hline\hline             
    Potential & $\Delta$R.A. & $\Delta$Dec. & $e$ & $\theta$ &  $r_{\rm core}$ & $r_{\rm cut}$ & $\sigma$ \\
    & [\arcsec] & [\arcsec] & & [deg]  & [kpc] & [kpc] & [km\ $s^{-1}$] \\
\hline\hline
   MACS0257 & & & & & & & \\
    \hline 
DM1 & $ -2.1^{+  0.3}_{ -0.2}$ & $  1.8^{+  0.3}_{ -0.2}$ & $ 0.59^{+ 0.02}_{-0.02}$ & $200.9^{+  1.1}_{ -1.2}$ & $62^{+2}_{-2}$ & $1014^{+145}_{-79}$ & $877^{+17}_{-16}$ \\
DM2 & $  8.5^{+  5.0}_{ -2.4}$ & $ -8.4^{+  0.8}_{ -2.6}$ & $ 0.87^{+ 0.03}_{-0.04}$ & $150.1^{+  2.1}_{ -0.8}$ & $189^{+6}_{-12}$ & $1093^{+111}_{-174}$ & $733^{+15}_{-22}$ \\
GAL1 & $[-14.1]$ & $[ 15.1]$ & $ 0.39^{+ 0.13}_{-0.14}$ & $  6.8^{+ 35.7}_{-34.5}$ & $[0]$ & $88^{+5}_{-10}$ & $184^{+6}_{-5}$ \\
GAL2 & $[-10.6]$ & $[ 17.6]$ & $[0.50]$ & $ 30.6^{+  8.2}_{-12.5}$ & $[0]$ & $[40]$ & $171^{+22}_{-16}$ \\
GAL3 & $[ 12.2]$ & $[ 19.6]$ & $ 0.17^{+ 0.09}_{-0.11}$ & $ 55.5^{+  9.6}_{ -7.0}$ & $[0]$ & $145^{+10}_{-31}$ & $322^{+8}_{-8}$ \\
GAL4 & $[ -0.2]$ & $[  0.0]$ & $[0.46]$ & $[ -9.6]$ & $[0]$ & $148^{+41}_{-27}$ & $349^{+3}_{-1}$ \\

L$^{*}$ galaxy &  & & & & $[0.15]$ & $51^{+4}_{-3}$ & $154^{+3}_{-2}$\\
    \hline\hline 
MACS0329     
& & & & & & & \\
    \hline  
DM1 & $  0.4^{+  0.1}_{ -0.2}$ & $ -0.4^{+  0.2}_{ -0.1}$ & $ 0.16^{+ 0.02}_{-0.01}$ & $ 64.7^{+  4.1}_{ -2.1}$ & $55^{+3}_{-4}$ & $[1000]$ & $959^{+14}_{-17}$ \\
DM2 & $ 43.2^{+  1.5}_{ -0.7}$ & $ 17.8^{+  1.2}_{ -1.5}$ & $[0.30]$ & $ 73.1^{+  5.0}_{ -4.4}$ & $7^{+10}_{-25}$ & $[1000]$ & $552^{+30}_{-40}$ \\
DM3 & $ 40.6^{+  2.2}_{ -1.0}$ & $ 67.9^{+  2.1}_{ -1.1}$ & $[0.30]$ & $ 50.6^{+  3.0}_{ -2.5}$ & $[25]$ & $[1000]$ & $573^{+40}_{-22}$ \\
GAL1 & $[0.0]$ & $[  0.0]$ & $[0.19]$ & $[-73.6]$ & $[0]$ & $[98]$ & $281^{+15}_{-30}$ \\
GAL2 & $[-12.7]$ & $[-39.9]$ & $[0.14]$ & $[ 56.9]$ & $[0]$ & $[41]$ & $218^{+4}_{-4}$ \\
L$^{*}$ galaxy &  & & & & $[0.15]$ & $[45]$ & $159^{+4}_{-4}$\\
\hline\hline
MACS0416
& & & & & & & \\
    \hline  
DM1 & $ -2.9^{+  0.3}_{ -0.3}$ & $  1.4^{+  0.3}_{ -0.2}$ & $ 0.78^{+ 0.01}_{-0.01}$ & $142.1^{+  0.4}_{ -0.4}$ & $59^{+2}_{-2}$ & $[1000]$ & $731^{+10}_{-11}$ \\
DM2 & $ 22.6^{+  0.3}_{ -0.2}$ & $-42.4^{+  0.4}_{ -0.6}$ & $ 0.69^{+ 0.01}_{-0.01}$ & $127.1^{+  0.2}_{ -0.3}$ & $92^{+2}_{-3}$ & $[1000]$ & $940^{+12}_{-11}$ \\
GAL1 & $[ 31.8]$ & $[-65.5]$ & $[0.04]$ & $[-40.4]$ & $[0]$ & $[62]$ & $137^{+10}_{-11}$ \\
GAL2 & $-37.2^{+  0.6}_{ -0.8}$ & $  7.8^{+  1.3}_{ -1.1}$ & $ 0.82^{+ 0.03}_{-0.03}$ & $118.5^{+  3.9}_{ -3.8}$ & $[25]$ & $[200]$ & $252^{+10}_{-7}$ \\
L$^{*}$ galaxy &  & & & & $[0.15]$ & $36^{+3}_{-2}$ & $137^{+2}_{-2}$\\
\hline
\hline 
Bullet Cluster
& & & & & & & \\
\hline 
DM1 & $  4.8^{+  0.2}_{ -0.1}$ & $  1.2^{+  0.5}_{ -0.5}$ & $ 0.68^{+ 0.03}_{-0.03}$ & $ 79.5^{+  0.2}_{ -0.7}$ & $138^{+8}_{-9}$ & $[1000]$ & $787^{+19}_{-25}$ \\
DM2 & $ 29.9^{+  0.0}_{ -0.2}$ & $ 26.3^{+  0.4}_{ -0.5}$ & $ 0.64^{+ 0.02}_{-0.02}$ & $ 55.8^{+  0.6}_{ -0.9}$ & $168^{+4}_{-4}$ & $[1000]$ & $1004^{+28}_{-21}$ \\
DM3 & $183.8^{+  0.4}_{ -0.4}$ & $ 49.1^{+  0.2}_{ -0.2}$ & $ 0.42^{+ 0.04}_{-0.03}$ & $ 11.8^{+  1.2}_{ -0.7}$ & $78^{+3}_{-5}$ & $[1000]$ & $885^{+23}_{-20}$ \\
GAL1 & $[  0.0]$ & $[  0.0]$ & $[0.26]$ & $[ 43.5]$ & $[0]$ & $[150]$ & $268^{+9}_{-13}$ \\
GAL2 & $[ 24.0]$ & $[ 29.1]$ & $[0.20]$ & $[ 37.4]$ & $[0]$ & $[112]$ & $118^{+12}_{-10}$ \\
GAL3 & $[ 51.9]$ & $[ 48.9]$ & $[0.13]$ & $[  9.9]$ & $[0]$ & $[60]$ & $164^{+2}_{-2}$ \\
GAL4 & $[  5.2]$ & $[ 23.0]$ & $ 0.16^{+ 0.02}_{-0.03}$ & $-77.1^{+ 10.0}_{-10.3}$ & $[0]$ & $73^{+10}_{-9}$ & $73^{+5}_{-6}$ \\
L$^{*}$ galaxy &  & & & & $[0.15]$ & $25^{+3}_{-2}$ & $165^{+2}_{-3}$\\
\hline
\hline 
MACS0940
& & & & & & & \\
    \hline 
DM1 & $  0.6^{+  0.8}_{ -0.7}$ & $  0.6^{+  1.4}_{ -0.2}$ & $ 0.46^{+ 0.19}_{-0.04}$ & $ 23.5^{+  2.0}_{ -1.2}$ & $31^{+79}_{-8}$ & $1386^{+565}_{-70}$ & $496^{+223}_{-42}$ \\
GAL1 & $[ -0.1]$ & $[  0.1]$ & $ 0.37^{+ 0.09}_{-0.06}$ & $[ -7.7]$ & $[0]$ & $[52]$ & $436^{+15}_{-30}$ \\
GAL2 & $[-11.8]$ & $[  3.1]$ & $ 0.66^{+ 0.07}_{-0.29}$ & $  5.9^{+ 19.9}_{-20.9}$ & $[0]$ & $[17]$ & $195^{+17}_{-14}$ \\
GAL3 & $[  5.9]$ & $[ -5.7]$ & $[0.00]$ & $ -5.7^{+ 87.0}_{-34.1}$ & $[0]$ & $68^{+34}_{-24}$ & $80^{+6}_{-24}$ \\
L$^{*}$ galaxy &  & & & & $[0.15]$ & $62^{+62}_{-94}$ & $162^{+28}_{-7}$\\
\hline
\hline 
MACS1206
& & & & & & & \\
    \hline 
DM1 & $ -0.1^{+  0.0}_{ -0.0}$ & $  0.7^{+  0.0}_{ -0.0}$ & $ 0.63^{+ 0.01}_{-0.01}$ & $ 19.9^{+  0.2}_{ -0.1}$ & $44^{+0}_{-1}$ & $[1000]$ & $888^{+7}_{-6}$ \\
DM2 & $  9.5^{+  0.5}_{ -0.2}$ & $  5.7^{+  0.4}_{ -0.3}$ & $ 0.70^{+ 0.02}_{-0.03}$ & $114.7^{+  0.7}_{ -0.5}$ & $94^{+3}_{-3}$ & $[1000]$ & $575^{+6}_{-9}$ \\
DM3 & $-32.5^{+  0.9}_{ -1.0}$ & $ -9.4^{+  0.2}_{ -0.2}$ & $ 0.40^{+ 0.02}_{-0.02}$ & $-15.5^{+  2.0}_{ -1.8}$ & $109^{+3}_{-3}$ & $[1000]$ & $592^{+8}_{-9}$ \\
GAL1 & $[ -0.1]$ & $[  0.0]$ & $[0.71]$ & $[ 14.4]$ & $1^{+0}_{-1}$ & $20^{+1}_{-1}$ & $355^{+11}_{-6}$ \\
GAL2 & $[ 35.8]$ & $[ 16.1]$ & $[0.23]$ & $133.8^{+ 69.7}_{-47.6}$ & $[0]$ & $4^{+1}_{-1}$ & $275^{+18}_{-11}$ \\
$\gamma$ &  &  &  $0.10^{+0.01}_{-0.01}$ & $101.4^{+  0.4}_{ -0.5}$ &  &  &  \\
L$^{*}$ galaxy &  & & & &  & $34^{+5}_{-1}$ & $198^{+5}_{-6}$\\
\hline

   \end{tabular}
    \caption{Best fit model parameters for the mass distribution in each cluster. From left to right: mass component, position relative to cluster centre ($\Delta$R.A. and $\Delta$Dec.), dPIE shape (ellipticity and orientation), core and cut radii, velocity dispersion. The final row 
    is the generic galaxy mass at the characteristic luminosity L$^*$, which is scaled to match each of cluster member galaxies.  Parameters in square brackets are fixed {\it a priori} in the model. }
    \label{tab:massmodel}
\end{table*}

\begin{table*}\ContinuedFloat
    \centering
    \begin{tabular}{c|c|c|c|c|c|c|c}
\hline\hline  
    Potential & $\Delta$R.A. & $\Delta$Dec. & $e$ & $\theta$ &  $r_{\rm core}$ & $r_{\rm cut}$ & $\sigma$ \\
    & [arcsec] & [arcsec] & & [deg]  & [kpc] & [kpc] & [km\ $s^{-1}$] \\
 \hline\hline             
    RXJ1347 & & & & & & & \\
\hline
DM1 & $  0.4^{+  0.1}_{ -0.1}$ & $  5.1^{+  0.4}_{ -0.2}$ & $ 0.76^{+ 0.02}_{-0.02}$ & $111.8^{+  0.5}_{ -0.7}$ & $37^{+1}_{-1}$ & $[1000]$ & $638^{+24}_{-24}$ \\
DM2 & $-13.6^{+  0.2}_{ -0.1}$ & $ -4.5^{+  0.2}_{ -0.4}$ & $ 0.78^{+ 0.00}_{-0.01}$ & $121.4^{+  0.2}_{ -0.1}$ & $78^{+2}_{-2}$ & $[1000]$ & $850^{+8}_{-4}$ \\
DM3 & $  2.0^{+  0.4}_{ -0.2}$ & $  0.3^{+  0.3}_{ -0.3}$ & $ 0.43^{+ 0.03}_{-0.02}$ & $ 60.4^{+  1.9}_{ -1.8}$ & $67^{+2}_{-2}$ & $[1000]$ & $739^{+23}_{-28}$ \\
GAL1 & $[  0.0]$ & $[ -0.0]$ & $[0.23]$ & $[-86.9]$ & $[0]$ & $84^{+13}_{-13}$ & $354^{+7}_{-5}$ \\
GAL2 & $[-17.8]$ & $[ -2.1]$ & $[0.30]$ & $[-64.1]$ & $[0]$ & $94^{+6}_{-4}$ & $364^{+2}_{-3}$ \\
GAL3 & $[ 15.5]$ & $[ 19.6]$ & $[0.06]$ & $[ 14.2]$ & $[0]$ & $88^{+31}_{-19}$ & $109^{+10}_{-7}$ \\
$\gamma$ &  &  &  & $ 64.4^{+  0.4}_{ -0.4}$ &  &  &  \\
L$^{*}$ galaxy &  & & & & $[0.15]$ & $81^{+9}_{-9}$ & $135^{+3}_{-4}$\\
\hline
\hline 
SMACS2031& & & & & & & \\
\hline
DM1 & $  0.4^{+  0.1}_{ -0.1}$ & $ -0.7^{+  0.1}_{ -0.1}$ & $ 0.31^{+ 0.02}_{-0.02}$ & $  4.4^{+  1.0}_{ -1.1}$ & $29^{+1}_{-1}$ & $[1000]$ & $624^{+12}_{-11}$ \\
DM2 & $ 61.2^{+  0.4}_{ -0.5}$ & $ 25.3^{+  0.4}_{ -0.3}$ & $ 0.46^{+ 0.05}_{-0.04}$ & $  6.4^{+  2.1}_{ -3.9}$ & $112^{+8}_{-7}$ & $[1000]$ & $1037^{+20}_{-19}$ \\
GAL1 & $[  0.1]$ & $[ -0.1]$ & $[0.09]$ & $[ -0.4]$ & $[0]$ & $128^{+12}_{-46}$ & $242^{+3}_{-4}$ \\
L$^{*}$ galaxy &  & & & & $[0.15]$ & $9^{+2}_{-1}$ & $161^{+19}_{-9}$\\
\hline
\hline 
SMACS2131& & & & & & & \\
\hline
DM1 & $ -3.2^{+  0.4}_{ -0.2}$ & $  3.2^{+  0.1}_{ -0.3}$ & $ 0.66^{+ 0.01}_{-0.01}$ & $155.2^{+  0.2}_{ -0.3}$ & $84^{+1}_{-1}$ & $[1000]$ & $866^{+15}_{-5}$ \\
DM2 & $ 21.2^{+  0.9}_{ -0.3}$ & $[ 17.0]$ & $ 0.53^{+ 0.02}_{-0.02}$ & $ 82.7^{+  6.4}_{ -6.9}$ & $95^{+14}_{-6}$ & $[1000]$ & $452^{+44}_{-12}$ \\
GAL1 & $[ -0.0]$ & $[  0.0]$ & $[0.11]$ & $ 59.0^{+  1.4}_{ -1.3}$ & $[0]$ & $155^{+3}_{-14}$ & $399^{+0}_{-1}$ \\
GAL2 & $[  6.7]$ & $[ -2.2]$ & $[0.76]$ & $  7.2^{+  9.7}_{ -6.4}$ & $[0]$ & $96^{+8}_{-17}$ & $93^{+18}_{-12}$ \\
L$^{*}$ galaxy &  & & & & $[0.15]$ & $93^{+6}_{-21}$ & $202^{+4}_{-1}$\\
\hline
\hline 
MACS2214& & & & & & & \\
\hline
DM1 & $ -1.2^{+  0.1}_{ -0.2}$ & $  0.7^{+  0.1}_{ -0.1}$ & $ 0.55^{+ 0.01}_{-0.01}$ & $147.5^{+  0.7}_{ -0.8}$ & $38^{+1}_{-1}$ & $[1000]$ & $903^{+15}_{-18}$ \\
DM2 & $-20.8^{+  0.2}_{ -0.2}$ & $ 17.2^{+  0.3}_{ -0.2}$ & $ 0.70^{+ 0.06}_{-0.05}$ & $112.0^{+  3.2}_{ -3.5}$ & $20^{+8}_{-5}$ & $[1000]$ & $299^{+11}_{-34}$ \\
DM3 & $ 20.8^{+  3.9}_{ -4.9}$ & $-22.8^{+  2.0}_{ -1.6}$ & $ 0.92^{+ 0.05}_{-0.03}$ & $ 22.6^{+  1.7}_{ -2.0}$ & $186^{+11}_{-13}$ & $[1000]$ & $753^{+42}_{-41}$ \\
GAL1 & $[  0.0]$ & $[  0.0]$ & $[0.20]$ & $151.1^{+ 34.0}_{-45.5}$ & $4^{+14}_{-3}$ & $9^{+16}_{-12}$ & $179^{+61}_{-28}$ \\
GAL2 & $[  8.2]$ & $[ 18.8]$ & $ 0.48^{+ 0.07}_{-0.04}$ & $[  0.0]$ & $[0]$ & $81^{+2}_{-3}$ & $169^{+5}_{-9}$ \\
GAL3 & $[ 19.9]$ & $[  2.0]$ & $ 0.42^{+ 0.06}_{-0.35}$ & $ 89.2^{+ 31.5}_{-53.1}$ & $[0]$ & $23^{+42}_{-8}$ & $73^{+8}_{-5}$ \\
L$^{*}$ galaxy &  & & & & $[0.15]$ & $46^{+3}_{0}$ & $111^{+2}_{-3}$\\
\hline
    \end{tabular}
    \caption{(continued).}
\end{table*}

\begin{figure*}
\includegraphics[width=14cm]{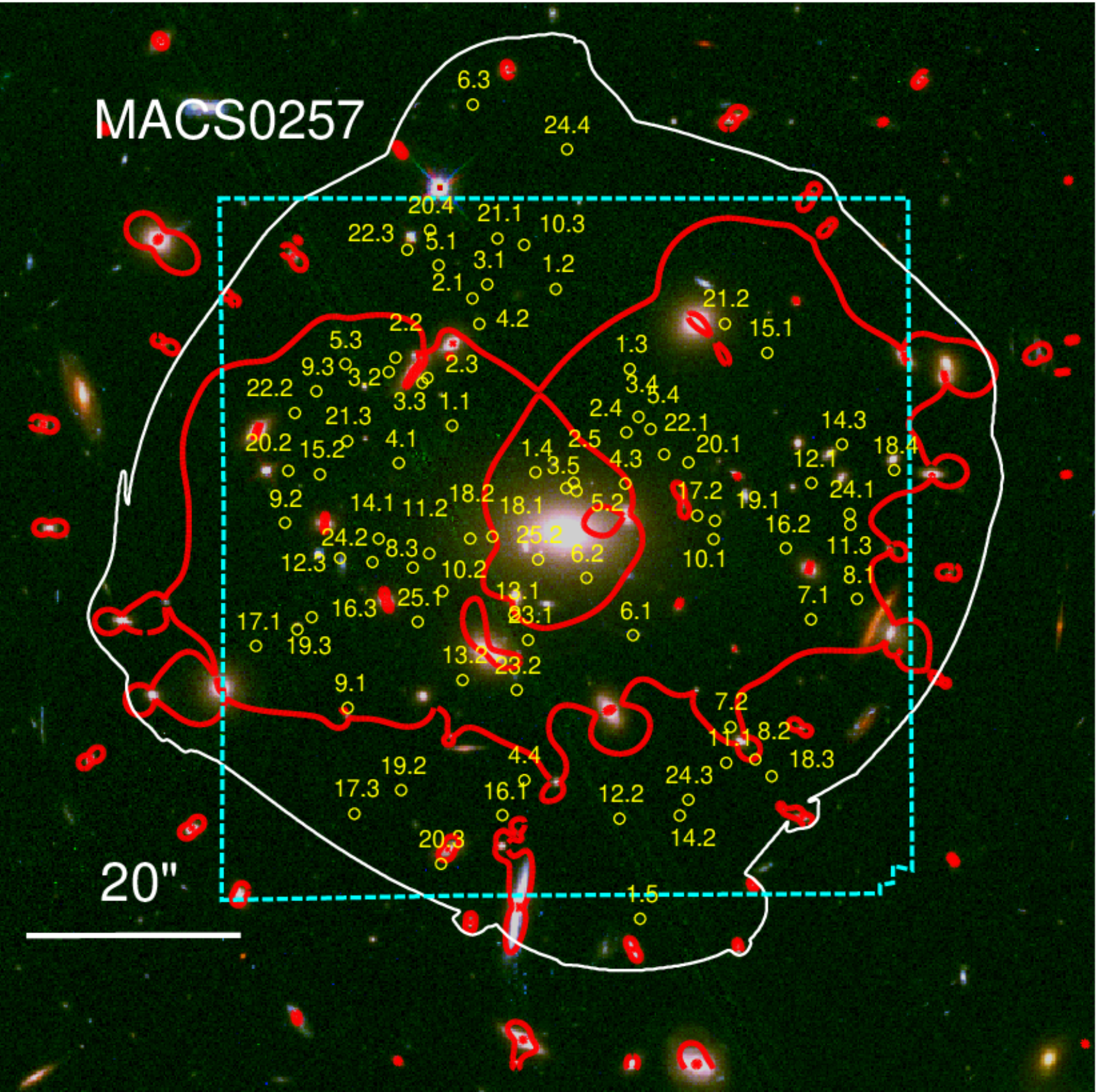}
\caption{\label{fig:MACS0257mul}Overview of all multiply-imaged systems used in the strong-lensing model of MACS0257. The cyan dashed line highlights the limits of the MUSE observations. The white line represents the limit of the region where we expect multiples images. The red line delineates the critical line for a source at $z=4$.}
\end{figure*}
\begin{figure*}\ContinuedFloat
\includegraphics[width=14cm]{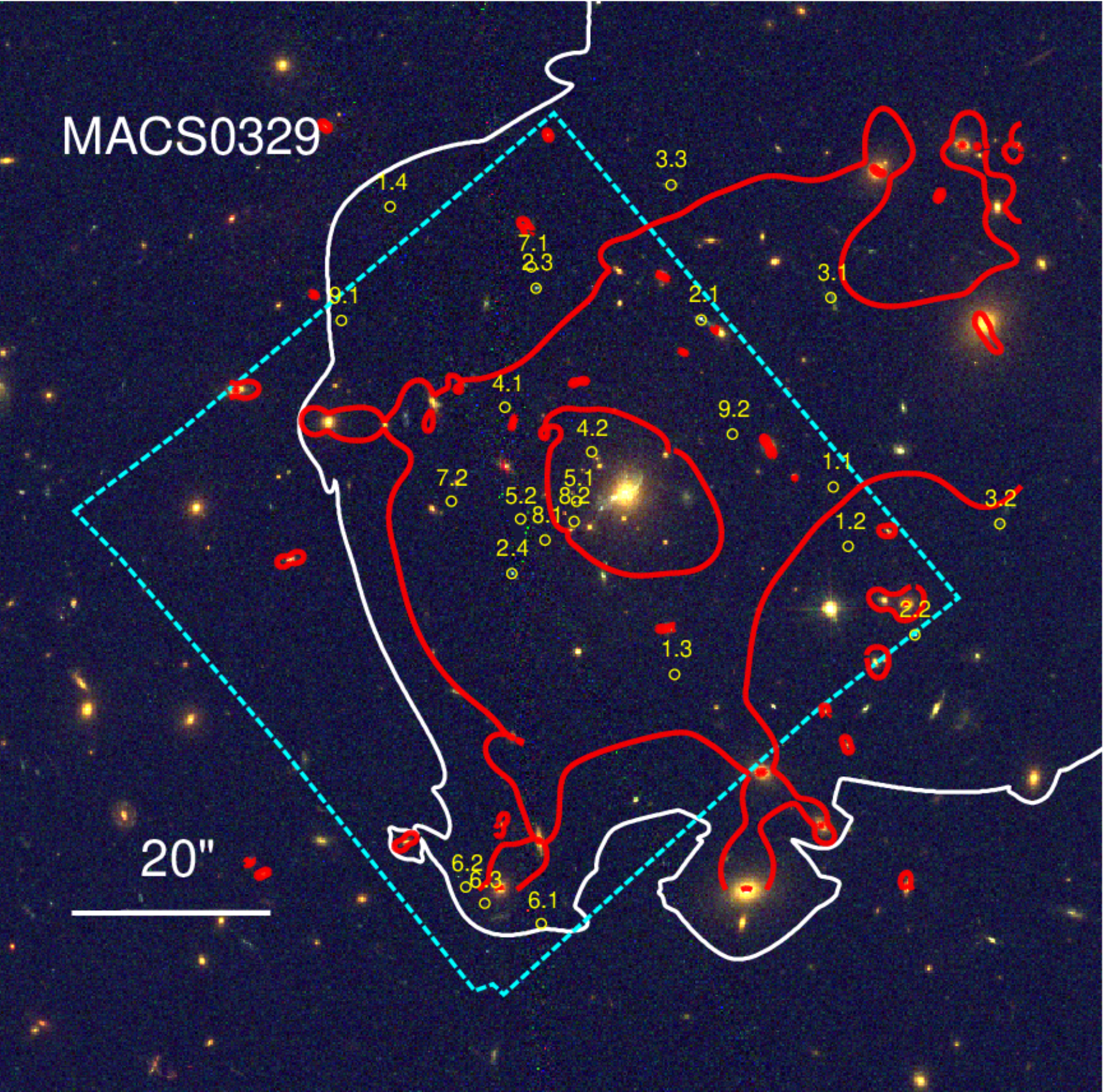}
%\caption{Same as Fig.\,\ref{fig:MACS0257mul} for MACS0329.}
\caption{\textbf{(continued)} Same figure for MACS0329.}
\end{figure*}
\begin{figure*}\ContinuedFloat
\includegraphics[width=14cm]{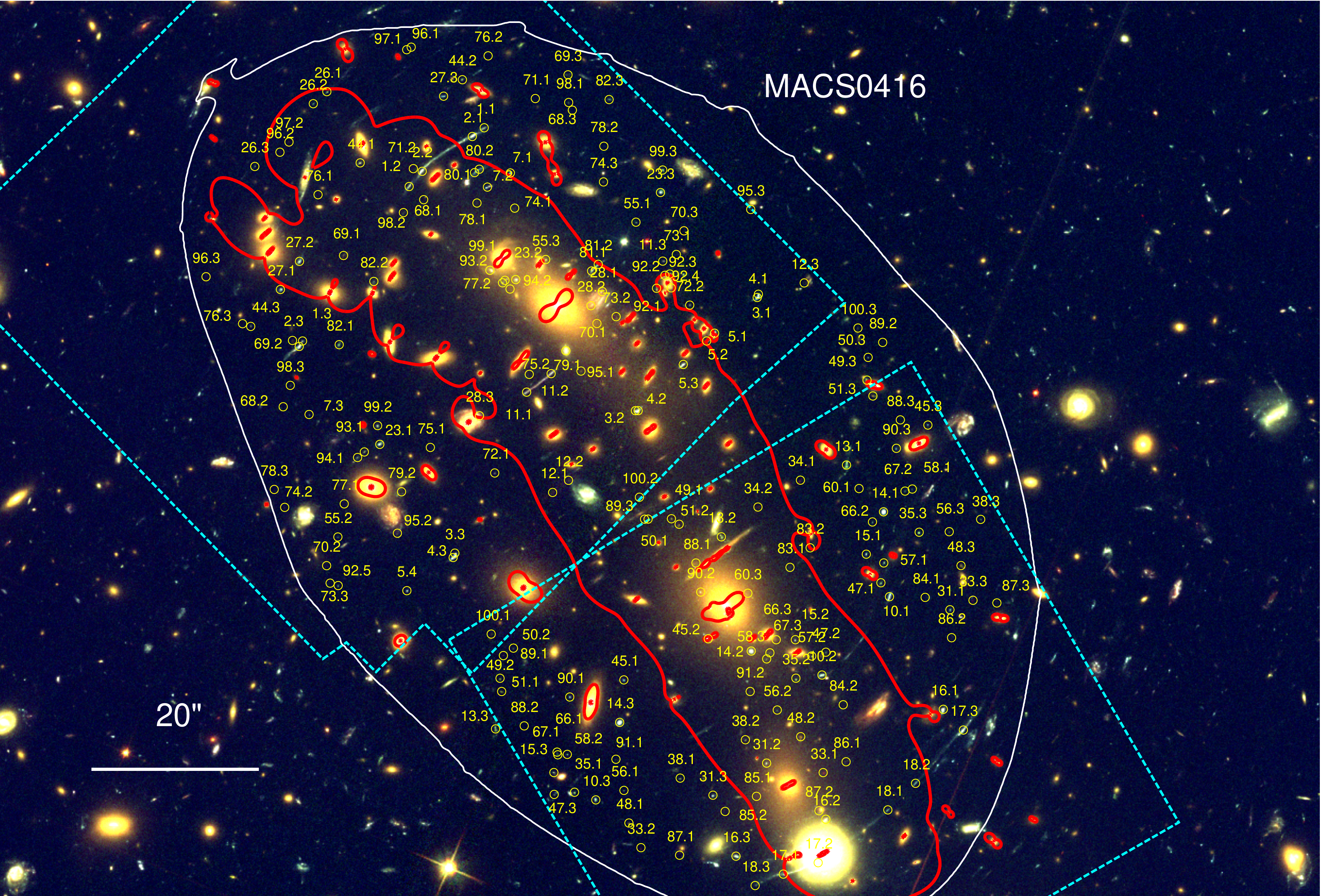}
%\caption{Same as Fig.\,\ref{fig:MACS0257mul} for MACS0416.}
\caption{\textbf{(continued)} Same figure for MACS0416.}
\end{figure*}
\begin{figure*}\ContinuedFloat
\includegraphics[width=14cm]{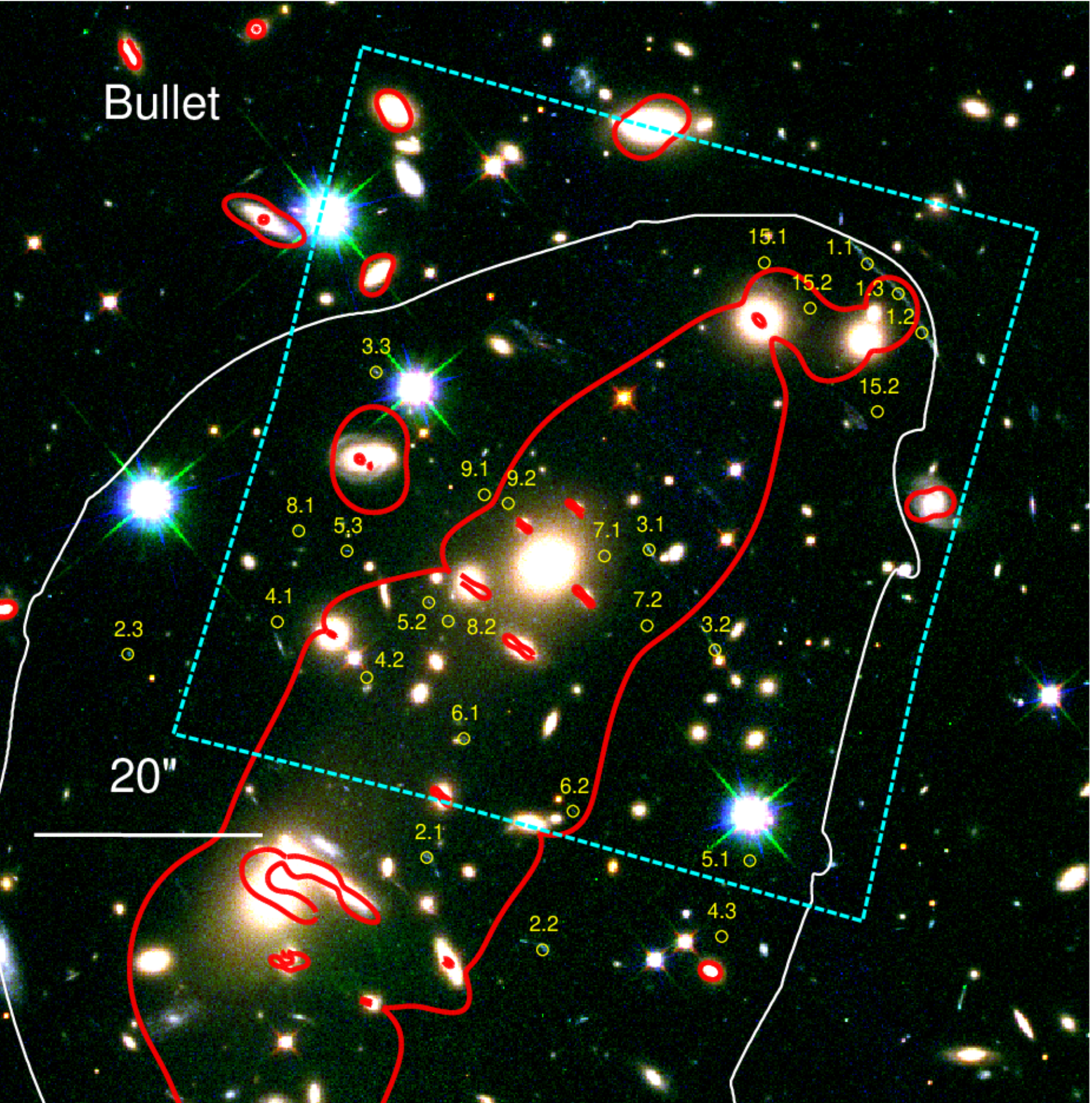}
\caption{Same as Fig.\,\ref{fig:MACS0257mul} for the Bullet Cluster.}
\end{figure*}
\begin{figure*}\ContinuedFloat
\includegraphics[width=14cm]{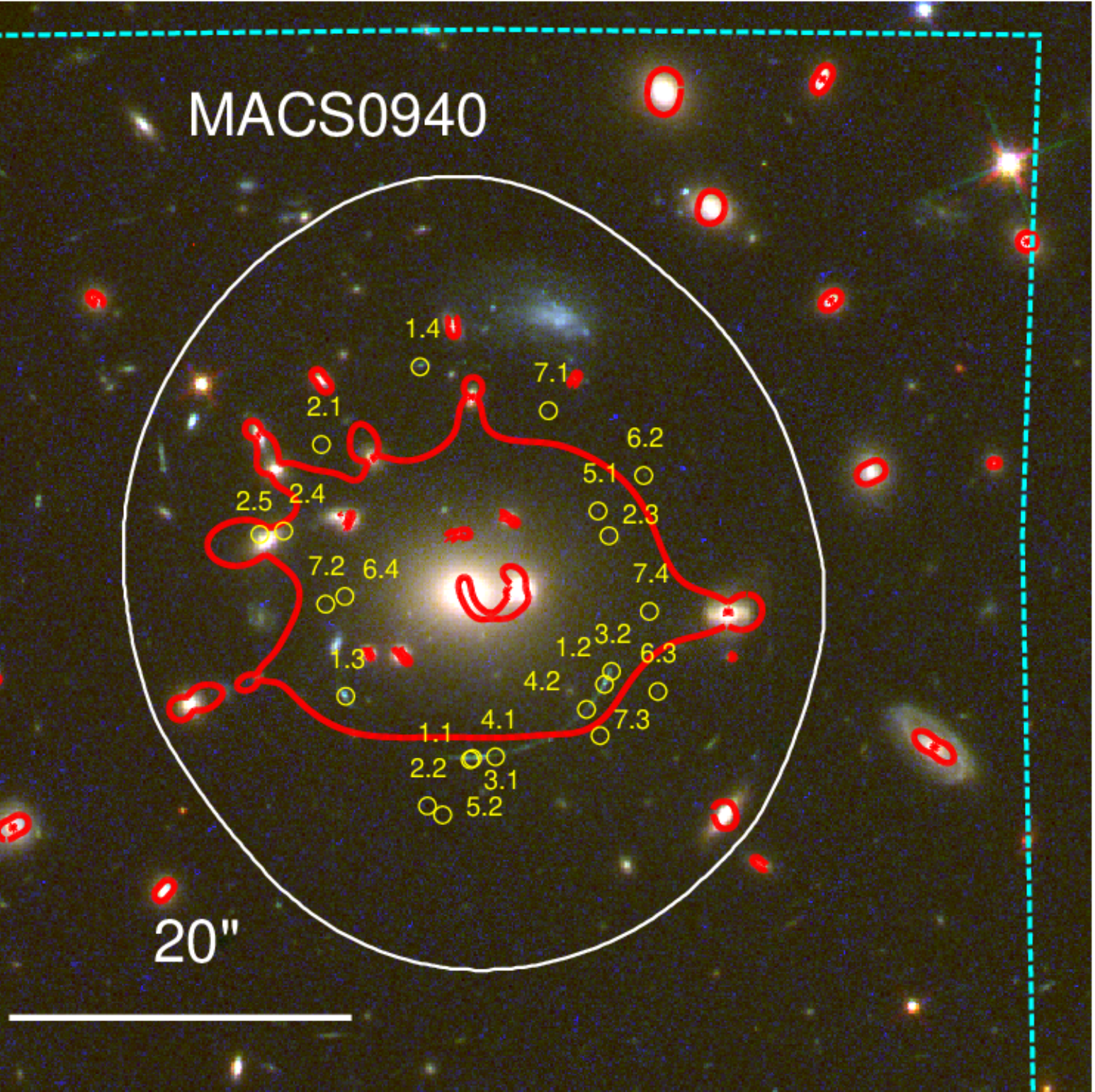}
%\caption{Same as Fig.\,\ref{fig:MACS0257mul} for MACS0940.}
\caption{\textbf{(continued)} Same figure for MACS0940.}
\end{figure*}
\begin{figure*}\ContinuedFloat
\includegraphics[width=14cm]{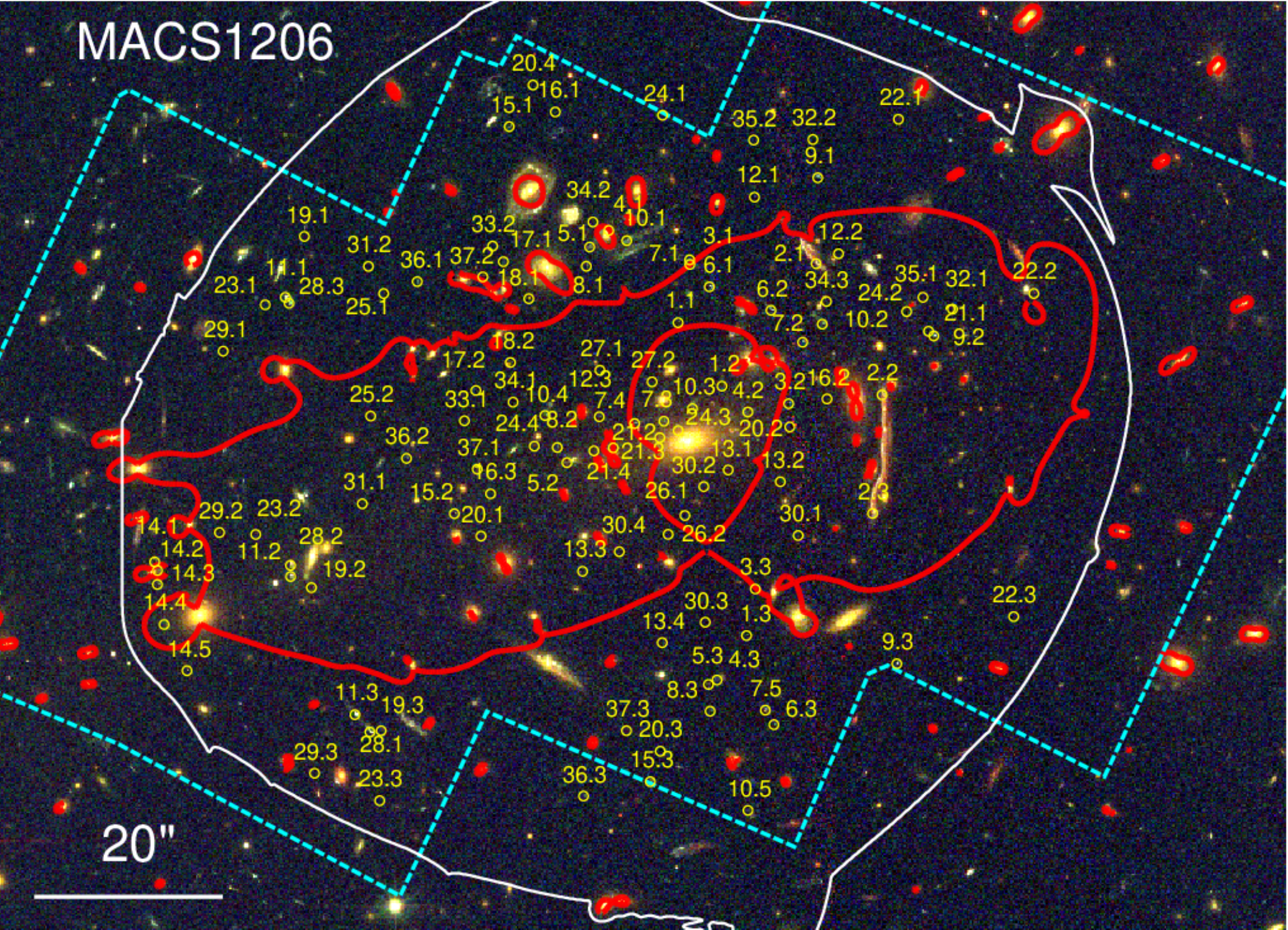}
%\caption{Same as Fig.\,\ref{fig:MACS0257mul} for MACS1206.}
\caption{\textbf{(continued)} Same figure for MACS1206.}
\end{figure*}
\begin{figure*}\ContinuedFloat
\includegraphics[width=14cm]{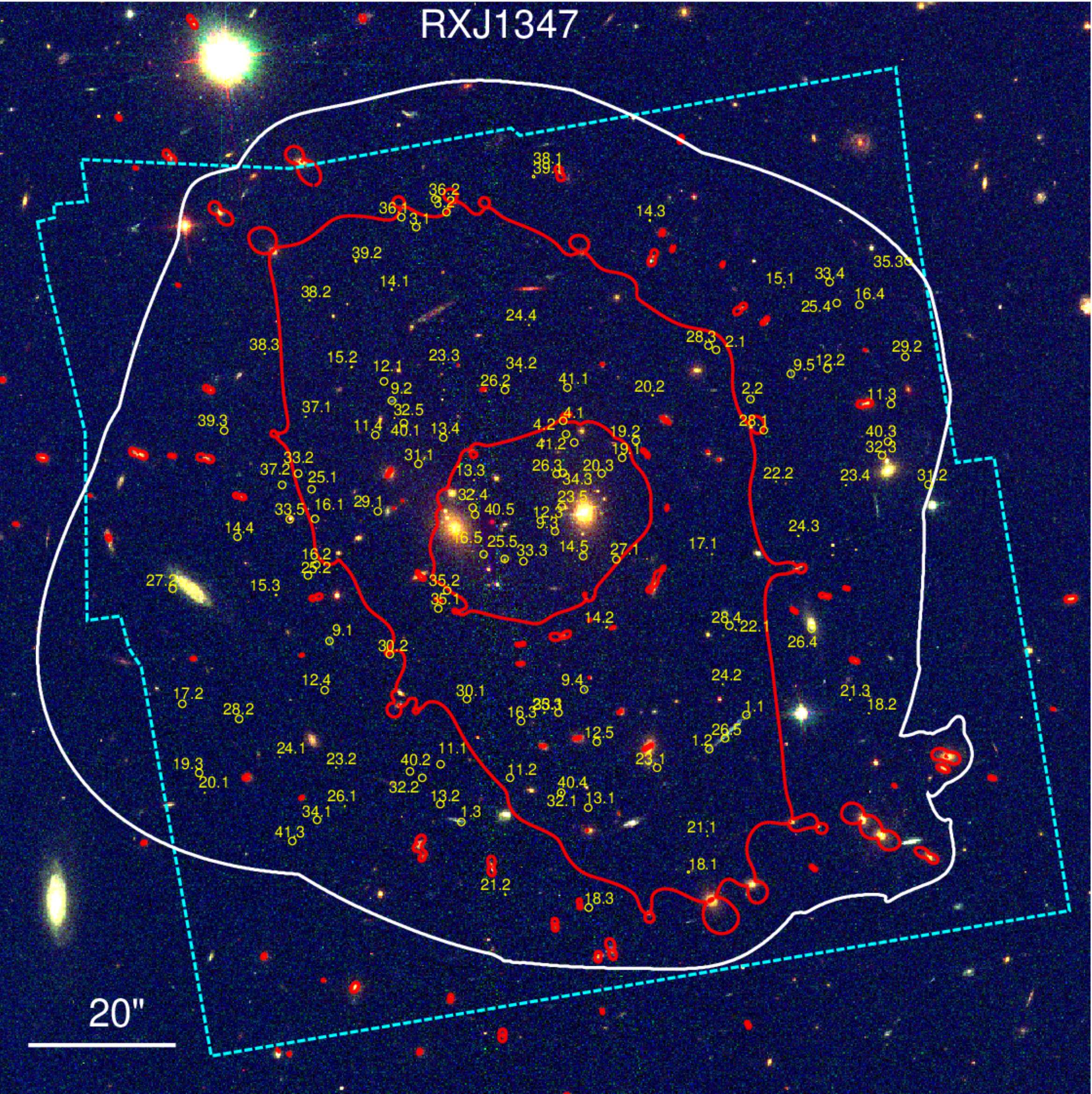}
%\caption{Same as Fig.\,\ref{fig:MACS0257mul} for RXJ1347.}
\caption{\textbf{(continued)} Same figure for RXJ1347.}
\end{figure*}
\begin{figure*}\ContinuedFloat
\includegraphics[width=14cm]{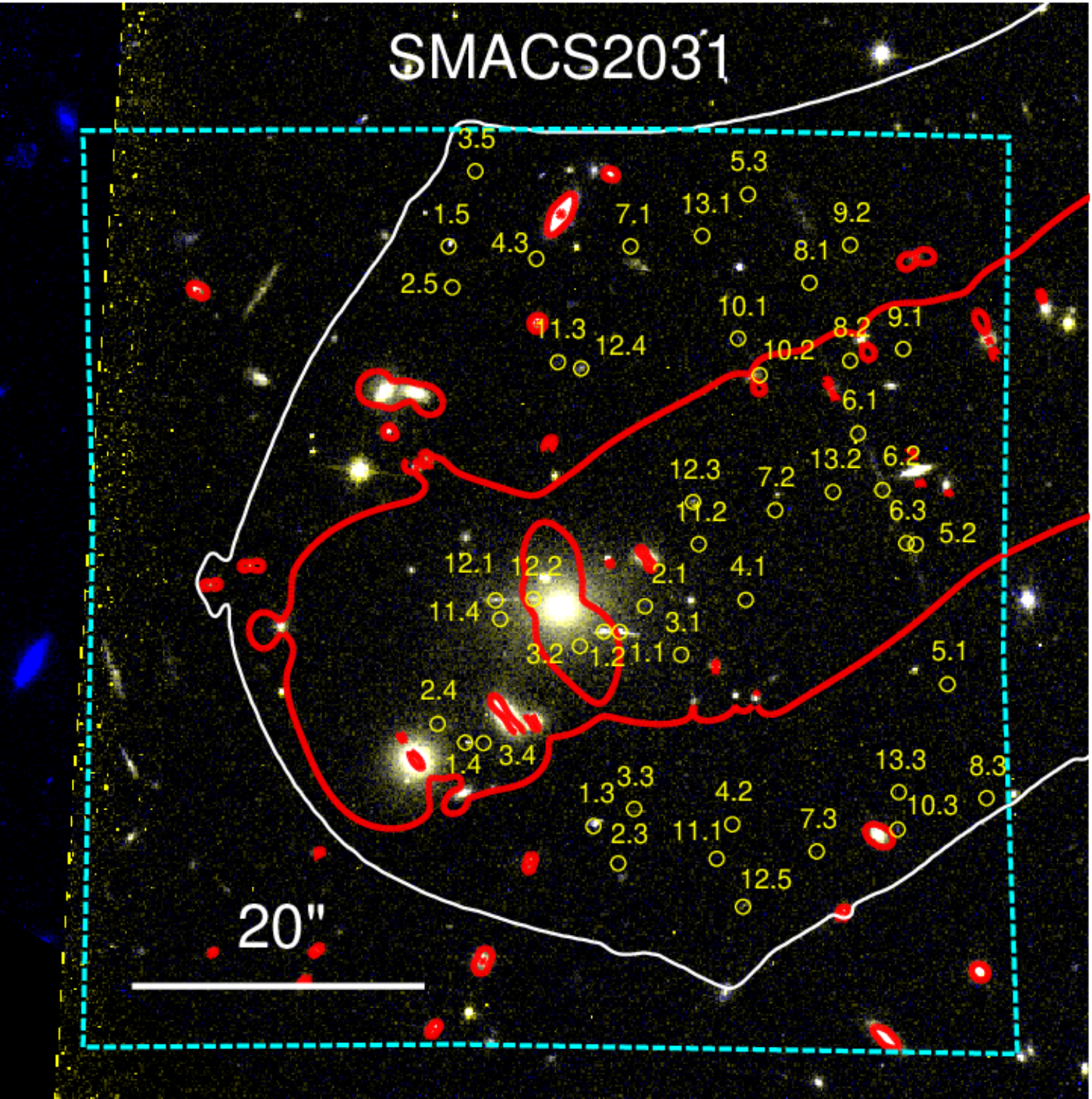}
%\caption{Same as Fig.\,\ref{fig:MACS0257mul} for SMACS2031.}
\caption{\textbf{(continued)} Same figure for SMACS2031.}
\end{figure*}
\begin{figure*}\ContinuedFloat
\includegraphics[width=14cm]{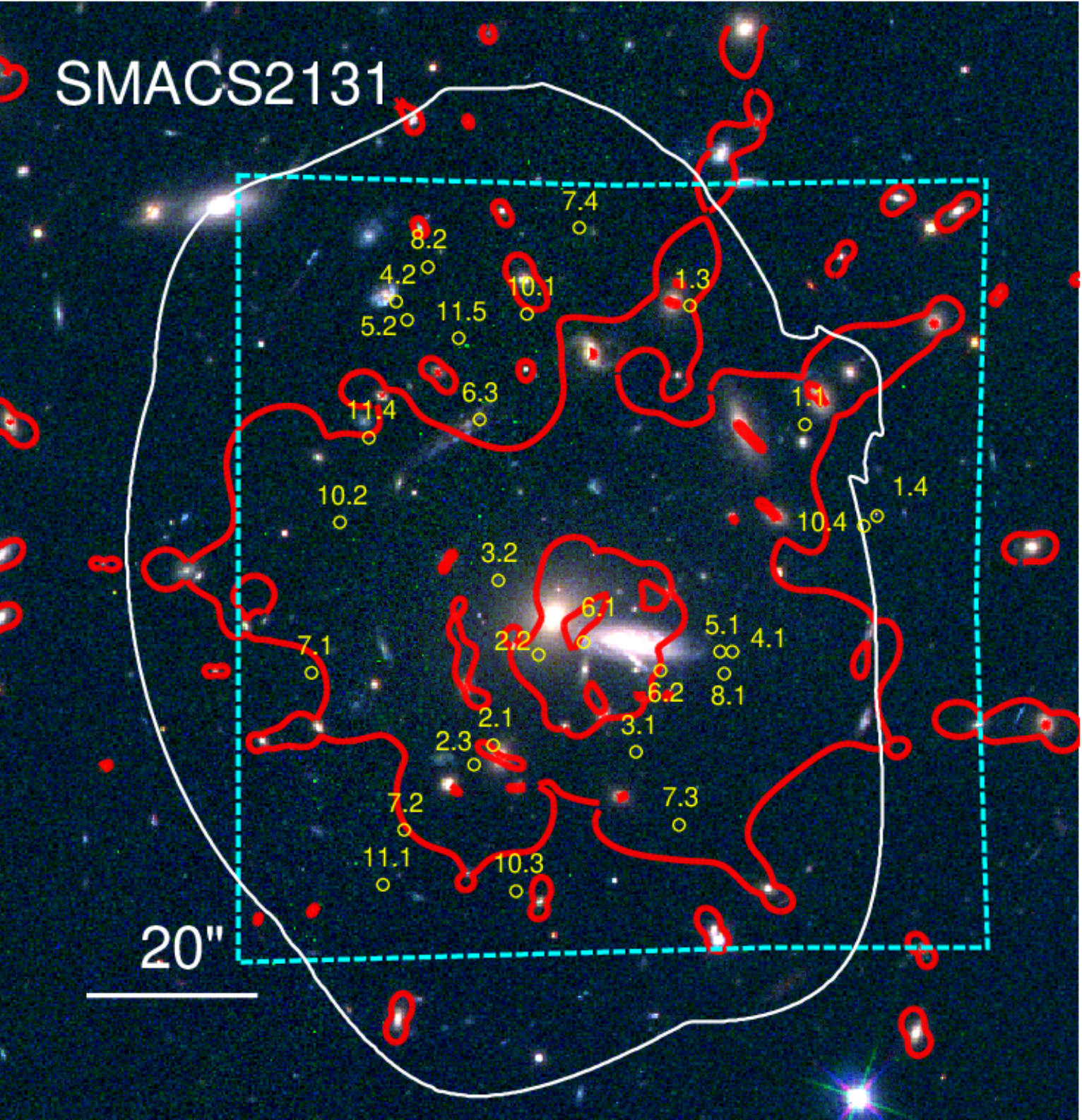}
%\caption{Same as Fig.\,\ref{fig:MACS0257mul} for SMACS2131.}
\caption{\textbf{(continued)} Same figure for SMACS2131.}
\end{figure*}
\begin{figure*}\ContinuedFloat
\includegraphics[width=14cm]{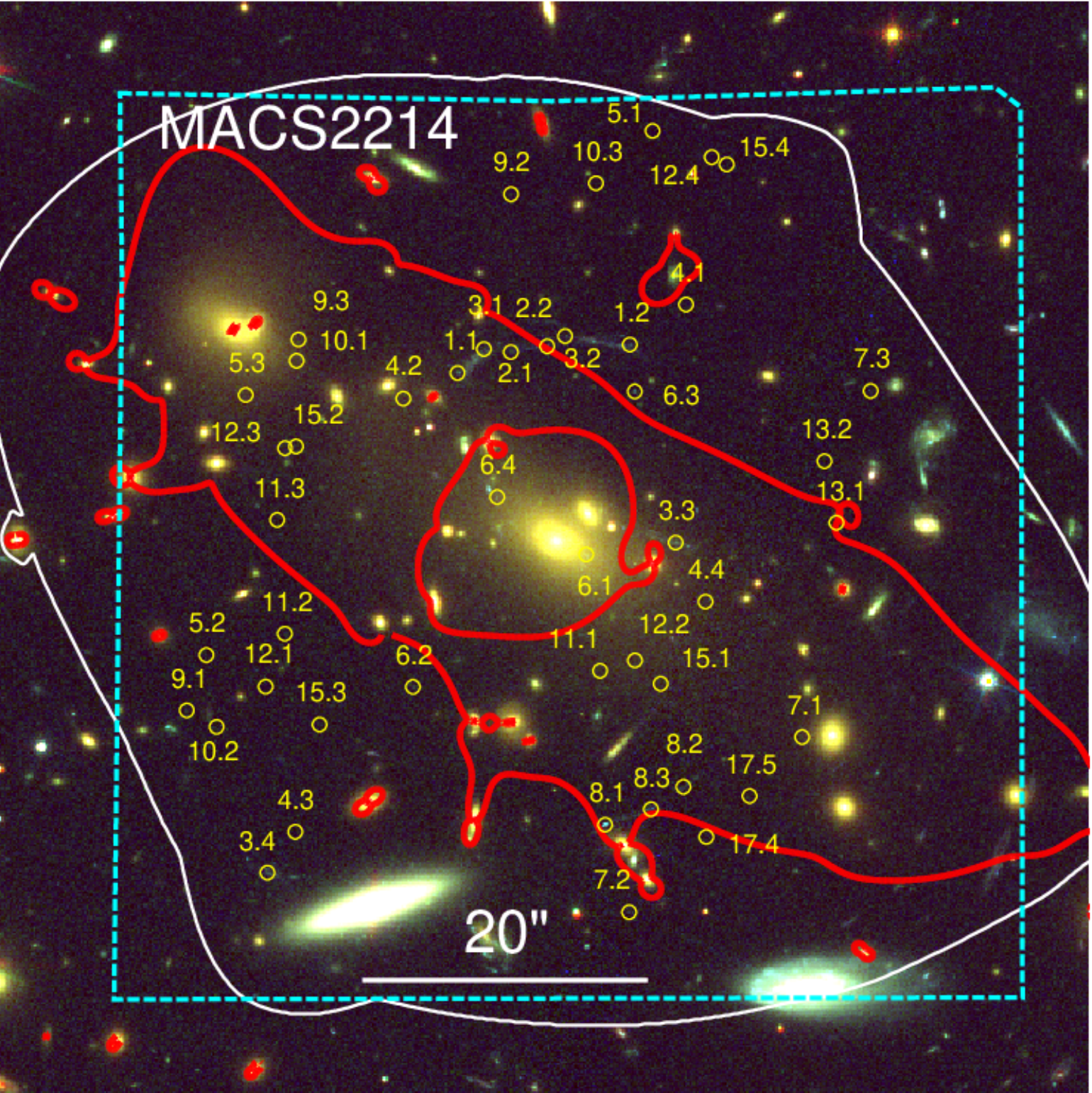}
%\caption{Same as Fig.\,\ref{fig:MACS0257mul} for MACS2214.}
\caption{\textbf{(continued)} Same figure for MACS2214.}
\end{figure*}

\clearpage

\bottomcaption{\label{tab:multiples} List of multiply imaged systems used as constraints in the MACS0257 mass model. From left to right: identification number for the multiple image, sky coordinates, redshift constraints from spectroscopy. Redshifts with error bars are not constrained with spectroscopy and are predictions from our lens model.}
\begin{supertabular}{lllll}
ID & $\alpha$ (J2000) & $\delta$ (J2000) & $z_{\rm system}$ \\
   & h  m  s & $^{\circ}$  $'$  $''$ & \\
\hline 
 1.1 & 02:57:41.84 & $-$22:09:07.78 & 4.69\\
 1.2 & 02:57:41.14 & $-$22:08:55.06 & 4.69\\
 1.3 & 02:57:40.65 & $-$22:09:02.51 & 4.69\\
 1.4 & 02:57:41.28 & $-$22:09:12.14 & 4.69\\
 1.5 & 02:57:40.58 & $-$22:09:53.75 & 4.69\\
 2.1 & 02:57:41.70 & $-$22:08:55.93 & 3.22\\
 2.2 & 02:57:42.22 & $-$22:09:01.45 & 3.22\\
 2.3 & 02:57:42.01 & $-$22:09:03.35 & 3.22\\
 2.4 & 02:57:40.67 & $-$22:09:08.42 & 3.22\\
 2.5 & 02:57:41.02 & $-$22:09:13.12 & 3.22\\
 3.1 & 02:57:41.60 & $-$22:08:54.63 & 3.85\\
 3.2 & 02:57:42.26 & $-$22:09:02.79 & 3.85\\
 3.3 & 02:57:42.04 & $-$22:09:03.80 & 3.85\\
 3.4 & 02:57:40.59 & $-$22:09:06.97 & 3.85\\
 3.5 & 02:57:41.07 & $-$22:09:13.62 & 3.85\\
 4.1 & 02:57:42.20 & $-$22:09:11.27 & 1.1690\\
 4.2 & 02:57:41.66 & $-$22:08:58.31 & 1.1690\\
 4.3 & 02:57:40.68 & $-$22:09:13.19 & 1.1690\\
 4.4 & 02:57:41.35 & $-$22:09:40.82 & 1.1690\\
 5.1 & 02:57:41.93 & $-$22:08:52.85 & 6.609\\
 5.2 & 02:57:41.00 & $-$22:09:13.89 & 6.609\\
 5.3 & 02:57:42.56 & $-$22:09:02.04 & 6.609\\
 5.4 & 02:57:40.51 & $-$22:09:08.11 & 6.609\\
 6.1 & 02:57:40.63 & $-$22:09:27.31 & 2.424\\
 6.2 & 02:57:40.94 & $-$22:09:21.98 & 2.424\\
 6.3 & 02:57:41.70 & $-$22:08:37.86 & 2.424\\
 7.1 & 02:57:39.43 & $-$22:09:25.85 & 4.69\\
 7.2 & 02:57:39.97 & $-$22:09:35.87 & 4.69\\
 8.1 & 02:57:39.12 & $-$22:09:23.93 & 6.080\\
 8.2 & 02:57:39.80 & $-$22:09:38.90 & 6.080\\
 8.3 & 02:57:42.10 & $-$22:09:21.02 & 6.080\\
 9.1 & 02:57:42.54 & $-$22:09:34.09 & 1.157\\
 9.2 & 02:57:42.96 & $-$22:09:16.84 & 1.157\\
 9.3 & 02:57:42.75 & $-$22:09:04.58 & 1.157\\
 10.1 & 02:57:40.08 & $-$22:09:18.37 & 1.169\\
 10.2 & 02:57:41.91 & $-$22:09:23.23 & 1.169\\
 10.3 & 02:57:41.36 & $-$22:08:50.94 & 1.169\\
 11.1 & 02:57:40.00 & $-$22:09:39.21 & 3.01\\
 11.2 & 02:57:41.99 & $-$22:09:19.72 & 3.01\\
 11.3 & 02:57:39.17 & $-$22:09:17.02 & 3.01\\
 12.1 & 02:57:39.43 & $-$22:09:13.13 & 3.807\\
 12.2 & 02:57:40.72 & $-$22:09:44.42 & 3.807\\
 12.3 & 02:57:42.59 & $-$22:09:20.14 & 3.807\\
 13.1 & 02:57:41.42 & $-$22:09:25.43 & 3.9545\\
 13.2 & 02:57:41.77 & $-$22:09:31.50 & 3.9545\\
 14.1 & 02:57:42.33 & $-$22:09:18.32 & 4.030\\
 14.2 & 02:57:40.31 & $-$22:09:44.13 & 4.030\\
 14.3 & 02:57:39.22 & $-$22:09:09.52 & 4.030\\
 15.1 & 02:57:39.72 & $-$22:09:01.01 & 6.467\\
 15.2 & 02:57:42.73 & $-$22:09:12.33 & 6.467\\
 16.1 & 02:57:41.50 & $-$22:09:44.09 & 6.225\\
 16.2 & 02:57:39.60 & $-$22:09:19.20 & 6.225\\
 16.3 & 02:57:42.78 & $-$22:09:25.64 & 6.225\\
 17.1 & 02:57:43.16 & $-$22:09:28.32 & 6.066\\
 17.2 & 02:57:40.19 & $-$22:09:16.17 & 6.066\\
 17.3 & 02:57:42.50 & $-$22:09:43.96 & 6.066\\
 18.1 & 02:57:41.57 & $-$22:09:18.13 & 5.277\\
 18.2 & 02:57:41.72 & $-$22:09:18.32 & 5.277\\
 18.3 & 02:57:39.69 & $-$22:09:40.47 & 5.277\\
 18.4 & 02:57:38.87 & $-$22:09:11.93 & 5.277\\
 19.1 & 02:57:40.08 & $-$22:09:16.65 & 3.223\\
 19.2 & 02:57:42.18 & $-$22:09:41.76 & 3.223\\
 19.3 & 02:57:42.88 & $-$22:09:26.79 & 3.223\\
 20.1 & 02:57:40.25 & $-$22:09:11.22 & 3.246\\
 20.2 & 02:57:42.94 & $-$22:09:11.99 & 3.246\\
 20.3 & 02:57:41.91 & $-$22:09:48.63 & 3.246\\
 20.4 & 02:57:41.99 & $-$22:08:49.57 & 3.246\\
 21.1 & 02:57:41.54 & $-$22:08:50.35 & 4.685\\
 21.2 & 02:57:40.01 & $-$22:08:58.30 & 4.685\\
 21.3 & 02:57:42.54 & $-$22:09:09.24 & 4.685\\
 22.1 & 02:57:40.42 & $-$22:09:10.47 & 5.32295\\
 22.2 & 02:57:42.90 & $-$22:09:06.61 & 5.32295\\
 22.3 & 02:57:42.14 & $-$22:08:51.40 & 5.32295\\
 23.1 & 02:57:41.33 & $-$22:09:27.77 & 5.557\\
 23.2 & 02:57:41.41 & $-$22:09:32.43 & 5.557\\
 24.1 & 02:57:39.17 & $-$22:09:16.02 & 4.69\\
 24.2 & 02:57:42.37 & $-$22:09:20.51 & 4.69\\
 24.3 & 02:57:40.25 & $-$22:09:42.65 & 4.69\\
 24.4 & 02:57:41.07 & $-$22:08:42.03 & 4.69\\
 25.1 & 02:57:42.07 & $-$22:09:26.07 & 4.4633\\
 25.2 & 02:57:41.26 & $-$22:09:20.27 & 4.4633\\
\hline
\end{supertabular}

\bottomcaption{Same as Table\,\ref{tab:multiples} for the MACS0329 mass model.}
\begin{supertabular}{lllll}
ID & $\alpha$ (J2000) & $\delta$ (J2000) & $z_{\rm system}$ \\
   & h  m  s & $^{\circ}$  $'$  $''$ & \\
\hline 
 1.1 & 03:29:40.17 & $-$02:11:45.71 & 6.18\\
 1.2 & 03:29:40.07 & $-$02:11:51.71 & 6.18\\
 1.3 & 03:29:41.24 & $-$02:12:04.66 & 6.18\\
 1.4 & 03:29:43.16 & $-$02:11:17.36 & 6.18\\
 2.1 & 03:29:41.06 & $-$02:11:28.82 & 2.14\\
 2.2 & 03:29:39.62 & $-$02:12:00.66 & 2.14\\
 2.3 & 03:29:42.17 & $-$02:11:25.61 & 2.14\\
 2.4 & 03:29:42.33 & $-$02:11:54.46 & 2.14\\
 3.1 & 03:29:40.18 & $-$02:11:26.56 & 2.42$\pm$0.08\\
 3.2 & 03:29:39.04 & $-$02:11:49.43 & ''\\
 3.3 & 03:29:41.26 & $-$02:11:15.16 & ''\\
 4.1 & 03:29:42.38 & $-$02:11:37.64 & 3.795\\
 4.2 & 03:29:41.80 & $-$02:11:42.15 & 3.795\\
 5.1 & 03:29:41.90 & $-$02:11:47.15 & 4.575\\
 5.2 & 03:29:42.28 & $-$02:11:48.93 & 4.575\\
 6.1 & 03:29:42.14 & $-$02:12:29.85 & 5.661\\
 6.2 & 03:29:42.65 & $-$02:12:26.15 & 5.661\\
 6.3 & 03:29:42.52 & $-$02:12:27.80 & 5.661\\
 7.1 & 03:29:42.20 & $-$02:11:23.44 & 6.0112\\
 7.2 & 03:29:42.74 & $-$02:11:47.14 & 6.0112\\
 8.1 & 03:29:42.11 & $-$02:11:51.07 & 3.8546\\
 8.2 & 03:29:41.91 & $-$02:11:49.19 & 3.8546\\
 9.1 & 03:29:43.48 & $-$02:11:28.86 & 6.0261\\
 9.2 & 03:29:40.85 & $-$02:11:40.35 & 6.0261\\
\hline
\end{supertabular}

\bottomcaption{Same as Table\,\ref{tab:multiples} for the MACS0416 mass model.}
\begin{supertabular}{lllll}
ID & $\alpha$ (J2000) & $\delta$ (J2000) & $z_{\rm system}$\\
   & h  m  s & $^{\circ}$  $'$  $''$ &\\
\hline 
 1.1 & 04:16:09.78 & $-$24:03:41.73 & 1.8960\\
 1.2 & 04:16:10.43 & $-$24:03:48.75 & 1.8960\\
 1.3 & 04:16:11.36 & $-$24:04:07.21 & 1.8960\\
 2.1 & 04:16:09.88 & $-$24:03:42.77 & 1.8925\\
 2.2 & 04:16:10.32 & $-$24:03:46.93 & 1.8925\\
 2.3 & 04:16:11.39 & $-$24:04:07.86 & 1.8925\\
 3.1 & 04:16:07.39 & $-$24:04:01.62 & 1.9885\\
 3.2 & 04:16:08.46 & $-$24:04:15.53 & 1.9885\\
 3.3 & 04:16:10.04 & $-$24:04:32.56 & 1.9885\\
 4.1 & 04:16:07.40 & $-$24:04:02.01 & 1.9885\\
 4.2 & 04:16:08.44 & $-$24:04:15.53 & 1.9885\\
 4.3 & 04:16:10.05 & $-$24:04:33.08 & 1.9885\\
 5.1 & 04:16:07.77 & $-$24:04:06.28 & 2.0970\\
 5.2 & 04:16:07.84 & $-$24:04:07.21 & 2.0970\\
 5.3 & 04:16:08.04 & $-$24:04:10.01 & 2.0970\\
 5.4 & 04:16:10.45 & $-$24:04:37.04 & 2.0970\\
 7.1 & 04:16:09.55 & $-$24:03:47.13 & 2.0854\\
 7.2 & 04:16:09.75 & $-$24:03:48.82 & 2.0854\\
 7.3 & 04:16:11.31 & $-$24:04:15.99 & 2.0854\\
 10.1 & 04:16:06.24 & $-$24:04:37.76 & 2.2982\\
 10.2 & 04:16:06.83 & $-$24:04:47.12 & 2.2982\\
 10.3 & 04:16:08.81 & $-$24:05:02.04 & 2.2982\\
 11.1 & 04:16:09.41 & $-$24:04:13.32 & 1.0060\\
 11.2 & 04:16:09.20 & $-$24:04:11.11 & 1.0060\\
 11.3 & 04:16:08.22 & $-$24:03:57.66 & 1.0060\\
 12.1 & 04:16:09.18 & $-$24:04:25.31 & 1.9530\\
 12.2 & 04:16:09.04 & $-$24:04:23.86 & 1.9530\\
 12.3 & 04:16:06.99 & $-$24:04:00.27 & 1.9530\\
 13.1 & 04:16:06.62 & $-$24:04:22.03 & 3.2226\\
 13.2 & 04:16:07.71 & $-$24:04:30.61 & 3.2226\\
 13.3 & 04:16:09.68 & $-$24:04:53.56 & 3.2226\\
 14.1 & 04:16:06.30 & $-$24:04:27.62 & 1.6350\\
 14.2 & 04:16:07.45 & $-$24:04:44.26 & 1.6350\\
 14.3 & 04:16:08.60 & $-$24:04:52.78 & 1.6350\\
 15.1 & 04:16:06.45 & $-$24:04:32.68 & 2.3340\\
 15.2 & 04:16:07.07 & $-$24:04:42.90 & 2.3340\\
 15.3 & 04:16:09.17 & $-$24:04:58.77 & 2.3340\\
 16.1 & 04:16:05.77 & $-$24:04:51.22 & 1.9644\\
 16.2 & 04:16:06.80 & $-$24:05:04.35 & 1.9644\\
 16.3 & 04:16:07.58 & $-$24:05:08.77 & 1.9644\\
 17.1 & 04:16:07.17 & $-$24:05:10.91 & 2.2181\\
 17.2 & 04:16:06.87 & $-$24:05:09.55 & 2.2181\\
 17.3 & 04:16:05.60 & $-$24:04:53.69 & 2.2181\\
 18.1 & 04:16:06.26 & $-$24:05:03.24 & 2.2210\\
 18.2 & 04:16:06.02 & $-$24:05:00.06 & 2.2210\\
 18.3 & 04:16:07.42 & $-$24:05:12.28 & 2.2210\\
 23.1 & 04:16:10.69 & $-$24:04:19.56 & 2.0960\\
 23.2 & 04:16:09.50 & $-$24:03:59.87 & 2.0960\\
 23.3 & 04:16:08.24 & $-$24:03:49.47 & 2.0960\\
 26.1 & 04:16:11.15 & $-$24:03:37.41 & 3.2380\\
 26.2 & 04:16:11.27 & $-$24:03:38.85 & 3.2380\\
 26.3 & 04:16:11.78 & $-$24:03:46.35 & 3.2380\\
 27.1 & 04:16:11.56 & $-$24:04:01.05 & 2.0960\\
 27.2 & 04:16:11.39 & $-$24:03:57.69 & 2.0960\\
 27.3 & 04:16:10.13 & $-$24:03:37.95 & 2.0960\\
 28.1 & 04:16:08.75 & $-$24:04:01.29 & 0.9394\\
 28.2 & 04:16:08.85 & $-$24:04:02.99 & 0.9394\\
 28.3 & 04:16:09.82 & $-$24:04:16.14 & 0.9394\\
 31.1 & 04:16:05.72 & $-$24:04:39.32 & 1.8178\\
 31.2 & 04:16:07.32 & $-$24:04:57.65 & 1.8178\\
 31.3 & 04:16:07.78 & $-$24:05:01.48 & 1.8178\\
 33.1 & 04:16:06.82 & $-$24:04:58.78 & 5.3651\\
 33.2 & 04:16:08.41 & $-$24:05:07.75 & 5.3651\\
 33.3 & 04:16:05.52 & $-$24:04:38.19 & 5.3651\\
 34.1 & 04:16:07.02 & $-$24:04:23.84 & 5.1060\\
 34.2 & 04:16:07.39 & $-$24:04:27.05 & 5.1060\\
 35.1 & 04:16:08.99 & $-$24:05:01.13 & 3.4900\\
 35.2 & 04:16:07.06 & $-$24:04:47.50 & 3.4900\\
 35.3 & 04:16:05.98 & $-$24:04:30.06 & 3.4900\\
 38.1 & 04:16:08.07 & $-$24:04:59.44 & 3.4400\\
 38.2 & 04:16:07.50 & $-$24:04:54.86 & 3.4400\\
 38.3 & 04:16:05.45 & $-$24:04:28.52 & 3.4400\\
 44.1 & 04:16:10.86 & $-$24:03:45.93 & 3.2905\\
 44.2 & 04:16:09.97 & $-$24:03:35.99 & 3.2905\\
 44.3 & 04:16:11.82 & $-$24:04:05.47 & 3.2905\\
 45.1 & 04:16:08.56 & $-$24:04:47.70 & 2.5420\\
 45.2 & 04:16:07.83 & $-$24:04:42.79 & 2.5420\\
 45.3 & 04:16:05.91 & $-$24:04:17.25 & 2.5420\\
 47.1 & 04:16:06.32 & $-$24:04:36.10 & 3.2510\\
 47.2 & 04:16:06.80 & $-$24:04:44.40 & 3.2510\\
 47.3 & 04:16:09.17 & $-$24:05:01.39 & 3.2510\\
 48.1 & 04:16:08.52 & $-$24:05:04.80 & 4.1210\\
 48.2 & 04:16:07.02 & $-$24:04:54.49 & 4.1210\\
 48.3 & 04:16:05.62 & $-$24:04:34.04 & 4.1210\\
 49.1 & 04:16:08.15 & $-$24:04:28.45 & 3.8696\\
 49.2 & 04:16:09.64 & $-$24:04:47.51 & 3.8696\\
 49.3 & 04:16:06.44 & $-$24:04:11.88 & 3.8696\\
 50.1 & 04:16:08.35 & $-$24:04:28.49 & 3.2195\\
 50.2 & 04:16:09.53 & $-$24:04:43.91 & 3.2195\\
 50.3 & 04:16:06.43 & $-$24:04:09.17 & 3.2195\\
 51.1 & 04:16:09.63 & $-$24:04:49.09 & 4.1020\\
 51.2 & 04:16:08.08 & $-$24:04:29.11 & 4.1020\\
 51.3 & 04:16:06.39 & $-$24:04:13.78 & 4.1020\\
 55.1 & 04:16:08.46 & $-$24:03:53.01 & 3.2910\\
 55.2 & 04:16:11.06 & $-$24:04:30.63 & 3.2910\\
 55.3 & 04:16:09.24 & $-$24:03:57.47 & 3.2910\\
 56.1 & 04:16:08.56 & $-$24:05:00.91 & 3.2910\\
 56.2 & 04:16:07.22 & $-$24:04:51.30 & 3.2910\\
 56.3 & 04:16:05.72 & $-$24:04:29.99 & 3.2910\\
 57.1 & 04:16:06.29 & $-$24:04:33.73 & 2.9259\\
 57.2 & 04:16:06.92 & $-$24:04:44.85 & 2.9259\\
 58.1 & 04:16:06.04 & $-$24:04:24.90 & 3.0802\\
 58.2 & 04:16:09.06 & $-$24:04:56.60 & 3.0802\\
 58.3 & 04:16:07.32 & $-$24:04:45.19 & 3.0802\\
 60.1 & 04:16:06.51 & $-$24:04:24.84 & 3.9230\\
 60.3 & 04:16:07.48 & $-$24:04:37.35 & 3.9230\\
 66.1 & 04:16:09.14 & $-$24:04:56.33 & 2.2300\\
 66.2 & 04:16:06.39 & $-$24:04:28.83 & 2.2300\\
 66.3 & 04:16:07.23 & $-$24:04:42.91 & 2.2300\\
 67.1 & 04:16:09.14 & $-$24:04:56.65 & 3.1094\\
 67.2 & 04:16:06.11 & $-$24:04:25.14 & 3.1094\\
 67.3 & 04:16:07.29 & $-$24:04:44.47 & 3.1094\\
 68.1 & 04:16:10.31 & $-$24:03:50.32 & 4.1138\\
 68.2 & 04:16:11.53 & $-$24:04:15.07 & 4.1138\\
 68.3 & 04:16:09.01 & $-$24:03:39.63 & 4.1138\\
 69.1 & 04:16:11.01 & $-$24:03:56.99 & 4.1150\\
 69.2 & 04:16:11.45 & $-$24:04:07.15 & 4.1150\\
 69.3 & 04:16:09.05 & $-$24:03:35.41 & 4.1150\\
 70.1 & 04:16:08.82 & $-$24:04:04.77 & 4.609\\
 70.2 & 04:16:11.16 & $-$24:04:34.05 & 4.609\\
 70.3 & 04:16:08.04 & $-$24:03:54.04 & 4.609\\
 71.1 & 04:16:09.33 & $-$24:03:38.24 & 6.147\\
 71.2 & 04:16:10.40 & $-$24:03:46.61 & 6.147\\
 72.1 & 04:16:09.69 & $-$24:04:22.98 & 1.14717\\
 72.2 & 04:16:07.99 & $-$24:04:02.92 & 1.14717\\
 73.1 & 04:16:08.10 & $-$24:03:56.87 & 4.07074\\
 73.2 & 04:16:08.63 & $-$24:04:04.30 & 4.07074\\
 73.3 & 04:16:11.12 & $-$24:04:36.14 & 4.07074\\
 74.1 & 04:16:09.51 & $-$24:03:51.34 & 4.30\\
 74.2 & 04:16:11.52 & $-$24:04:27.09 & 4.30\\
 74.3 & 04:16:08.74 & $-$24:03:48.20 & 4.30\\
 75.1 & 04:16:10.25 & $-$24:04:19.93 & 4.30\\
 75.2 & 04:16:09.39 & $-$24:04:11.19 & 4.30\\
 76.1 & 04:16:11.23 & $-$24:03:49.73 & 5.0996\\
 76.2 & 04:16:09.75 & $-$24:03:33.14 & 5.0996\\
 76.3 & 04:16:11.89 & $-$24:04:05.14 & 5.0996\\
 77.1 & 04:16:11.00 & $-$24:04:26.67 & 5.998\\
 77.2 & 04:16:09.55 & $-$24:04:01.00 & 5.998\\
 78.1 & 04:16:09.84 & $-$24:03:50.72 & 6.06645\\
 78.2 & 04:16:08.74 & $-$24:03:43.93 & 6.06645\\
 78.3 & 04:16:11.61 & $-$24:04:24.95 & 6.06645\\
 79.1 & 04:16:09.07 & $-$24:04:08.38 & 2.99113\\
 79.2 & 04:16:10.50 & $-$24:04:25.23 & 2.99113\\
 80.1 & 04:16:09.86 & $-$24:03:47.09 & 2.243\\
 80.2 & 04:16:09.82 & $-$24:03:46.68 & 2.243\\
 81.1 & 04:16:08.84 & $-$24:03:58.84 & 1.827\\
 81.2 & 04:16:08.78 & $-$24:03:58.05 & 1.827\\
 82.1 & 04:16:11.05 & $-$24:04:07.65 & 2.922\\
 82.2 & 04:16:10.74 & $-$24:04:00.09 & 2.922\\
 82.3 & 04:16:08.69 & $-$24:03:38.36 & 2.922\\
 83.1 & 04:16:07.11 & $-$24:04:34.26 & 3.0750\\
 83.2 & 04:16:06.94 & $-$24:04:31.92 & 3.0750\\
 84.1 & 04:16:05.93 & $-$24:04:37.84 & 4.5300\\
 84.2 & 04:16:06.65 & $-$24:04:50.67 & 4.5300\\
 85.1 & 04:16:07.40 & $-$24:05:01.62 & 5.9730\\
 85.2 & 04:16:07.68 & $-$24:05:03.42 & 5.9730\\
 86.1 & 04:16:06.62 & $-$24:04:57.49 & 3.9230\\
 86.2 & 04:16:05.70 & $-$24:04:42.66 & 3.9230\\
 87.1 & 04:16:08.08 & $-$24:05:08.65 & 5.6380\\
 87.2 & 04:16:06.86 & $-$24:05:03.34 & 5.6380\\
 87.3 & 04:16:05.31 & $-$24:04:38.51 & 5.6380\\
 88.1 & 04:16:07.93 & $-$24:04:33.73 & 4.5018\\
 88.2 & 04:16:09.43 & $-$24:04:53.20 & 4.5018\\
 88.3 & 04:16:06.15 & $-$24:04:16.63 & 4.5018\\
 89.1 & 04:16:09.61 & $-$24:04:44.80 & 5.1060\\
 89.2 & 04:16:06.30 & $-$24:04:07.37 & 5.1060\\
 89.3 & 04:16:08.38 & $-$24:04:28.50 & 5.1060\\
 90.1 & 04:16:09.03 & $-$24:04:49.75 & 2.2800\\
 90.2 & 04:16:07.89 & $-$24:04:37.20 & 2.2800\\
 90.3 & 04:16:06.18 & $-$24:04:20.03 & 2.2800\\
 91.1 & 04:16:08.63 & $-$24:04:57.19 & 3.7153\\
 91.2 & 04:16:07.46 & $-$24:04:49.09 & 3.7153\\
 92.1 & 04:16:08.28 & $-$24:04:00.91 & 3.2240\\
 92.2 & 04:16:08.20 & $-$24:03:59.36 & 3.2240\\
 92.3 & 04:16:08.16 & $-$24:03:59.21 & 3.2240\\
 92.4 & 04:16:08.15 & $-$24:04:00.86 & 3.2240\\
 92.5 & 04:16:11.05 & $-$24:04:36.43 & 3.2240\\
 93.1 & 04:16:10.83 & $-$24:04:20.47 & 3.2883\\
 93.2 & 04:16:09.62 & $-$24:04:00.26 & 3.2883\\
 94.1 & 04:16:10.88 & $-$24:04:21.13 & 3.2883\\
 94.2 & 04:16:09.60 & $-$24:03:59.96 & 3.2883\\
 95.1 & 04:16:08.94 & $-$24:04:10.80 & 4.0690\\
 95.2 & 04:16:10.54 & $-$24:04:30.18 & 4.0690\\
 95.3 & 04:16:07.45 & $-$24:03:51.49 & 4.0690\\
 96.1 & 04:16:10.42 & $-$24:03:32.10 & 6.1480\\
 96.2 & 04:16:11.56 & $-$24:03:44.65 & 6.1480\\
 96.3 & 04:16:12.21 & $-$24:03:59.53 & 6.1480\\
 97.1 & 04:16:10.46 & $-$24:03:32.40 & 6.1480\\
 97.2 & 04:16:11.48 & $-$24:03:43.43 & 6.1480\\
 98.1 & 04:16:09.04 & $-$24:03:38.71 & 3.6100\\
 98.2 & 04:16:10.48 & $-$24:03:51.87 & 3.6100\\
 98.3 & 04:16:11.47 & $-$24:04:12.51 & 3.6100\\
 99.1 & 04:16:09.73 & $-$24:03:58.76 & 2.281\\
 99.2 & 04:16:10.71 & $-$24:04:17.32 & 2.281\\
 99.3 & 04:16:08.22 & $-$24:03:46.78 & 2.281\\
 100.1 & 04:16:09.72 & $-$24:04:42.23 & 3.9680\\
 100.2 & 04:16:08.43 & $-$24:04:25.88 & 3.9680\\
 100.3 & 04:16:06.52 & $-$24:04:05.66 & 3.9680\\
\hline
\end{supertabular}

\bottomcaption{Same as \ref{tab:multiples} for the Bullet cluster mass model.}
\begin{supertabular}{lllll}
ID & $\alpha$ (J2000) & $\delta$ (J2000) & $z_{\rm system}$ \\
   & h  m  s & $^{\circ}$  $'$  $''$ & \\
\hline 
 1.1 & 06:58:31.90 & $-$55:56:30.21 & 3.24\\
 1.2 & 06:58:31.33 & $-$55:56:36.26 & 3.24\\
 1.3 & 06:58:31.58 & $-$55:56:32.83 & 3.24\\
 2.1 & 06:58:36.50 & $-$55:57:22.48 & 2.35$\pm$0.06\\
 2.2 & 06:58:35.29 & $-$55:57:30.59 & ''\\
 2.3 & 06:58:39.64 & $-$55:57:04.64 & ''\\
 3.1 & 06:58:34.18 & $-$55:56:55.39 & 3.2541\\
 3.2 & 06:58:33.49 & $-$55:57:04.19 & 3.2541\\
 3.3 & 06:58:37.04 & $-$55:56:39.81 & 3.2541\\
 4.1 & 06:58:38.07 & $-$55:57:01.79 & 2.8\\
 4.2 & 06:58:37.14 & $-$55:57:06.66 & 2.8\\
 4.3 & 06:58:33.42 & $-$55:57:29.39 & 2.8\\
 5.1 & 06:58:33.12 & $-$55:57:22.72 & 2.92$\pm$0.06\\
 5.2 & 06:58:36.49 & $-$55:57:00.05 & ''\\
 5.3 & 06:58:37.34 & $-$55:56:55.54 & ''\\
 6.1 & 06:58:36.12 & $-$55:57:12.02 & 1.24$\pm$0.02\\
 6.2 & 06:58:34.97 & $-$55:57:18.39 & ''\\
 7.1 & 06:58:34.65 & $-$55:56:55.95 & 1.67$\pm$0.04\\
 7.2 & 06:58:34.20 & $-$55:57:02.07 & ''\\
 8.1 & 06:58:37.85 & $-$55:56:53.78 & 3.34$\pm$0.25\\
 8.2 & 06:58:36.28 & $-$55:57:01.70 & ''\\
 9.1 & 06:58:35.91 & $-$55:56:50.58 & 5.27$\pm$0.33\\
 9.2 & 06:58:35.66 & $-$55:56:51.33 & ''\\
 10.1 & 06:58:15.14 & $-$55:56:22.60 & 2.99\\
 10.2 & 06:58:14.73 & $-$55:56:33.50 & 2.99\\
 10.3 & 06:58:14.87 & $-$55:56:51.45 & 2.99\\
 11.1 & 06:58:14.86 & $-$55:56:46.00 & 3.13$\pm$0.12\\
 11.2 & 06:58:14.78 & $-$55:56:39.25 & ''\\
 11.3 & 06:58:15.55 & $-$55:56:17.25 & ''\\
 12.1 & 06:58:16.89 & $-$55:56:38.55 & 1.69$\pm$0.06\\
 12.2 & 06:58:16.62 & $-$55:56:45.50 & ''\\
 12.3 & 06:58:16.78 & $-$55:56:19.58 & ''\\
 13.1 & 06:58:15.76 & $-$55:56:23.40 & 3.45$\pm$0.33\\
 13.2 & 06:58:15.41 & $-$55:56:31.37 & ''\\
 14.1 & 06:58:15.38 & $-$55:56:43.51 & 2.05$\pm$0.08\\
 14.2 & 06:58:15.32 & $-$55:56:41.56 & ''\\
 14.3 & 06:58:15.94 & $-$55:56:17.41 & ''\\
 15.1 & 06:58:32.98 & $-$55:56:30.13 & 3.537\\
 15.2 & 06:58:32.50 & $-$55:56:34.12 & 3.537\\
 15.3 & 06:58:31.80 & $-$55:56:43.21 & 3.537\\
\hline
\end{supertabular}

\bottomcaption{Same as Table\,\ref{tab:multiples} for the MACS0940 mass model.}
\begin{supertabular}{lllll}
ID & $\alpha$ (J2000) & $\delta$ (J2000) & $z_{\rm system}$ \\
   & h  m  s & $^{\circ}$  $'$  $''$ & \\
\hline 
 1.1 & 09:40:53.68 & 07:44:15.69 & 4.03\\
 1.2 & 09:40:53.16 & 07:44:20.03 & 4.03\\
 1.3 & 09:40:54.18 & 07:44:19.33 & 4.03\\
 1.4 & 09:40:53.88 & 07:44:38.62 & 4.03\\
 3.1 & 09:40:53.69 & 07:44:15.61 & 4.03\\
 3.2 & 09:40:53.13 & 07:44:20.79 & 4.03\\
 4.1 & 09:40:53.59 & 07:44:15.79 & 4.03\\
 4.2 & 09:40:53.23 & 07:44:18.56 & 4.03\\
 2.1 & 09:40:54.27 & 07:44:34.07 & 5.70\\
 2.2 & 09:40:53.86 & 07:44:12.91 & 5.70\\
 2.3 & 09:40:53.14 & 07:44:28.73 & 5.70\\
 2.4 & 09:40:54.42 & 07:44:29.01 & 5.70\\
 2.5 & 09:40:54.52 & 07:44:28.81 & 5.70\\
 5.1 & 09:40:53.18 & 07:44:30.18 & 5.70\\
 5.2 & 09:40:53.80 & 07:44:12.39 & 5.70\\
 6.2 & 09:40:53.00 & 07:44:32.25 & 5.491\\
 6.3 & 09:40:52.95 & 07:44:19.61 & 5.491\\
 6.4 & 09:40:54.18 & 07:44:25.18 & 5.491\\
 7.1 & 09:40:53.38 & 07:44:36.05 & 3.57\\
 7.2 & 09:40:54.26 & 07:44:24.73 & 3.57\\
 7.3 & 09:40:53.18 & 07:44:17.03 & 3.57\\
 7.4 & 09:40:52.98 & 07:44:24.30 & 3.57\\
\hline
\end{supertabular}

\bottomcaption{Same as Table\, \ref{tab:multiples} for the MACS1206mass model.}
\begin{supertabular}{lllll}
ID & $\alpha$ (J2000) & $\delta$ (J2000) & $z_{\rm system}$ \\
   & h  m  s & $^{\circ}$  $'$  $''$ & \\
\hline 
 1.1 & 12:06:12.22 & $-$08:47:50.72 & 1.0121\\
 1.2 & 12:06:11.90 & $-$08:47:57.46 & 1.0121\\
 1.3 & 12:06:11.73 & $-$08:48:23.96 & 1.0121\\
 2.1 & 12:06:11.23 & $-$08:47:44.45 & 1.0369\\
 2.2 & 12:06:10.76 & $-$08:47:58.39 & 1.0369\\
 2.3 & 12:06:10.82 & $-$08:48:10.99 & 1.0369\\
 3.1 & 12:06:12.14 & $-$08:47:44.04 & 1.0433\\
 3.2 & 12:06:11.43 & $-$08:47:59.32 & 1.0433\\
 3.3 & 12:06:11.67 & $-$08:48:19.01 & 1.0433\\
 4.1 & 12:06:12.72 & $-$08:47:40.92 & 1.4248\\
 4.2 & 12:06:11.72 & $-$08:48:00.21 & 1.4248\\
 4.3 & 12:06:11.94 & $-$08:48:28.67 & 1.4248\\
 5.1 & 12:06:12.85 & $-$08:47:42.68 & 1.4254\\
 5.2 & 12:06:13.02 & $-$08:48:05.59 & 1.4254\\
 5.3 & 12:06:12.00 & $-$08:48:29.15 & 1.4254\\
 6.1 & 12:06:11.99 & $-$08:47:46.90 & 1.4255\\
 6.2 & 12:06:11.55 & $-$08:47:49.41 & 1.4255\\
 6.3 & 12:06:11.53 & $-$08:48:33.42 & 1.4255\\
 7.1 & 12:06:12.14 & $-$08:47:44.53 & 1.4257\\
 7.2 & 12:06:11.33 & $-$08:47:52.79 & 1.4257\\
 7.3 & 12:06:12.32 & $-$08:48:01.18 & 1.4257\\
 7.4 & 12:06:12.53 & $-$08:48:01.43 & 1.4257\\
 7.5 & 12:06:11.59 & $-$08:48:31.88 & 1.4257\\
 8.1 & 12:06:12.88 & $-$08:47:44.72 & 1.4864\\
 8.2 & 12:06:13.09 & $-$08:48:03.97 & 1.4864\\
 8.3 & 12:06:11.99 & $-$08:48:31.99 & 1.4864\\
 9.1 & 12:06:11.22 & $-$08:47:35.32 & 1.9600\\
 9.2 & 12:06:10.39 & $-$08:47:52.12 & 1.9600\\
 9.3 & 12:06:10.65 & $-$08:48:26.95 & 1.9600\\
 10.1 & 12:06:12.59 & $-$08:47:42.00 & 2.5393\\
 10.2 & 12:06:11.18 & $-$08:47:50.88 & 2.5393\\
 10.3 & 12:06:12.12 & $-$08:47:59.85 & 2.5393\\
 10.4 & 12:06:13.17 & $-$08:48:00.58 & 2.5393\\
 10.5 & 12:06:11.72 & $-$08:48:42.53 & 2.5393\\
 11.1 & 12:06:15.04 & $-$08:47:48.02 & 3.0358\\
 11.2 & 12:06:15.00 & $-$08:48:17.68 & 3.0358\\
 11.3 & 12:06:14.54 & $-$08:48:32.36 & 3.0358\\
 12.1 & 12:06:11.67 & $-$08:47:37.38 & 3.3890\\
 12.2 & 12:06:11.07 & $-$08:47:43.39 & 3.3890\\
 12.3 & 12:06:12.78 & $-$08:48:00.71 & 3.3890\\
 13.1 & 12:06:11.86 & $-$08:48:06.36 & 3.3961\\
 13.2 & 12:06:11.49 & $-$08:48:07.61 & 3.3961\\
 13.3 & 12:06:12.90 & $-$08:48:17.13 & 3.3961\\
 13.4 & 12:06:12.33 & $-$08:48:24.71 & 3.3961\\
 14.1 & 12:06:15.97 & $-$08:48:16.13 & 3.7531\\
 14.2 & 12:06:15.95 & $-$08:48:17.04 & 3.7531\\
 14.3 & 12:06:15.95 & $-$08:48:18.53 & 3.7531\\
 14.4 & 12:06:15.91 & $-$08:48:22.78 & 3.7531\\
 14.5 & 12:06:15.74 & $-$08:48:27.69 & 3.7531\\
 15.1 & 12:06:13.43 & $-$08:47:29.89 & 3.7611\\
 15.2 & 12:06:13.82 & $-$08:48:11.00 & 3.7611\\
 15.3 & 12:06:12.42 & $-$08:48:39.47 & 3.7611\\
 16.1 & 12:06:13.10 & $-$08:47:28.33 & 3.7617\\
 16.2 & 12:06:11.15 & $-$08:47:58.82 & 3.7617\\
 16.3 & 12:06:13.56 & $-$08:48:08.90 & 3.7617\\
 17.1 & 12:06:13.47 & $-$08:47:44.23 & 3.8224\\
 17.2 & 12:06:13.67 & $-$08:47:57.92 & 3.8224\\
 18.1 & 12:06:13.29 & $-$08:47:48.17 & 4.0400\\
 18.2 & 12:06:13.42 & $-$08:47:54.94 & 4.0400\\
 19.1 & 12:06:14.90 & $-$08:47:41.55 & 4.0520\\
 19.2 & 12:06:14.85 & $-$08:48:18.86 & 4.0520\\
 19.3 & 12:06:14.35 & $-$08:48:34.07 & 4.0520\\
 20.1 & 12:06:13.63 & $-$08:48:13.39 & 4.0553\\
 20.2 & 12:06:11.42 & $-$08:48:01.80 & 4.0553\\
 20.3 & 12:06:12.35 & $-$08:48:36.24 & 4.0553\\
 20.4 & 12:06:13.26 & $-$08:47:25.47 & 4.0553\\
 21.1 & 12:06:10.42 & $-$08:47:51.63 & 4.0718\\
 21.2 & 12:06:12.35 & $-$08:48:02.93 & 4.0718\\
 21.3 & 12:06:12.68 & $-$08:48:04.01 & 4.0718\\
 21.4 & 12:06:12.82 & $-$08:48:04.33 & 4.0718\\
 22.1 & 12:06:10.64 & $-$08:47:29.11 & 4.2913\\
 22.2 & 12:06:09.67 & $-$08:47:47.62 & 4.2913\\
 22.3 & 12:06:09.81 & $-$08:48:21.94 & 4.2913\\
 23.1 & 12:06:15.18 & $-$08:47:48.81 & 4.7293\\
 23.2 & 12:06:15.25 & $-$08:48:13.21 & 4.7293\\
 23.3 & 12:06:14.36 & $-$08:48:41.49 & 4.7293\\
 24.1 & 12:06:12.33 & $-$08:47:28.68 & 5.6984\\
 24.2 & 12:06:10.58 & $-$08:47:49.54 & 5.6984\\
 24.3 & 12:06:12.22 & $-$08:48:02.18 & 5.6984\\
 24.4 & 12:06:13.25 & $-$08:48:03.84 & 5.6984\\
 25.1 & 12:06:14.33 & $-$08:47:47.62 & 5.7927\\
 25.2 & 12:06:14.42 & $-$08:48:00.64 & 5.7927\\
 26.1 & 12:06:12.17 & $-$08:48:11.20 & 6.0106\\
 26.2 & 12:06:12.29 & $-$08:48:13.21 & 6.0106\\
 27.1 & 12:06:12.78 & $-$08:47:55.77 & 6.0601\\
 27.2 & 12:06:12.41 & $-$08:47:57.00 & 6.0601\\
 28.1 & 12:06:14.43 & $-$08:48:34.20 & 3.0385\\
 28.2 & 12:06:15.00 & $-$08:48:16.50 & 3.0385\\
 28.3 & 12:06:15.01 & $-$08:47:48.65 & 3.0385\\
 29.1 & 12:06:15.48 & $-$08:47:53.74 & 5.0174\\
 29.2 & 12:06:15.51 & $-$08:48:13.01 & 5.0174\\
 29.3 & 12:06:14.83 & $-$08:48:38.58 & 5.0174\\
 30.1 & 12:06:11.36 & $-$08:48:13.32 & 4.104\\
 30.2 & 12:06:12.04 & $-$08:48:08.13 & 4.104\\
 30.3 & 12:06:12.02 & $-$08:48:22.54 & 4.104\\
 30.4 & 12:06:12.64 & $-$08:48:15.04 & 4.104\\
 31.1 & 12:06:14.48 & $-$08:48:09.95 & 3.011\\
 31.2 & 12:06:14.44 & $-$08:47:44.74 & 3.011\\
 32.1 & 12:06:10.25 & $-$08:47:49.28 & 3.161\\
 32.2 & 12:06:11.25 & $-$08:47:31.25 & 3.161\\
 33.1 & 12:06:13.75 & $-$08:48:01.14 & 3.207\\
 33.2 & 12:06:13.55 & $-$08:47:42.58 & 3.207\\
 34.1 & 12:06:13.40 & $-$08:47:59.16 & 4.047\\
 34.2 & 12:06:12.83 & $-$08:47:40.02 & 4.047\\
 34.3 & 12:06:11.16 & $-$08:47:48.46 & 4.047\\
 35.1 & 12:06:10.46 & $-$08:47:48.02 & 4.050\\
 35.2 & 12:06:11.68 & $-$08:47:31.32 & 4.050\\
 36.1 & 12:06:14.09 & $-$08:47:46.35 & 2.504\\
 36.2 & 12:06:14.17 & $-$08:48:05.13 & 2.504\\
 36.3 & 12:06:12.90 & $-$08:48:41.00 & 2.504\\
 37.1 & 12:06:13.66 & $-$08:48:06.28 & 1.69$\pm$0.01\\
 37.2 & 12:06:13.62 & $-$08:47:45.85 & '' \\
 37.3 & 12:06:12.59 & $-$08:48:34.06 & '' \\
\hline
\end{supertabular}

\bottomcaption{Same as Table\,\ref{tab:multiples} for the RXJ1347 mass model.}
\begin{supertabular}{lllll}
ID & $\alpha$ (J2000) & $\delta$ (J2000) & $z_{\rm system}$ \\
   & h  m  s & $^{\circ}$  $'$  $''$ & \\
\hline 
 1.1 & 13:47:29.09 & $-$11:45:37.22 & 1.75\\
 1.2 & 13:47:29.44 & $-$11:45:41.93 & 1.75\\
 1.3 & 13:47:31.75 & $-$11:45:51.93 & 1.75\\
 2.1 & 13:47:29.37 & $-$11:44:47.16 & 4.086\\
 2.2 & 13:47:29.05 & $-$11:44:53.93 & 4.086\\
 3.1 & 13:47:32.18 & $-$11:44:30.32 & 3.67\\
 3.2 & 13:47:31.89 & $-$11:44:28.28 & 3.67\\
 4.1 & 13:47:30.80 & $-$11:44:56.89 & 2.32$\pm$0.20\\
 4.2 & 13:47:30.78 & $-$11:44:58.73 & ''\\
 9.1 & 13:47:32.99 & $-$11:45:27.09 & 2.127\\
 9.2 & 13:47:32.40 & $-$11:44:54.13 & 2.127\\
 9.3 & 13:47:30.88 & $-$11:45:12.04 & 2.127\\
 9.4 & 13:47:30.61 & $-$11:45:33.74 & 2.127\\
 9.5 & 13:47:28.68 & $-$11:44:50.44 & 2.127\\
 11.1 & 13:47:31.95 & $-$11:45:44.01 & 4.847\\
 11.2 & 13:47:31.30 & $-$11:45:45.86 & 4.847\\
 11.3 & 13:47:27.74 & $-$11:44:54.61 & 4.847\\
 11.4 & 13:47:32.56 & $-$11:44:58.84 & 4.847\\
 12.1 & 13:47:32.48 & $-$11:44:51.49 & 3.704\\
 12.2 & 13:47:28.34 & $-$11:44:49.73 & 3.704\\
 12.3 & 13:47:30.97 & $-$11:45:10.86 & 3.704\\
 12.4 & 13:47:33.03 & $-$11:45:33.84 & 3.704\\
 12.5 & 13:47:30.49 & $-$11:45:40.93 & 3.704\\
 13.1 & 13:47:30.57 & $-$11:45:49.96 & 3.705\\
 13.2 & 13:47:31.95 & $-$11:45:49.48 & 3.705\\
 13.3 & 13:47:31.64 & $-$11:45:05.02 & 3.705\\
 13.4 & 13:47:31.92 & $-$11:44:59.19 & 3.705\\
 14.1 & 13:47:32.40 & $-$11:44:38.89 & 3.681\\
 14.2 & 13:47:30.40 & $-$11:45:25.21 & 3.681\\
 14.3 & 13:47:29.99 & $-$11:44:29.41 & 3.681\\
 14.4 & 13:47:33.85 & $-$11:45:12.80 & 3.681\\
 14.5 & 13:47:30.62 & $-$11:45:15.45 & 3.681\\
 15.1 & 13:47:28.74 & $-$11:44:38.53 & 3.993\\
 15.2 & 13:47:32.78 & $-$11:44:49.53 & 3.993\\
 15.3 & 13:47:33.48 & $-$11:45:20.78 & 3.993\\
 16.1 & 13:47:33.12 & $-$11:45:10.34 & 4.7445\\
 16.2 & 13:47:33.11 & $-$11:45:16.64 & 4.7445\\
 16.3 & 13:47:31.20 & $-$11:45:38.10 & 4.7445\\
 16.4 & 13:47:28.04 & $-$11:44:40.96 & 4.7445\\
 16.5 & 13:47:31.55 & $-$11:45:15.20 & 4.7445\\
 17.1 & 13:47:29.41 & $-$11:45:15.20 & 5.008\\
 17.2 & 13:47:34.36 & $-$11:45:35.71 & 5.008\\
 18.1 & 13:47:29.63 & $-$11:45:58.79 & 4.947\\
 18.2 & 13:47:27.95 & $-$11:45:37.01 & 4.947\\
 18.3 & 13:47:30.56 & $-$11:46:03.67 & 4.947\\
 19.1 & 13:47:30.25 & $-$11:45:01.97 & 4.038\\
 19.2 & 13:47:30.12 & $-$11:44:59.61 & 4.038\\
 19.3 & 13:47:34.20 & $-$11:45:45.19 & 4.038\\
 20.1 & 13:47:34.15 & $-$11:45:47.92 & 5.769\\
 20.2 & 13:47:29.97 & $-$11:44:53.38 & 5.769\\
 20.3 & 13:47:30.44 & $-$11:45:04.11 & 5.769\\
 21.1 & 13:47:29.46 & $-$11:45:53.84 & 4.384\\
 21.2 & 13:47:31.34 & $-$11:46:01.89 & 4.384\\
 21.3 & 13:47:28.12 & $-$11:45:35.17 & 4.384\\
 22.1 & 13:47:29.19 & $-$11:45:25.58 & 3.995\\
 22.2 & 13:47:28.81 & $-$11:45:05.24 & 3.995\\
 23.1 & 13:47:29.93 & $-$11:45:44.49 & 4.382\\
 23.2 & 13:47:32.93 & $-$11:45:44.47 & 4.382\\
 23.3 & 13:47:31.92 & $-$11:44:49.06 & 4.382\\
 23.4 & 13:47:28.16 & $-$11:45:05.73 & 4.382\\
 23.5 & 13:47:30.82 & $-$11:45:08.63 & 4.382\\
 24.1 & 13:47:33.45 & $-$11:45:43.01 & 3.7508\\
 24.2 & 13:47:29.31 & $-$11:45:33.04 & 3.7508\\
 24.3 & 13:47:28.61 & $-$11:45:12.69 & 3.7508\\
 24.4 & 13:47:31.12 & $-$11:44:43.76 & 3.7508\\
 25.1 & 13:47:33.15 & $-$11:45:06.29 & 4.084\\
 25.2 & 13:47:33.19 & $-$11:45:18.11 & 4.084\\
 25.3 & 13:47:30.98 & $-$11:45:36.92 & 4.084\\
 25.4 & 13:47:28.25 & $-$11:44:40.74 & 4.084\\
 25.5 & 13:47:31.35 & $-$11:45:15.83 & 4.084\\
 26.1 & 13:47:32.84 & $-$11:45:49.78 & 3.5675\\
 26.2 & 13:47:31.35 & $-$11:44:52.65 & 3.5675\\
 26.3 & 13:47:30.86 & $-$11:45:04.19 & 3.5675\\
 26.4 & 13:47:28.48 & $-$11:45:25.39 & 3.5675\\
 26.5 & 13:47:29.29 & $-$11:45:40.45 & 3.5675\\
 27.1 & 13:47:30.31 & $-$11:45:15.90 & 5.723\\
 27.2 & 13:47:34.45 & $-$11:45:19.92 & 5.723\\
 28.1 & 13:47:28.93 & $-$11:44:58.18 & 4.086\\
 28.2 & 13:47:33.83 & $-$11:45:37.79 & 4.086\\
 28.3 & 13:47:29.45 & $-$11:44:46.58 & 4.086\\
 28.4 & 13:47:29.25 & $-$11:45:24.99 & 4.086\\
 29.1 & 13:47:32.53 & $-$11:45:09.31 & 3.703\\
 29.2 & 13:47:27.61 & $-$11:44:48.12 & 3.703\\
 30.1 & 13:47:31.70 & $-$11:45:35.05 & 3.065\\
 30.2 & 13:47:32.42 & $-$11:45:28.92 & 3.065\\
 31.1 & 13:47:32.16 & $-$11:45:02.80 & 5.421\\
 31.2 & 13:47:27.39 & $-$11:45:05.62 & 5.421\\
 32.1 & 13:47:30.82 & $-$11:45:47.90 & 4.163\\
 32.2 & 13:47:32.12 & $-$11:45:45.86 & 4.163\\
 32.3 & 13:47:27.82 & $-$11:45:01.63 & 4.163\\
 32.4 & 13:47:31.65 & $-$11:45:08.79 & 4.163\\
 32.5 & 13:47:32.29 & $-$11:44:57.20 & 4.163\\
 33.1 & 13:47:30.85 & $-$11:45:36.93 & 5.673\\
 33.2 & 13:47:33.28 & $-$11:45:04.12 & 5.673\\
 33.3 & 13:47:31.17 & $-$11:45:16.17 & 5.673\\
 33.4 & 13:47:28.32 & $-$11:44:37.88 & 5.673\\
 33.5 & 13:47:33.36 & $-$11:45:10.41 & 5.673\\
 34.1 & 13:47:33.10 & $-$11:45:51.63 & 4.947\\
 34.2 & 13:47:31.20 & $-$11:44:50.07 & 4.947\\
 34.3 & 13:47:30.80 & $-$11:45:04.04 & 4.947\\
 35.1 & 13:47:31.97 & $-$11:45:22.65 & 6.567\\
 35.2 & 13:47:31.89 & $-$11:45:20.18 & 6.567\\
 35.3 & 13:47:27.58 & $-$11:44:35.04 & 6.567\\
 36.1 & 13:47:32.31 & $-$11:44:28.97 & 4.038\\
 36.2 & 13:47:32.00 & $-$11:44:26.49 & 4.038\\
 37.1 & 13:47:33.21 & $-$11:44:56.34 & 4.298\\
 37.2 & 13:47:33.43 & $-$11:45:05.69 & 4.298\\
 38.1 & 13:47:30.97 & $-$11:44:22.23 & 3.682\\
 38.2 & 13:47:33.17 & $-$11:44:40.38 & 3.682\\
 38.3 & 13:47:33.59 & $-$11:44:47.70 & 3.682\\
 39.1 & 13:47:31.08 & $-$11:44:23.41 & 4.660\\
 39.2 & 13:47:32.74 & $-$11:44:34.97 & 4.660\\
 39.3 & 13:47:33.97 & $-$11:44:58.24 & 4.660\\
 40.1 & 13:47:32.38 & $-$11:44:56.52 & 4.835\\
 40.2 & 13:47:32.23 & $-$11:45:44.98 & 4.835\\
 40.3 & 13:47:27.77 & $-$11:44:59.74 & 4.835\\
 40.4 & 13:47:30.82 & $-$11:45:47.90 & 4.835\\
 40.5 & 13:47:31.63 & $-$11:45:09.76 & 4.835\\
 41.1 & 13:47:30.76 & $-$11:44:52.35 & 4.875\\
 41.2 & 13:47:30.70 & $-$11:44:59.87 & 4.875\\
 41.3 & 13:47:33.33 & $-$11:45:54.54 & 4.875\\
\hline
\end{supertabular}

\bottomcaption{Same as Table\, \ref{tab:multiples} for the SMACS2031 mass model.}
\begin{supertabular}{lllll}
ID & $\alpha$ (J2000) & $\delta$ (J2000) & $z_{\rm system}$ \\
   & h  m  s & $^{\circ}$  $'$  $''$ & \\
\hline 
 1.1 & 20:31:52.90 & $-$40:37:32.62 & 3.5077 & \\
 1.2 & 20:31:52.99 & $-$40:37:32.60 & 3.5077 & \\
 1.3 & 20:31:53.06 & $-$40:37:45.95 & 3.5077 & \\
 1.4 & 20:31:53.83 & $-$40:37:40.21 & 3.5077 & \\
 1.5 & 20:31:53.93 & $-$40:37:06.17 & 3.5077 & \\
 2.1 & 20:31:52.75 & $-$40:37:30.87 & 3.5077 & \\
 2.3 & 20:31:52.91 & $-$40:37:48.49 & 3.5077 & \\
 2.4 & 20:31:53.99 & $-$40:37:38.93 & 3.5077 & \\
 2.5 & 20:31:53.91 & $-$40:37:08.98 & 3.5077 & \\
 3.1 & 20:31:52.53 & $-$40:37:34.18 & 5.6231 & \\
 3.2 & 20:31:53.14 & $-$40:37:33.60 & 5.6231 & \\
 3.3 & 20:31:52.81 & $-$40:37:44.77 & 5.6231 & \\
 3.4 & 20:31:53.72 & $-$40:37:40.25 & 5.6231 & \\
 3.5 & 20:31:53.77 & $-$40:37:01.01 & 5.6231 & \\
 4.1 & 20:31:52.14 & $-$40:37:30.42 & 3.34 & \\
 4.2 & 20:31:52.22 & $-$40:37:45.82 & 3.34 & \\
 4.3 & 20:31:53.40 & $-$40:37:07.02 & 3.34 & \\
 5.1 & 20:31:50.92 & $-$40:37:36.22 & 3.723 & \\
 5.2 & 20:31:51.11 & $-$40:37:26.63 & 3.723 & \\
 5.3 & 20:31:52.12 & $-$40:37:02.62 & 3.723 & \\
 6.1 & 20:31:51.46 & $-$40:37:19.01 & 1.4249 & \\
 6.2 & 20:31:51.32 & $-$40:37:22.89 & 1.4249 & \\
 6.3 & 20:31:51.17 & $-$40:37:26.54 & 1.4249 & \\
 7.1 & 20:31:52.83 & $-$40:37:06.17 & 5.2397 & \\
 7.2 & 20:31:51.96 & $-$40:37:24.27 & 5.2397 & \\
 7.3 & 20:31:51.71 & $-$40:37:47.67 & 5.2397 & \\
 8.1 & 20:31:51.76 & $-$40:37:08.68 & 5.6128 & \\
 8.2 & 20:31:51.51 & $-$40:37:14.02 & 5.6128 & \\
 8.3 & 20:31:50.69 & $-$40:37:44.03 & 5.6128 & \\
 9.1 & 20:31:51.19 & $-$40:37:13.22 & 6.4085 & \\
 9.2 & 20:31:51.51 & $-$40:37:06.08 & 6.4085 & \\
 10.1 & 20:31:52.18 & $-$40:37:12.50 & 3.8561 & \\
 10.2 & 20:31:52.06 & $-$40:37:15.01 & 3.8561 & \\
 10.3 & 20:31:51.22 & $-$40:37:46.18 & 3.8561 & \\
 11.1 & 20:31:52.31 & $-$40:37:48.19 & 2.2556 & \\
 11.2 & 20:31:52.42 & $-$40:37:26.60 & 2.2556 & \\
 11.3 & 20:31:53.27 & $-$40:37:14.12 & 2.2556 & \\
 11.4 & 20:31:53.62 & $-$40:37:31.75 & 2.2556 & \\
 12.1 & 20:31:53.65 & $-$40:37:30.45 & 3.414 & \\
 12.2 & 20:31:53.42 & $-$40:37:30.37 & 3.414 & \\
 12.3 & 20:31:52.46 & $-$40:37:23.71 & 3.414 & \\
 12.4 & 20:31:53.13 & $-$40:37:14.55 & 3.414 & \\
 12.5 & 20:31:52.16 & $-$40:37:51.47 & 3.414 & \\
 13.1 & 20:31:52.40 & $-$40:37:05.46 & 4.73 & \\
 13.2 & 20:31:51.61 & $-$40:37:23.02 & 4.73 & \\
 13.3 & 20:31:51.22 & $-$40:37:43.62 & 4.73 & \\
\hline
\end{supertabular}

\bottomcaption{Same as Table\,\ref{tab:multiples} for the SMACS2131 mass model. \label{tab:multiples2}.}
\begin{supertabular}{lllll}
ID & $\alpha$ (J2000) & $\delta$ (J2000) & $z_{\rm system}$ \\
   & h  m  s & $^{\circ}$  $'$  $''$ & \\
\hline 
 1.1 & 21:31:02.97 & $-$40:19:04.59 & 5.718\\
 1.3 & 21:31:03.82 & $-$40:18:54.54 & 5.718\\
 1.4 & 21:31:02.44 & $-$40:19:12.32 & 5.718\\
 11.1 & 21:31:06.09 & $-$40:19:43.48 & 5.718\\
 11.4 & 21:31:06.19 & $-$40:19:05.71 & 5.718\\
 11.5 & 21:31:05.53 & $-$40:18:57.28 & 5.718\\
 2.1 & 21:31:05.27 & $-$40:19:31.69 & 3.478\\
 2.2 & 21:31:04.94 & $-$40:19:24.05 & 3.478\\
 2.3 & 21:31:05.41 & $-$40:19:33.34 & 3.478\\
 3.1 & 21:31:04.22 & $-$40:19:32.26 & 0.994\\
 3.2 & 21:31:05.23 & $-$40:19:17.77 & 0.994\\
 4.1 & 21:31:03.51 & $-$40:19:23.78 & 4.971\\
 4.2 & 21:31:05.99 & $-$40:18:54.20 & 4.971\\
 5.1 & 21:31:03.60 & $-$40:19:23.78 & 3.353\\
 5.2 & 21:31:05.91 & $-$40:18:55.75 & 3.353\\
 6.1 & 21:31:04.61 & $-$40:19:22.99 & 1.265\\
 6.2 & 21:31:04.04 & $-$40:19:25.39 & 1.265\\
 6.3 & 21:31:05.37 & $-$40:19:04.15 & 1.265\\
 7.1 & 21:31:06.62 & $-$40:19:25.57 & 4.820\\
 7.2 & 21:31:05.93 & $-$40:19:38.85 & 4.820\\
 7.3 & 21:31:03.90 & $-$40:19:38.43 & 4.820\\
 7.4 & 21:31:04.64 & $-$40:18:47.95 & 4.820\\
 8.1 & 21:31:03.57 & $-$40:19:25.65 & 4.45465\\
 8.2 & 21:31:05.75 & $-$40:18:51.28 & 4.45465\\
 10.1 & 21:31:05.02 & $-$40:18:55.25 & 4.59\\
 10.2 & 21:31:06.41 & $-$40:19:12.85 & 4.59\\
 10.3 & 21:31:05.10 & $-$40:19:44.03 & 4.59\\
 10.4 & 21:31:02.53 & $-$40:19:13.17 & 4.59\\
\hline
\end{supertabular}

\bottomcaption{Same as Table\, \ref{tab:multiples} for the MACS2214 mass model.}
\begin{supertabular}{lllll}
ID & $\alpha$ (J2000) & $\delta$ (J2000) & $z_{\rm system}$ \\
   & h  m  s & $^{\circ}$  $'$  $''$ & \\
\hline 
 1.1 & 22:14:57.76 & -14:00:01.08 & 3.01\\
 1.2 & 22:14:56.93 & -13:59:59.09 & 3.01\\
 2.1 & 22:14:57.51 & -13:59:59.61 & 3.01\\
 2.2 & 22:14:57.33 & -13:59:59.21 & 3.01\\
 3.1 & 22:14:57.64 & -13:59:59.40 & 3.01\\
 3.2 & 22:14:57.24 & -13:59:58.49 & 3.01\\
 3.3 & 22:14:56.71 & -14:00:12.99 & 3.01\\
 3.4 & 22:14:58.68 & -14:00:36.15 & 3.01\\
 4.1 & 22:14:56.66 & -13:59:56.27 & 3.14\\
 4.2 & 22:14:58.02 & -14:00:02.88 & 3.14\\
 4.3 & 22:14:58.55 & -14:00:33.29 & 3.14\\
 4.4 & 22:14:56.57 & -14:00:17.13 & 3.14\\
 5.1 & 22:14:56.82 & -13:59:44.09 & 6.62\\
 5.2 & 22:14:58.97 & -14:00:20.88 & 6.62\\
 5.3 & 22:14:58.78 & -14:00:02.62 & 6.62\\
 6.1 & 22:14:57.14 & -14:00:13.82 & 1.16\\
 6.2 & 22:14:57.98 & -14:00:23.11 & 1.16\\
 6.3 & 22:14:56.91 & -14:00:02.38 & 1.16\\
 6.4 & 22:14:57.57 & -14:00:09.79 & 1.16\\
 7.1 & 22:14:56.10 & -14:00:26.63 & 2.95\\
 7.2 & 22:14:56.94 & -14:00:38.93 & 2.95\\
 7.3 & 22:14:55.77 & -14:00:02.34 & 2.95\\
 8.1 & 22:14:57.05 & -14:00:32.79 & 2.98\\
 8.2 & 22:14:56.67 & -14:00:30.13 & 2.98\\
 8.3 & 22:14:56.83 & -14:00:31.70 & 2.98\\
 9.1 & 22:14:59.07 & -14:00:24.76 & 3.66\\
 9.2 & 22:14:57.51 & -13:59:48.53 & 3.66\\
 9.3 & 22:14:58.53 & -13:59:58.73 & 3.66\\
 10.1 & 22:14:58.54 & -14:00:00.22 & 4.34\\
 10.2 & 22:14:58.92 & -14:00:25.92 & 4.34\\
 10.3 & 22:14:57.10 & -13:59:47.76 & 4.34\\
 11.1 & 22:14:57.08 & -14:00:21.99 & 4.81\\
 11.2 & 22:14:58.60 & -14:00:19.38 & 4.81\\
 11.3 & 22:14:58.63 & -14:00:11.39 & 4.81\\
 12.1 & 22:14:58.69 & -14:00:23.10 & 5.09\\
 12.2 & 22:14:56.91 & -14:00:21.26 & 5.09\\
 12.3 & 22:14:58.59 & -14:00:06.36 & 5.09\\
 12.4 & 22:14:56.54 & -13:59:45.93 & 5.09\\
 13.1 & 22:14:55.94 & -14:00:11.59 & 6.47\\
 13.2 & 22:14:55.99 & -14:00:07.27 & 6.47\\
 15.1 & 22:14:56.78 & -14:00:22.90 & 5.57\\
 15.2 & 22:14:58.54 & -14:00:06.24 & 5.57\\
 15.3 & 22:14:58.43 & -14:00:25.76 & 5.57\\
 15.4 & 22:14:56.47 & -13:59:46.44 & 5.57\\
 17.4 & 22:14:56.56 & -14:00:33.65 & 2.95\\
 17.5 & 22:14:56.36 & -14:00:30.77 & 2.95\\
\hline
\end{supertabular}

\end{appendix}

%%%%%%%%%%%%%%%%%%%%%%%%%%%%%%%%%%%%%%%%%%%%%%%%%%

%%%%%%%%%%%%%%%%% APPENDICES %%%%%%%%%%%%%%%%%%%%%

%\input{appendices.tex}

%%%%%%%%%%%%%%%%%%%%%%%%%%%%%%%%%%%%%%%%%%%%%%%%%%
%
%-------------------------------------------------------------------

\end{document}